\documentclass[useAMS,usenatbib,usegraphicx]{mn2e}
\usepackage[T1]{fontenc}
\usepackage[latin1]{inputenc}
\usepackage{color}
\newcounter{subfigure}

\title[IFU spectroscopy of 10 ETG nuclei: II]
  {IFU spectroscopy of 10 early type galactic nuclei: II - Nuclear emission line properties}
\author[Ricci et al.]
  {T.V.~Ricci,$^1$\thanks{tvricci@iag.usp.br}
  J.E.~Steiner,$^1$ R.B.~Menezes$^1$ \\
  $^1$Instituto de Astronomia, Geof\'isica e Ci\^encias Atmosf\'ericas, Universidade de S\~ao Paulo, 05508-900, S\~ao Paulo, Brasil }
\date{Released 2002 Xxxxx XX}

\pagerange{\pageref{firstpage}--\pageref{lastpage}} \pubyear{2002}

\def\LaTeX{L\kern-.36em\raise.3ex\hbox{a}\kern-.15em
    T\kern-.1667em\lower.7ex\hbox{E}\kern-.125emX}

\begin{document}

\label{firstpage}

\maketitle

\begin{abstract}
Although it is well known that massive galaxies have central black holes, most of them accreting at low Eddington ratios, many important questions still remain open. Among them, are the nature of the ionizing source, the characteristics and frequencies of the broad line region and of the dusty torus. We report observations of 10 early-type galactic nuclei, observed with the IFU/GMOS spectrograph on the Gemini South telescope, analysed with standard techniques for spectral treatment and compared with results obtained with principal component analysis Tomography (Paper I).  We performed spectral synthesis of each spaxel of the data cubes and subtracted the stellar component from the original cube, leaving a data cube with emission lines only.  The emission lines were decomposed in multi-Gaussian components. We show here that, for eight galaxies previously known to have emission lines, the narrow line region can be decomposed in two components with distinct line widths. In addition to this, broad H$\alpha$ emission was detected in six galaxies. The two galaxies not previously known to have emission lines show weak H$\alpha$+[N II] lines. All 10 galaxies may be classified as low-ionization nuclear emission regions in diagnostic diagrams and seven of them have bona fide active galactic nuclei with luminosities between 10$^{40}$ and 10$^{43}$ erg s$^{-1}$.  Eddington ratios are always $<$ 10$^{-3}$. 
\end{abstract}

\begin{keywords}
Techniques: imaging spectroscopy - galaxies: active - galaxies: elliptical and lenticular, cD - galaxies: nuclei - galaxies: Seyfert - galaxies: stellar content
\end{keywords}

\section{Introduction} \label{sec:intro}

Nuclear gas emission emerging from early-type galaxies (ETGs) in the local Universe is associated with low-luminosity active galactic nuclei (LLAGN) in 2/3 of the cases \citep{2008ARA&A..46..475H}, where most of these objects are classified as low ionization nuclear emission regions (LINERs; \citealt{1980A&A....87..152H}). In order to distinguish, at optical wavelengths, between LINERs, Seyfert galaxies, starburst regions and transition objects (TOs), the so-called diagnostic diagrams or BPT diagrams \citep{1981PASP...93....5B}, which compare different emission line ratios, are commonly used. Most recent diagnostic diagrams (e.g., \citealt{1997ApJS..112..315H,2003MNRAS.346.1055K,2006MNRAS.372..961K}) compare the [O III]$\lambda$5007/H$\beta$ ratio with the [O I]$\lambda$6300/H$\alpha$, [N II]$\lambda$6581/H$\alpha$ and ([S II]$\lambda$6716 + [S II]$\lambda$6731)/H$\alpha$ line ratios. The reason for such choices is that these ratios are composed of emission lines that are close enough in wavelength, and therefore, are less affected by dust reddening effects \citep{1987ApJS...63..295V}.  

Photoionization of LINERs by AGNs was proposed by \citet{1983ApJ...264..105F} and \citet{1983ApJ...269L..37H}. However, other mechanisms may be responsible for LINER emission (e.g. shockwaves - \citealt{1980A&A....87..152H}; post-asymptotic giant branch star populations (pAGBs) - \citealt{1994A&A...292...13B}). In the optical, detection of a broad-line region (BLR), specially in the H$\alpha$ line, is a typical feature of AGNs \citep{1997ApJS..112..391H,2008ARA&A..46..475H}. However, the BLR is not detected in several LINERs \citep{1997ApJS..112..391H}. One hypothesis is that the BLR is very weak in these objects and demands an accurate subtraction of the star light from the spectra \citep{1997ApJS..112..391H}. Although there are no reasons to discard the unified models \citep{1993ARA&A..31..473A} in LINERs (see, for instance, \citealt{1999ApJ...525..673B}), BLRs seem to be intrinsically absent, even if their radio or X-ray emissions are typically non-stellar \citep{2008ARA&A..46..475H}. Because of the low bolometric luminosities and Eddington ratios, which are characteristics of LINERs, the formation of BLRs may not occur in some objects \citep{2000ApJ...530L..65N,2006ApJ...648L.101E,2008ARA&A..46..475H,2011A&A...525L...8C}. 

This is the second of a series of papers whose goal is to detect and characterize the nuclear and circumnuclear (scales of an order of 100 pc) gas emissions in a sample of 10 ETGs. In Ricci et al. (2014, hereafter Paper I), we analysed this sample with the principal component analysis (PCA) Tomography technique only \citep{1997ApJ...475..173H,2009MNRAS.395...64S}. This methodology consists in applying PCA to data cubes. With PCA Tomography, one is able to detect spatial and spectral correlations along data cubes, allowing an efficient and statistically optimized extraction of information from the observed region. In Paper I, we concluded that at least eight galaxies contain an AGN in their central region. In addition, we showed that seven galaxies possess gas discs in their circumnuclear environment. In one object, the circumnuclear gas component seemed to have an ionization cone structure. Stellar discs, also in the circumnuclear regions, were detected in seven galaxies. 

In order to validate the results presented in Paper I, we will analyse the data cubes of the sample galaxies with techniques that are well established in the literature. In fact, in this work and in Paper III, we will study data cubes with the stellar components properly subtracted. The subtraction was carried out with the stellar population synthesis in each spectrum of the data cubes. With the gas cubes (i.e., data cubes with the gas component only), one is allowed to analyse and characterize the emission lines contained along the observed field of view (FOV) of the galaxies of the sample.  

In this paper, we intend to detect and characterize the nuclear emission lines of the 10 ETGs, to unveil LLAGNs in these objects. In Paper I, the eigenspectra related to the AGNs were very similar to LINER spectra. However, eigenspectra display correlations between the wavelengths and, thus, are not adequate to measure emission line fluxes and their respective ratios. Hence, in Paper I, it was not possible to use diagnostic diagrams to classify the emissions accurately. Besides, some parameters, such as the colour excess E(B-V) and the luminosity of the lines are essential for a correct characterization of these regions. Nevertheless, our main intention is not to show that results obtained with PCA Tomography may arise from known methodologies but rather to use both techniques as complementary tools. Joint analysis of the results from this work and from Paper I may highlight useful information that would not be possible to extract using only one of the procedures discussed above.   

Section \ref{dados_PaperII} presents a brief summary of the general characteristics of the data cubes, in addition to a description of the stellar population synthesis. In section \ref{sec:resuts}, we analyse the spectra extracted from the nuclear regions of the galaxies of the sample. Finally, in section \ref{sec:conc}, we discuss our results and present the main conclusions of this work.

\section{Spectroscopic data and spectral synthesis} \label{dados_PaperII}

A detailed description of the sample studied in this work was presented in Paper I. In short, it is composed by 10 massive ($\sigma >$ 200 km s$^{-1}$) ETGs from the local Universe (d $<$ 31 Mpc). These galaxies were observed with an integral field spectrograph operating on the Gemini-South telescope (programmes GS-2008A-Q-51 and GS-2008B-Q-21). This instrument allows us to obtain 2D spectra of the central region of these galaxies, with a FOV of 3.5 arcsec x 5 arcsec and a spectral resolution $<\sim$ 1 arcsec. The spectra cover from the H$\beta$ line to the [S II]$\lambda\lambda$6716, 6731 doublet. Data treatment and reduction procedures were also shown in Paper I. High- and low-frequency noises were removed from the data cubes. However, for this work, we did not perform high-frequency noise filtering in spectral dimension, since the spectral line decomposition may be affected by systematical errors (e.g. high-frequency kinematical features). The exceptions are NGC 1399 and NGC 1404, because their emission lines are too weak (see section \ref{cD_cases}) and are only evident after the high-frequency noise filtering is applied to the spectra of both data cubes. We also applied the Richardson Lucy deconvolution technique to the spatial dimensions of the data cubes, in addition to the correction of the differential atmospheric refraction effect. For more details about the procedures discussed above, see Paper I.

The starlight subtraction from the galaxies is an essential step to study the emission lines emerging from LLAGNs. In Paper I, we isolated the gas emission from the stellar component with PCA Tomography applied to the data cubes of the galaxies. However, as mentioned in section \ref{sec:intro}, eigenspectra are not adequate to measure emission line fluxes. Thus, an accurate fit of the stellar population in the spectra of the data cubes is necessary for the correct characterization of the emission lines in these objects. In light of this, we performed a stellar population synthesis in each spectrum of the data cubes of the sample galaxies using the {\sc starlight} code \citep{2005MNRAS.358..363C}. In other words, we built data cubes with stellar components only. We used a simple stellar population (SSP) library described by \citet{2009MNRAS.398L..44W}. This library was created with MILES spectral energy distributions (SEDs; \citealt{2010MNRAS.404.1639V}) with solar abundances ([Fe/H] = 0.0 and [$\alpha$/Fe] = 0.0), whose fluxes were recalibrated, pixel by pixel, to non-solar abundances by means of the theoretical SSP models proposed by \citet{2007MNRAS.382..498C}. This library contains 120 SSPs with a resolution of 2.51 \AA\ \citep{2011A&A...532A..95F,2011A&A...531A.109B}, ages between 3 and 12 Gyr with steps of 1 Gyr and abundances [Fe/H] = -0.5, -0.25, 0.0 and 0.2 and [$\alpha$/Fe] = 0.0, 0.2 and 0.4. SSP libraries built as described above should result in a more accurate fit to stellar population spectra of ETGs, since they are composed by observed stellar spectra in the optical region and are not biased against non-solar abundances. Because the sample is composed of massive ETGs and considering the fact that the stellar velocity dispersion in this type of galaxies is correlated with their stellar ages, metalicities and the presence of $\alpha$ elements \citep{2005ApJ...621..673T}, the use of the SSP library proposed by \citet{2009MNRAS.398L..44W} is quite adequate for the galaxies of the sample. In appendix \ref{starlight_results}, we show fitting results for the galaxies' centre [high-signal-to-noise (S/N) ratio] and for the upper-right region of the FOV (low S/N ratio) of the data cubes. 

The spectral synthesis procedure for NGC 2663 was different than that for the other galaxies of the sample. Since we were not obtaining reliable results for the red region of the spectra, we ran the {\sc starlight} code twice for this galaxy's data cube: one for wavelengths $<$ 6130\AA\ and the other one for wavelengths $>$ 6130\AA. Doing this, the fitting results around the H$\alpha$ line were reasonable. As we were interested in removing the starlight from the spectra, not in the results of the stellar populations, this procedure was consistent with our goals. 

The gas cubes of the eight galaxies of the sample, whose AGNs were previously detected by PCA Tomography in Paper I, revealed emission lines along the FOV. Since no emission lines are seen in the spectral range between 5500 and 5700\AA, we estimated the standard deviation of the gas cubes (and hence also of the original data cubes) within this wavelength range. All spectra of each gas cube were used for this calculation. For the data cubes drawn from the programme GS-2008A-Q-51, we calculated $\sigma_\lambda$ = 2.1$\times$10$^{-18}$ erg s$^{-1}$cm$^{-2}$\AA$^{-1}$, while for the data cubes belonging to the programme GS-2008B-Q-21, $\sigma_\lambda$ = 5$\times$10$^{-19}$ erg s$^{-1}$cm$^{-2}$\AA$^{-1}$.

In this paper, we will only analyze the nuclear spectra extracted from the galaxies. The emission lines originated in the circumnuclear regions of the objects will be discussed in Paper III. 

\section{results} \label{sec:resuts}

\subsection{Nuclear emission line properties} \label{line_properties}

We extracted the nuclear spectra from the gas cube of each galaxy of the sample. This was done through a weighted sum of all spectra of the gas data cubes, whose weights were given by a 2D Gaussian function with peak intensity of 1 and full width at half-maximum (FWHM) equal to the FWHM of the Gaussian point spread function (PSF) of the images of the data cubes (Paper I). These functions were centred at positions $x_{cen}$ and $y_{cen}$ of the point-like objects detected on the [O I]$\lambda$6300 line image (see Paper I for an explanation on why we used this image). In NGC 1399 and NGC 1404, where no [O I] lines were detected, both functions were centred in the region of maximum intensity of the image, which corresponds to the [N II]$\lambda\lambda$6548, 6581+H$\alpha$ lines.

The nuclear emission lines were decomposed as a sum of Gaussian functions by means of a Gauss Newton algorithm, which is an application of the least-squares method in non-linear functions. The statistical uncertanties were estimated with a Monte Carlo simulation by applying the Gauss Newton algorithm 100 times to the fitting result of the observed data added to random noise. These noises were drawn from a normalized Gaussian distribution centred in zero and with $\sigma$ equal to the standard deviation of the residual between fitted and observed data. The errors associated with each parameter of the Gaussian functions were ascertained by the standard deviation of their measurements obtained from the results of all 100 simulated fitting procedures. 

\begin{figure*}
\begin{center}
\includegraphics[width=70mm,height=55mm]{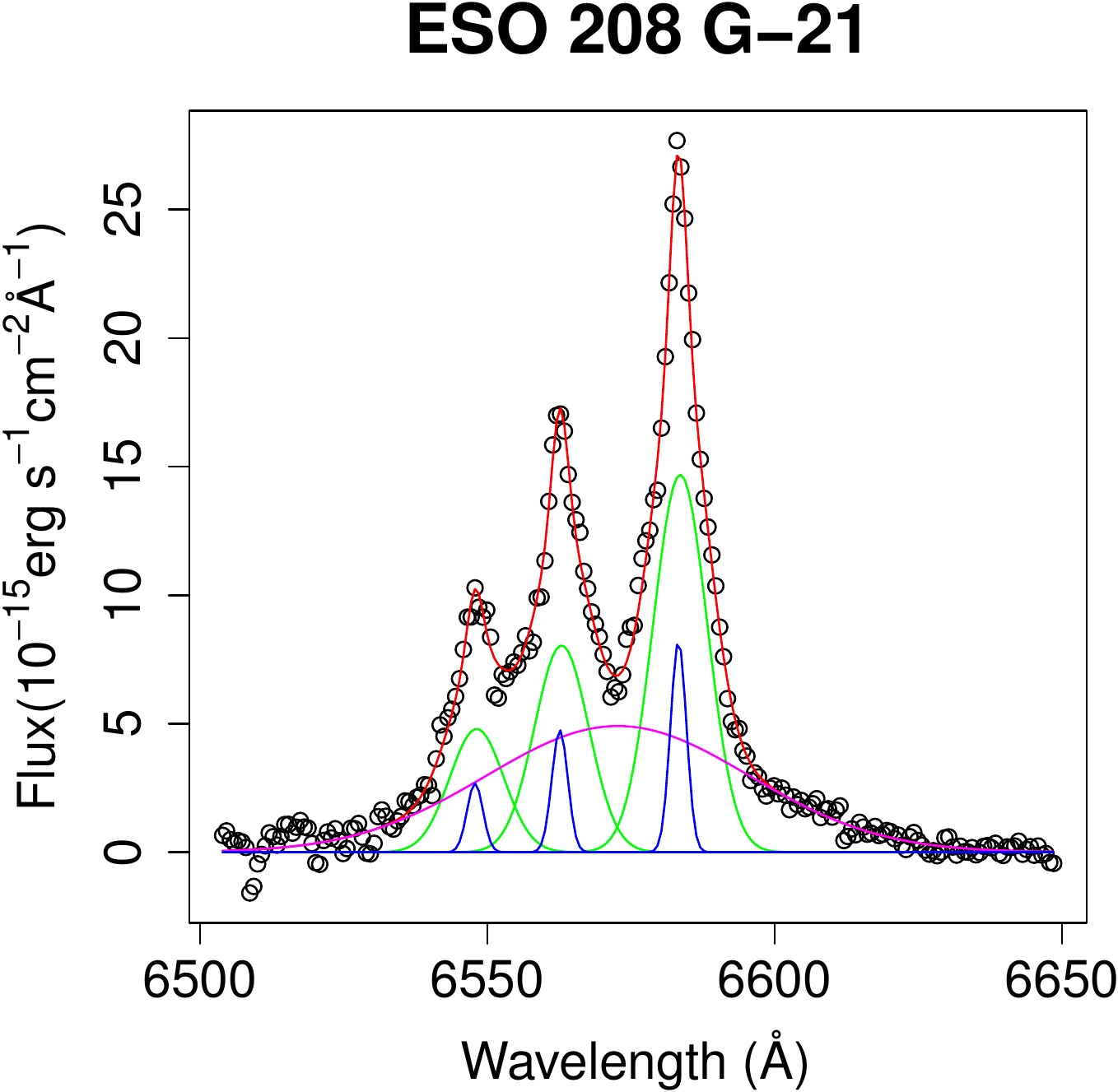}
\includegraphics[width=70mm,height=55mm]{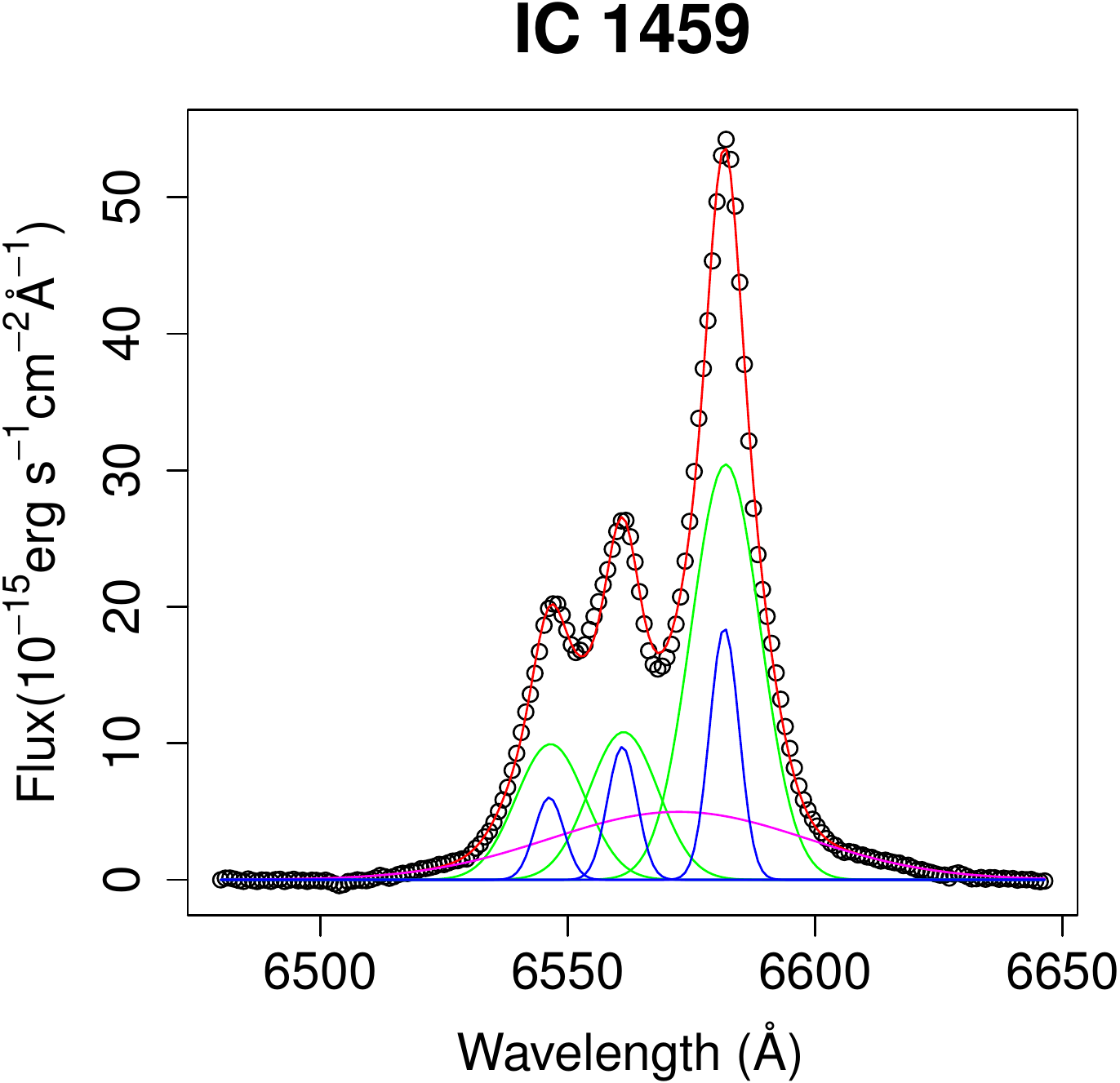}
\includegraphics[width=70mm,height=55mm]{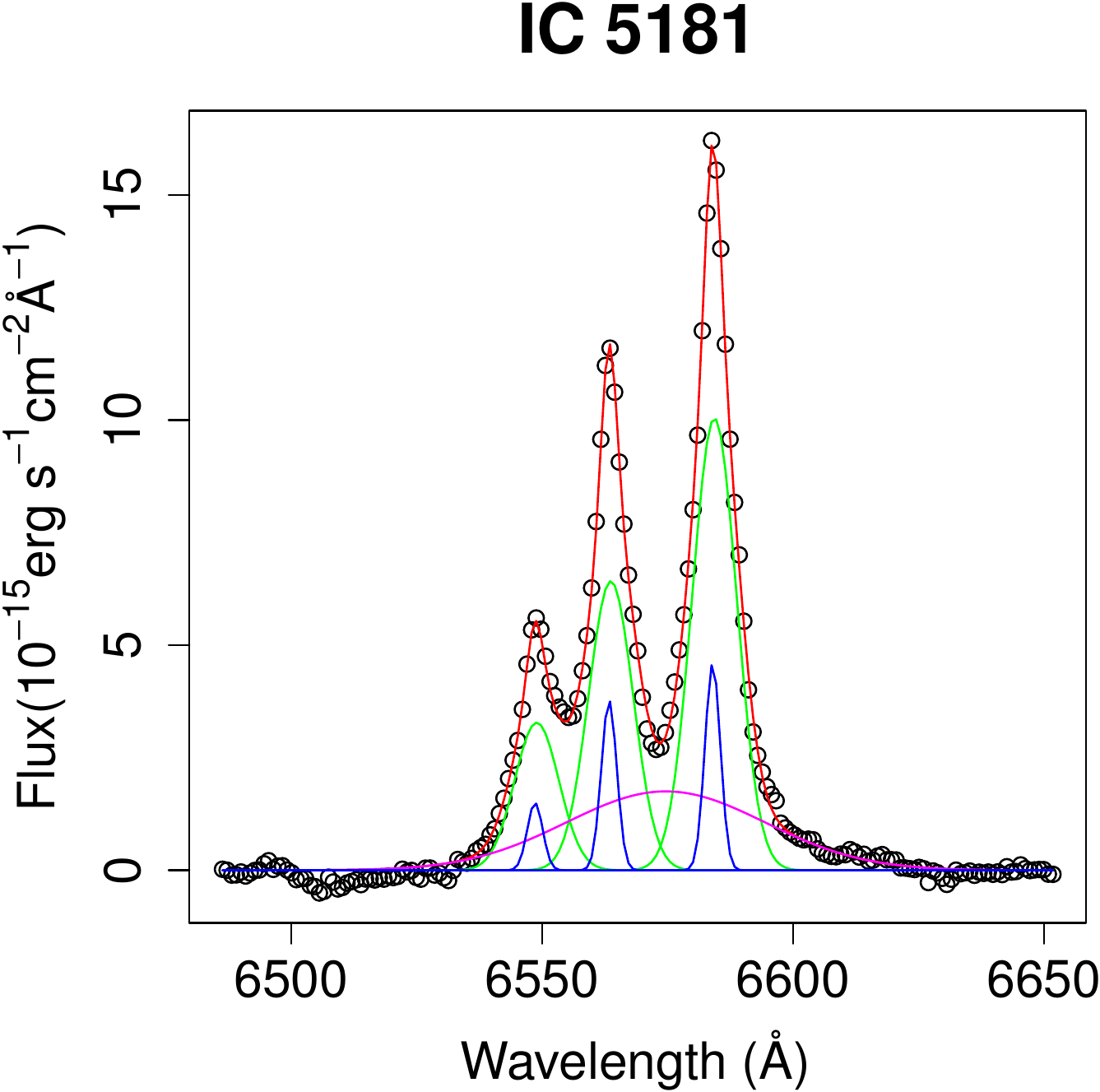}
\includegraphics[width=70mm,height=55mm]{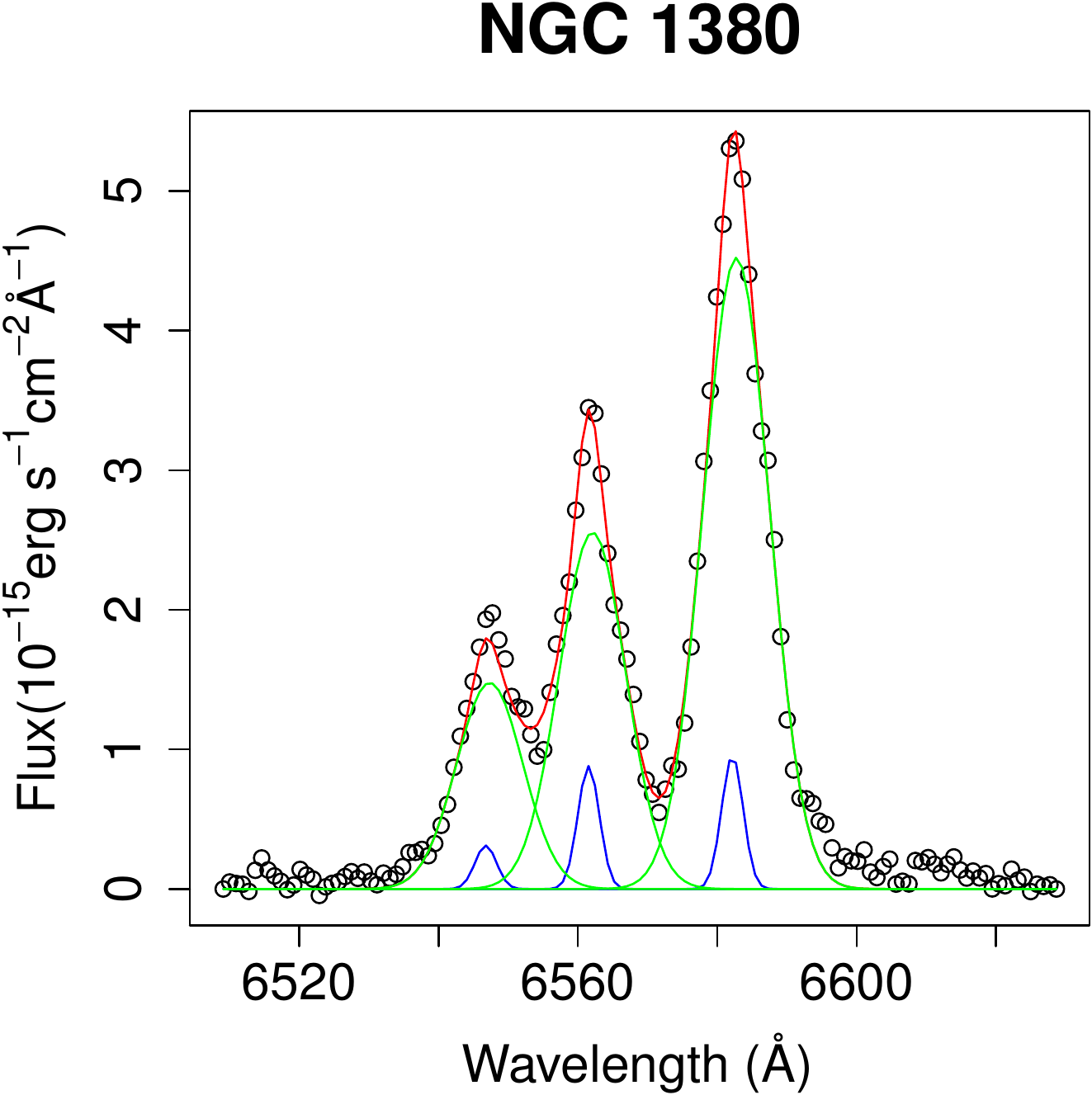}
\includegraphics[width=70mm,height=55mm]{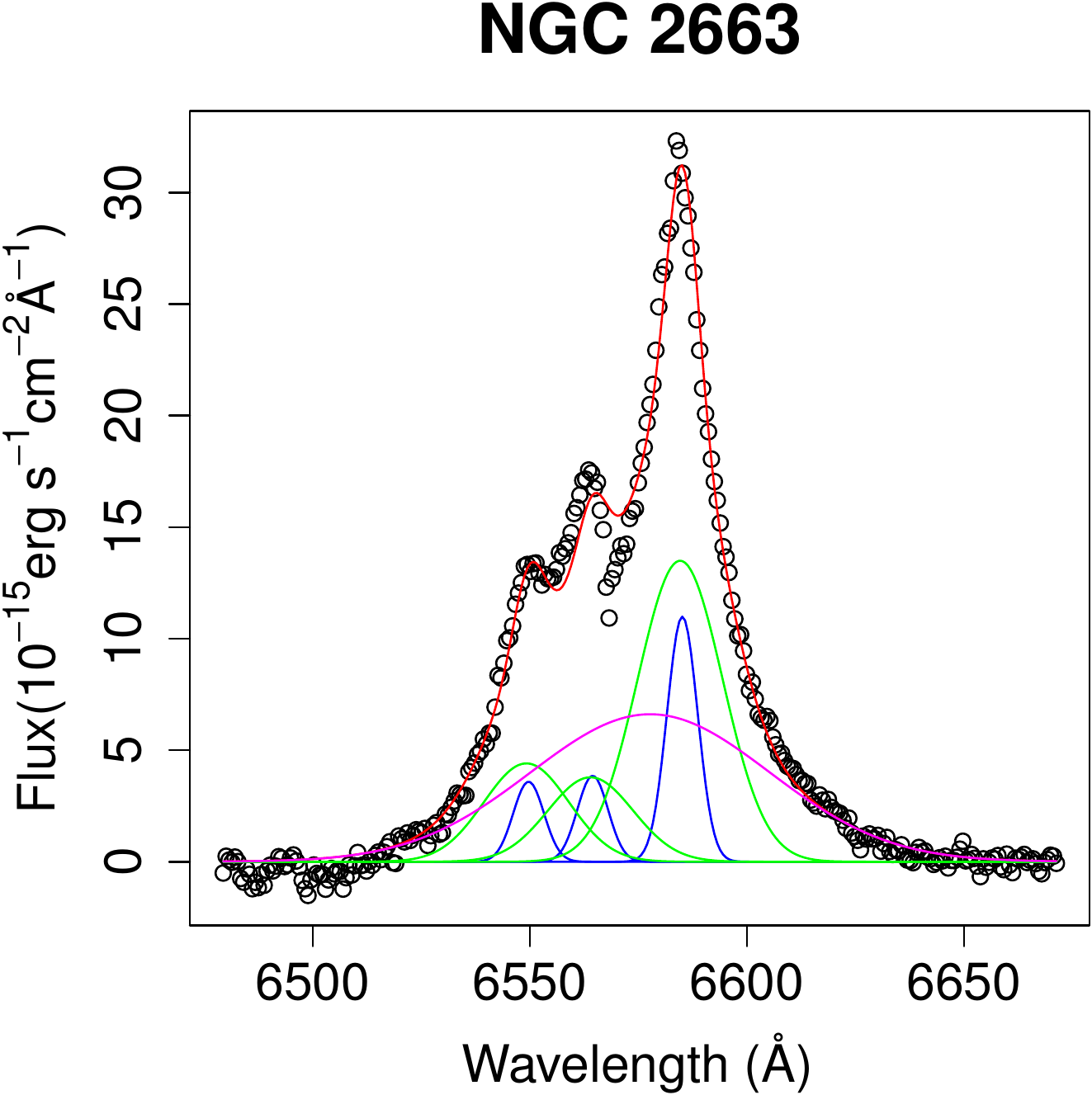}
\includegraphics[width=70mm,height=55mm]{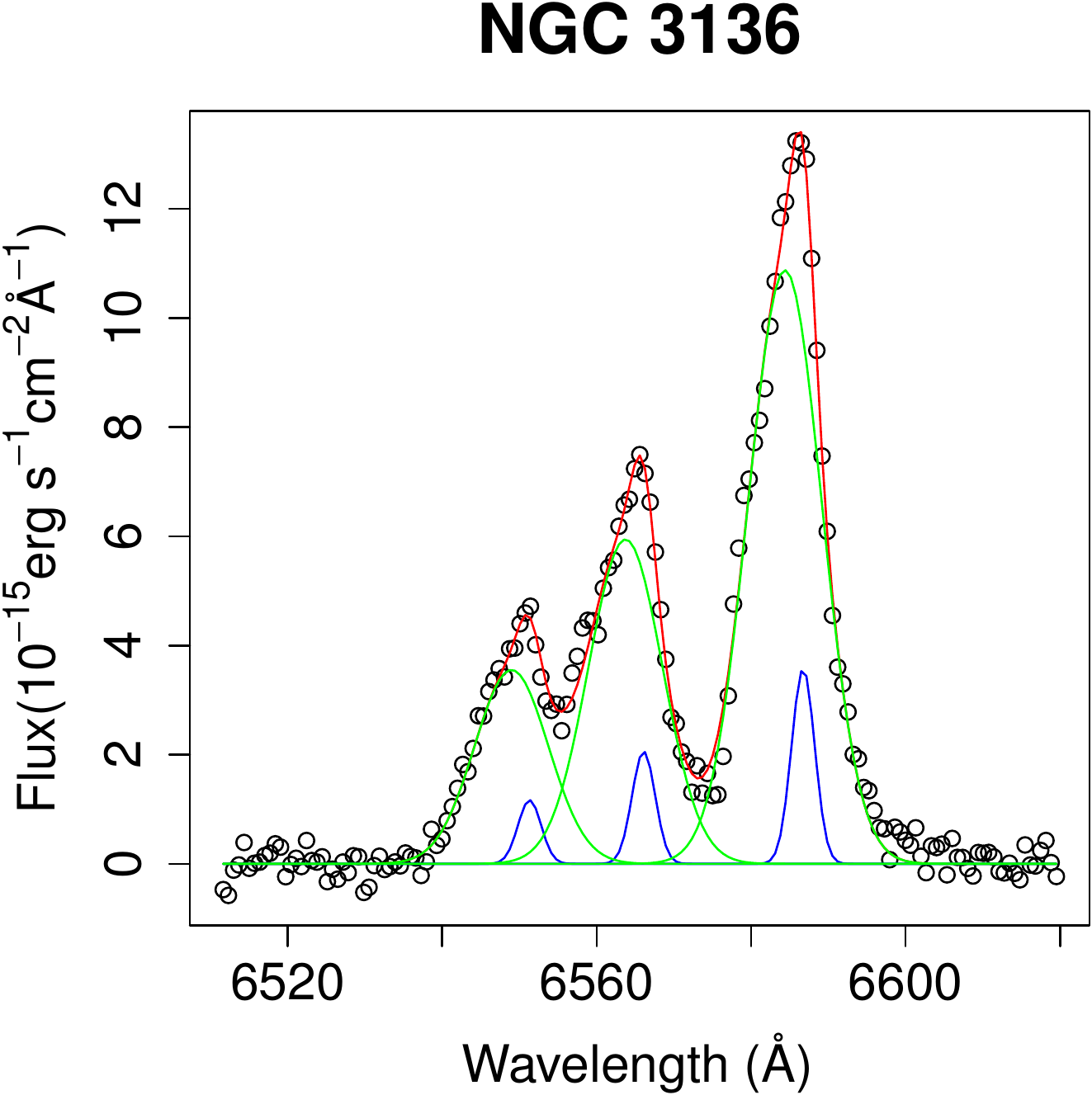}
\includegraphics[width=70mm,height=55mm]{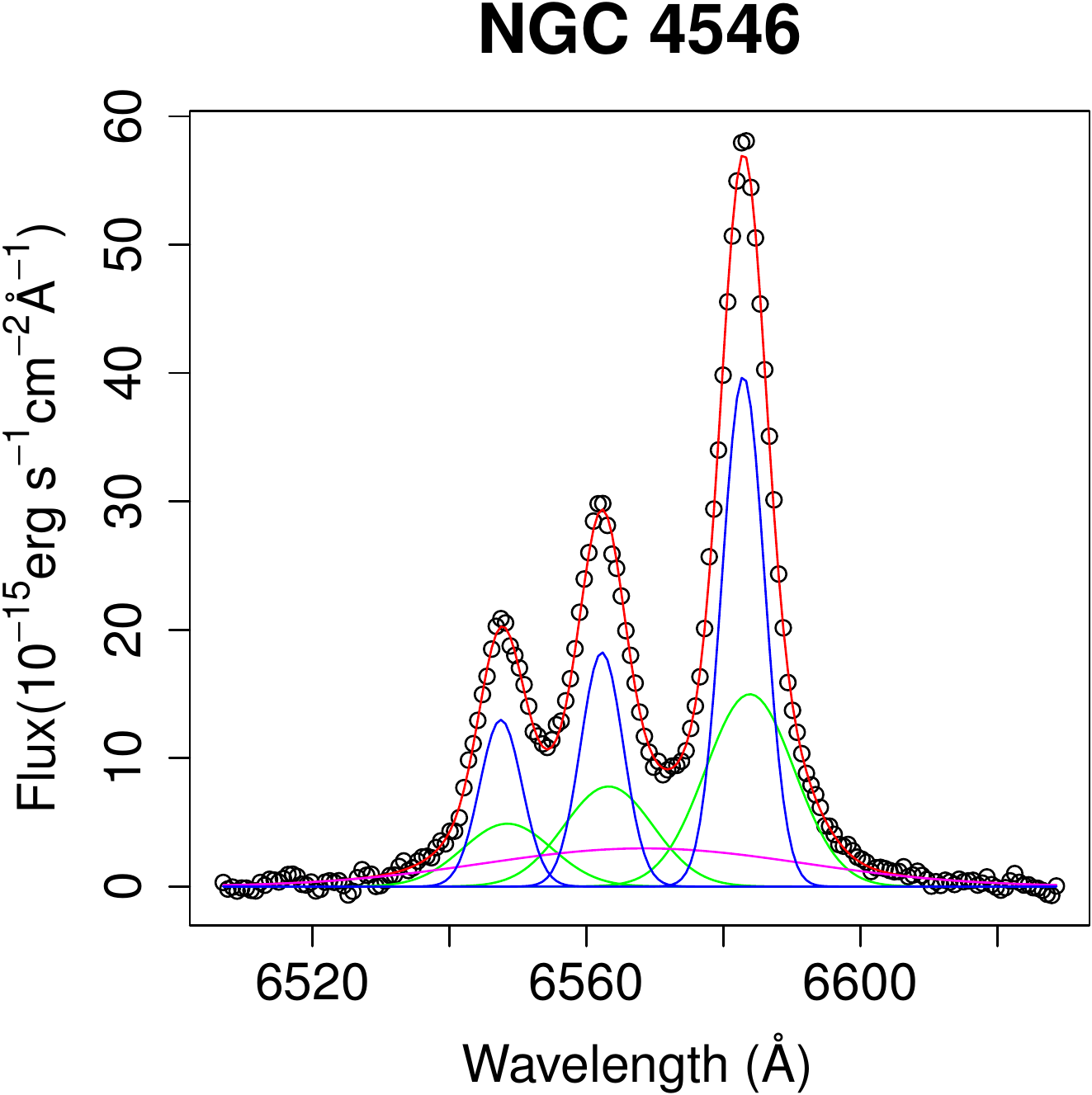}
\includegraphics[width=70mm,height=55mm]{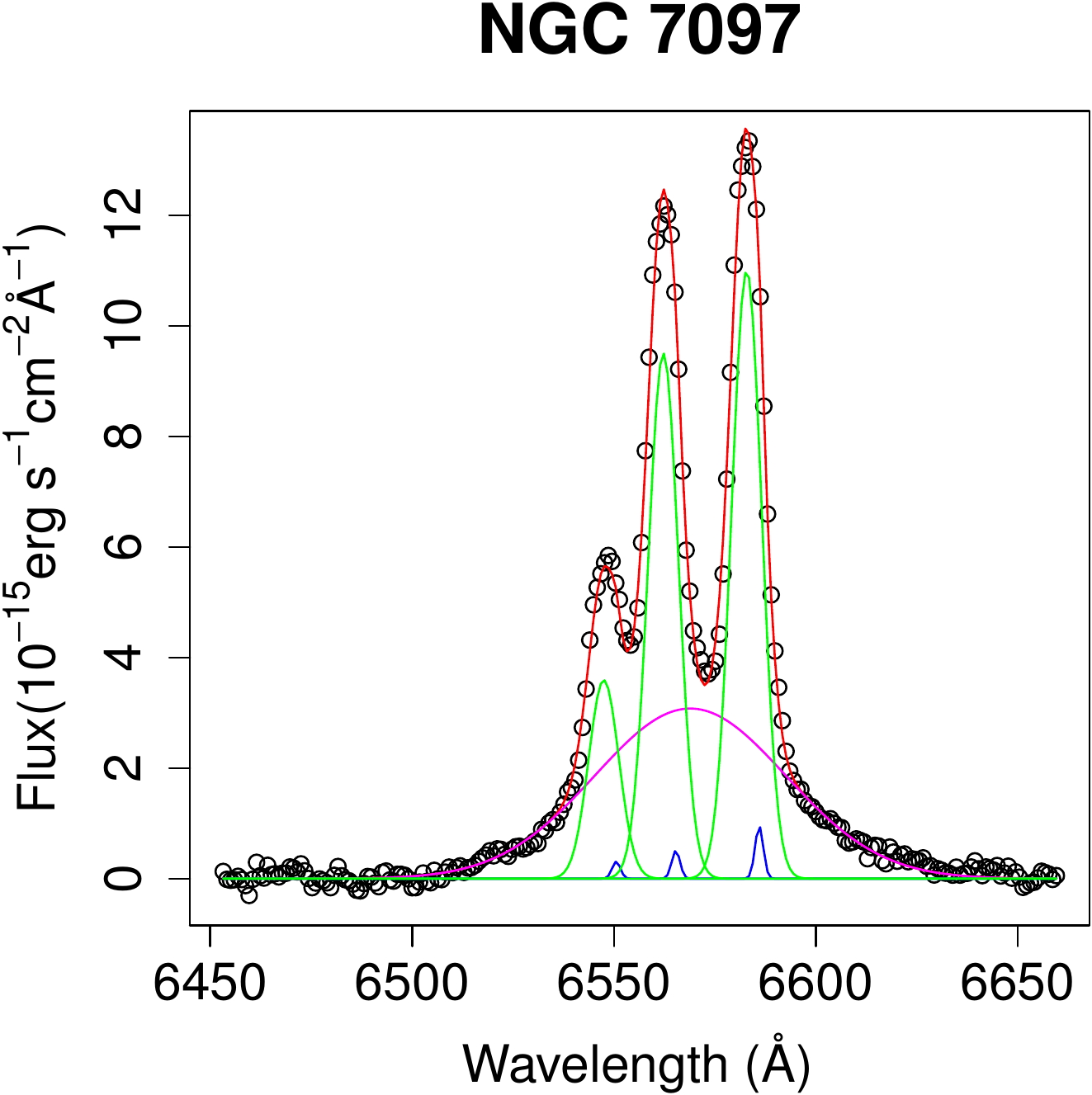}

\caption{H$\alpha$ and [N II]$\lambda \lambda$6548, 6583 emission lines detected in eight galaxies of the sample. The Gaussians correspond to the line decompositions. Gaussians from set 1 are shown in blue, those from set 2 are shown in green. The magenta Gaussian is associated with the broad component of H$\alpha$. The sum of all Gaussians is shown in red.\label{perfil_NII_Ha}
}
\end{center}
\end{figure*}

We began the fitting procedure for the H$\alpha$ + [N II]$\lambda \lambda$6548, 6583 lines in the nuclear spectra of seven galaxies of the sample. We decomposed the lines in this spectral range as a sum of seven Gaussian functions, one of which corresponds to the broad component of the H$\alpha$ line. The other six Gaussian functions describe the narrow components of each of the three lines. In this case, there are two sets of three Gaussians. In each set, the Gaussian functions have the same FWHM and are separated on the H$\alpha$ and [N II] lines by their respective wavelength in the rest frame\footnote{This publication has made use of the Atomic Line List, maintained by Peter van Hoof and hosted by the department of Physics and Astronomy of the University of Kentucky. See http://www.pa.uky.edu/~peter/newpage/}, given by $\lambda\lambda_{[N\ II]}$ = 6583.4, 6548.0 \AA\ and $\lambda_{H\alpha}$ = 6562.8 \AA. In other words, there is only one radial velocity measurement and also only one velocity dispersion measurement per set. Set 1 is related to the Gaussian functions with the lowest FWHM values. Hence, the Gaussian functions of set 2 have the largest FWHMs. The radial velocity of set 1 is measured as $V_r([N II]\lambda6583)_1$ = ($\lambda_{obs} - 6583.4)\times c/6583.4$, where $\lambda_{obs}$ corresponds to the peak of the Gaussian related to the [N II]$\lambda$6583 line and $c$ is the speed of light. In set 2, the radial velocities are measured with respect to $V_r$([N II]$\lambda$6583)$_1$, the reason being that lowest FWHM values must be related to more external zones of the narrow line region (NLR) and, therefore, are less affected by the gravity of the supermassive black hole (SMBH). Hence, the radial velocities estimated with set 1 must be a reasonable indicator of the redshifts of the galaxies. The real values of the FWHM are given by $FWHM^2_{real} = FWHM^2_{observed} - FWHM^2_{instrumental}$, whose $FWHM_{instrumental}$ values are related to the spectral resolution of the data cubes, shown in Paper I. In both sets, the theoretical ratio [N II]$\lambda6583$/[N II]$\lambda6548$ = 3.06 \citep{2006agna.book.....O} was fixed for the line decomposition. The fitting results for H$\alpha$ and [N II] lines in eight galaxies of the sample are presented in Fig. \ref{perfil_NII_Ha}.

\begin{figure*}
\begin{center}
\includegraphics[width=70mm,height=55mm]{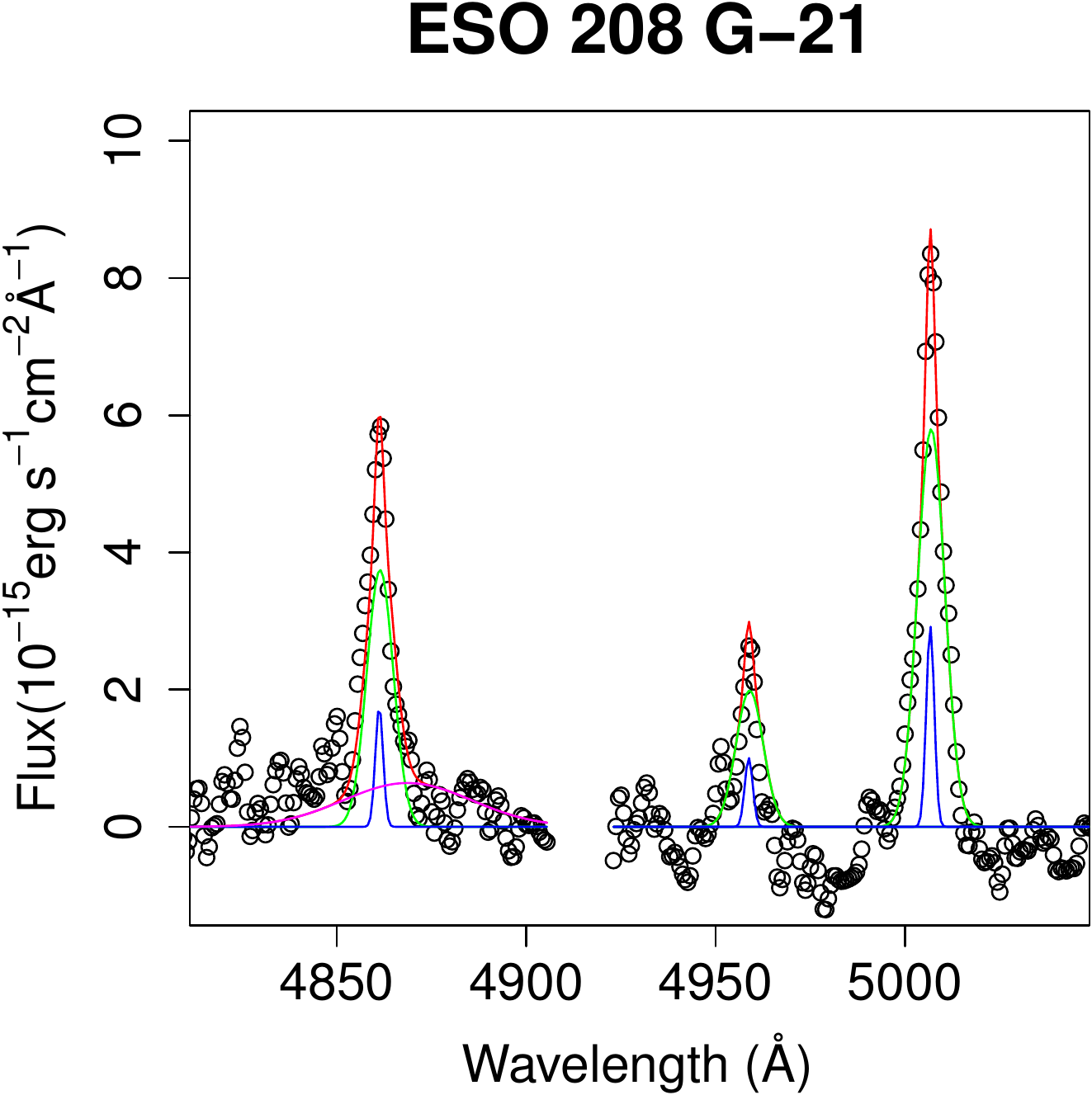}
\includegraphics[width=70mm,height=55mm]{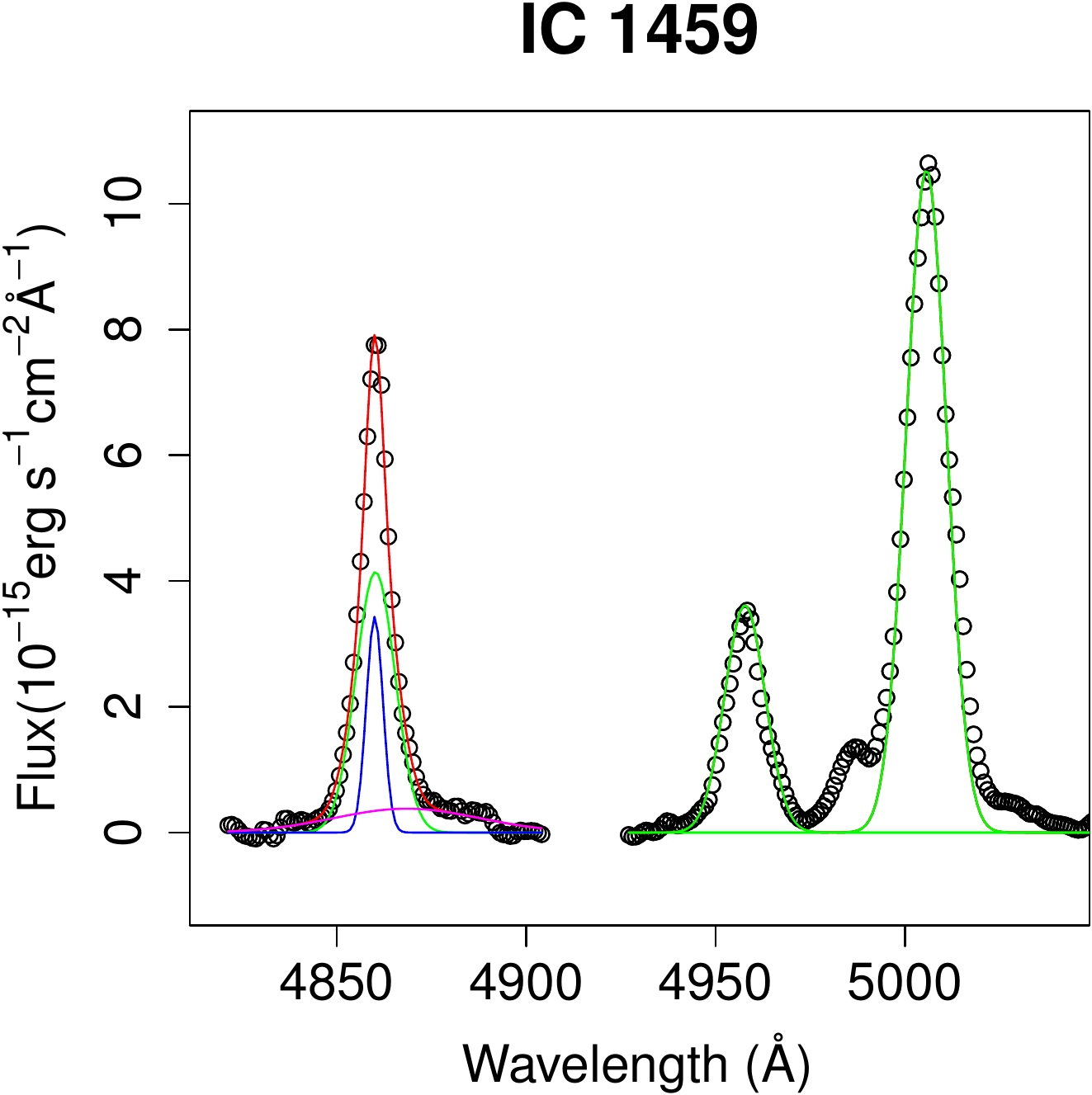}
\includegraphics[width=70mm,height=55mm]{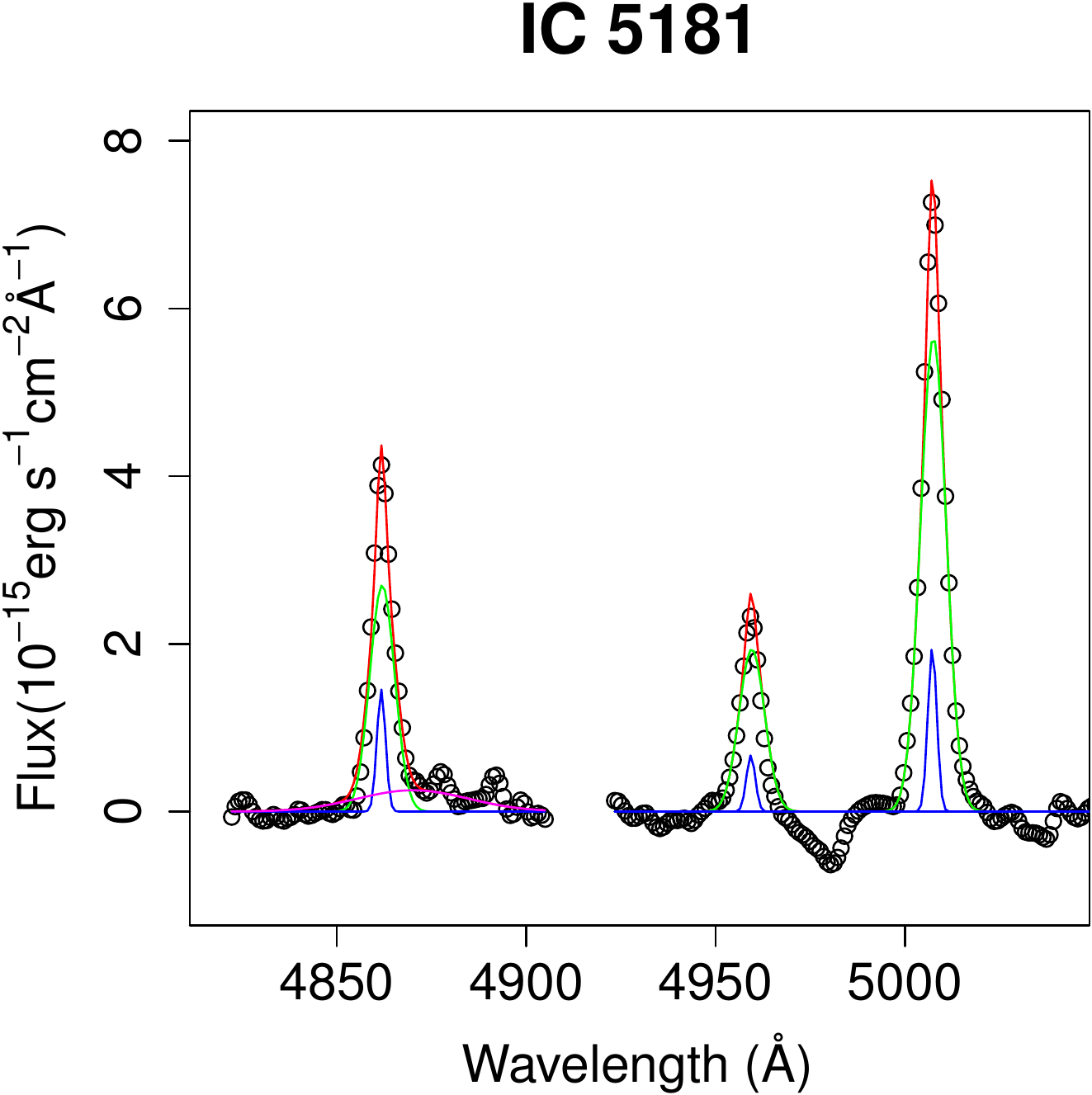}
\includegraphics[width=70mm,height=55mm]{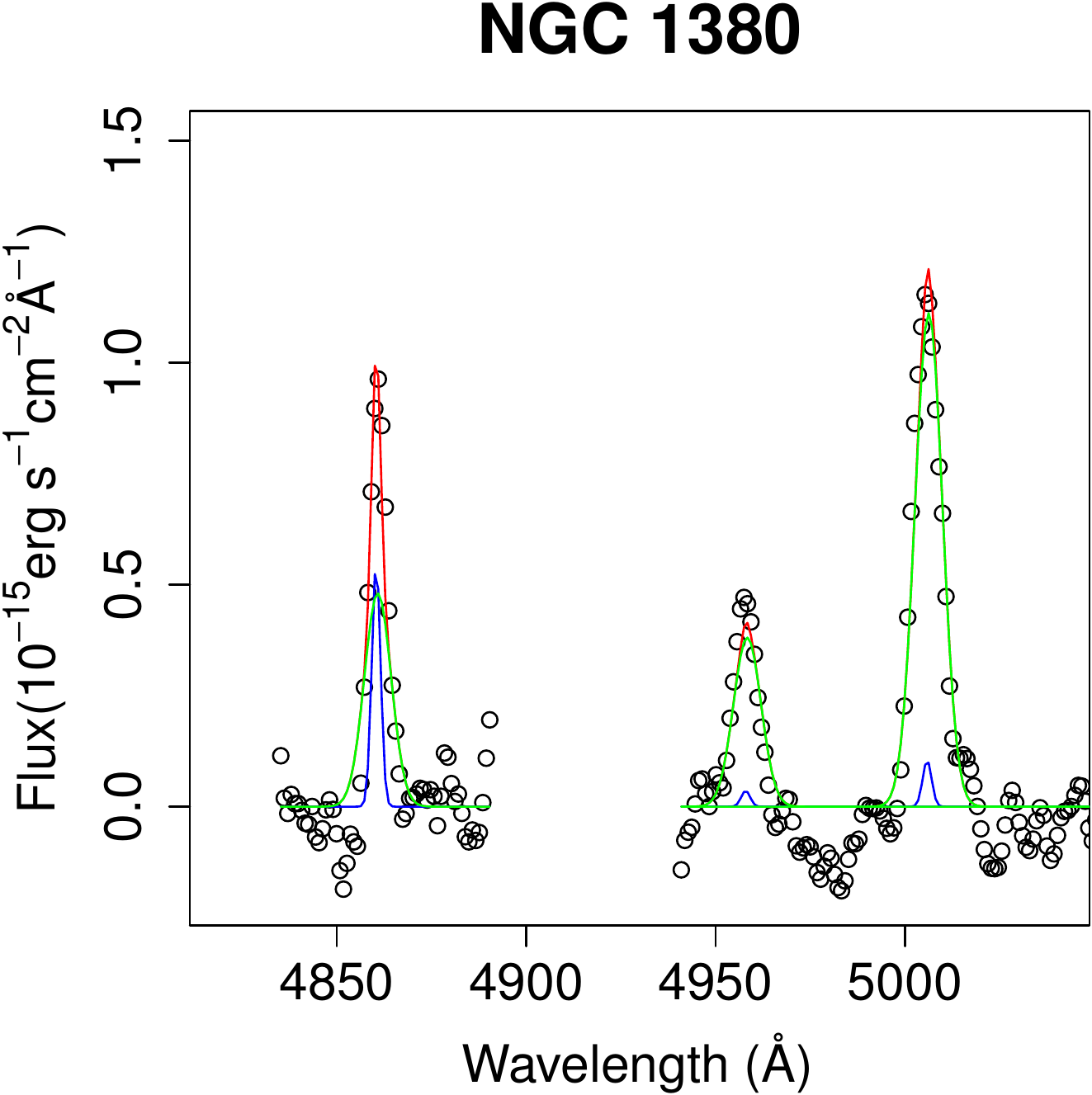}
\includegraphics[width=70mm,height=55mm]{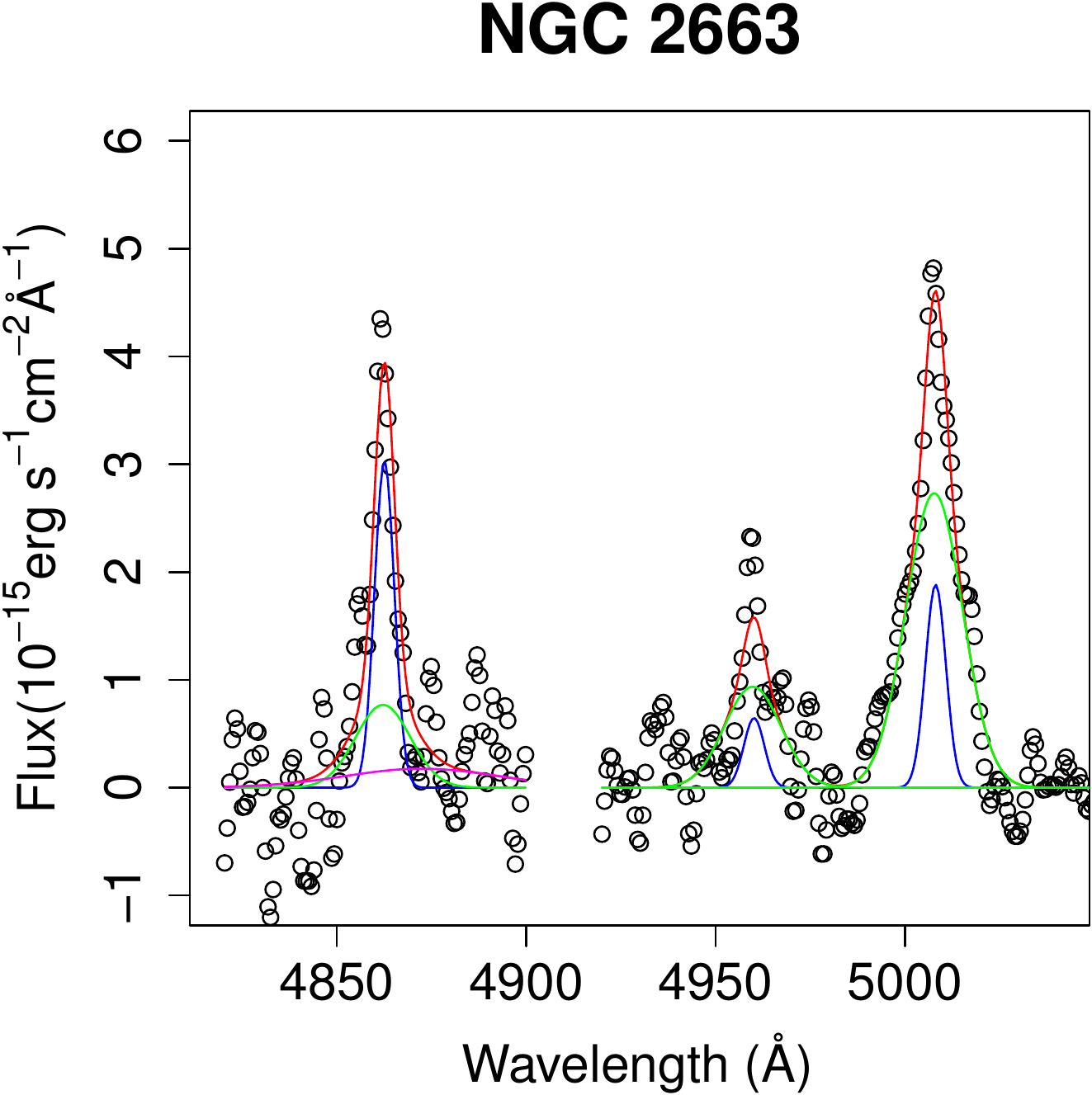}
\includegraphics[width=70mm,height=55mm]{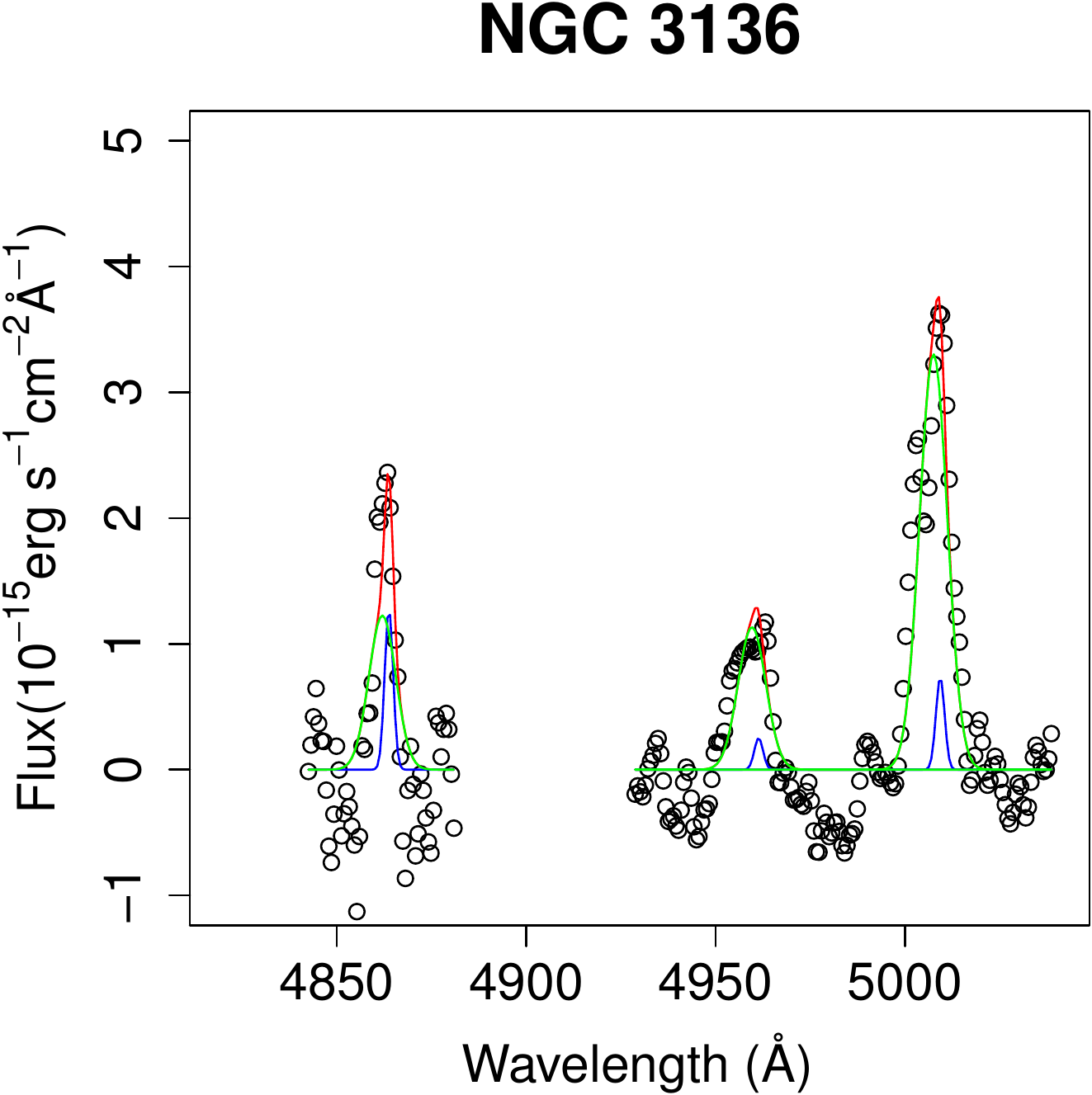}
\includegraphics[width=70mm,height=55mm]{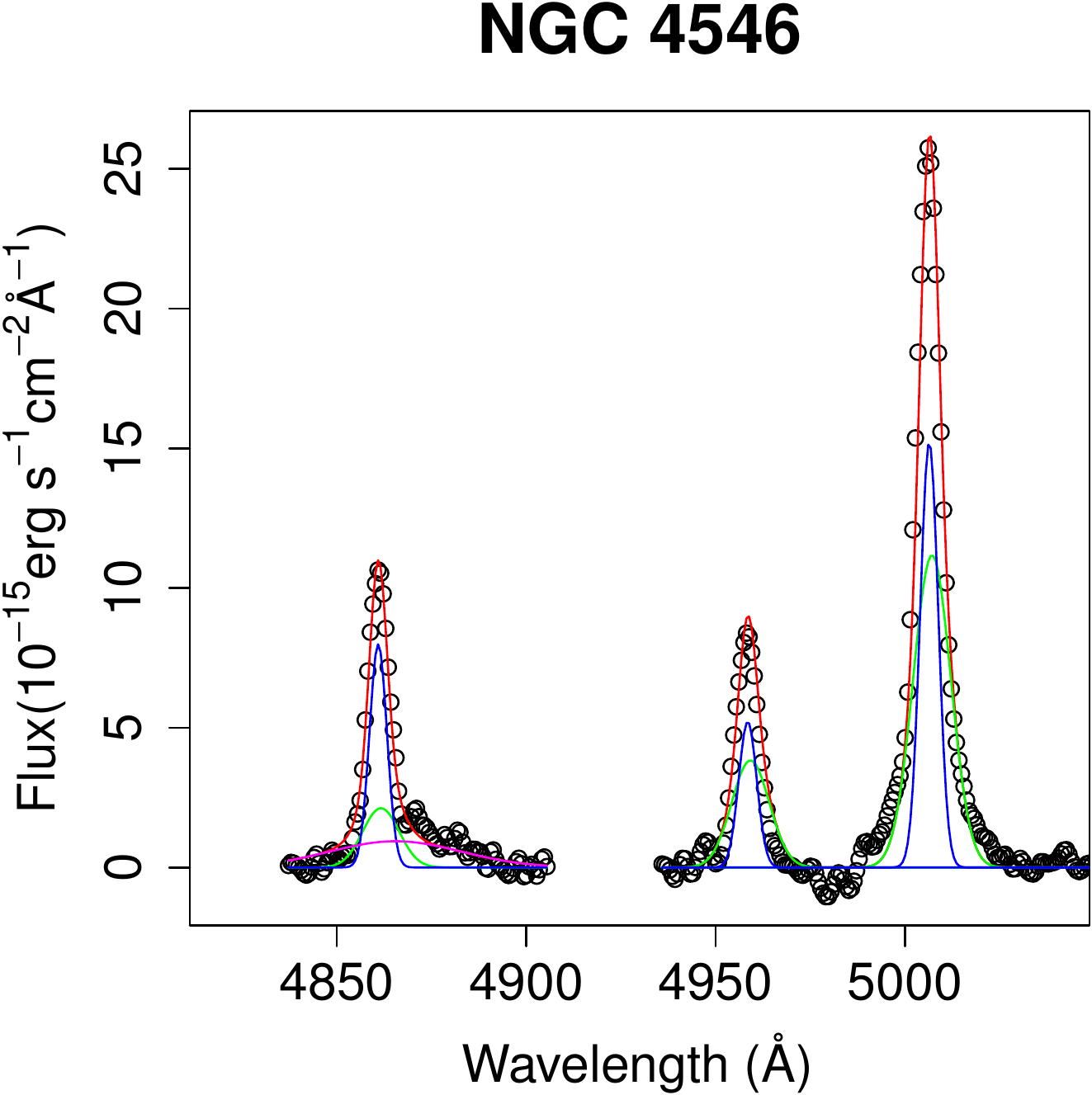}
\includegraphics[width=70mm,height=55mm]{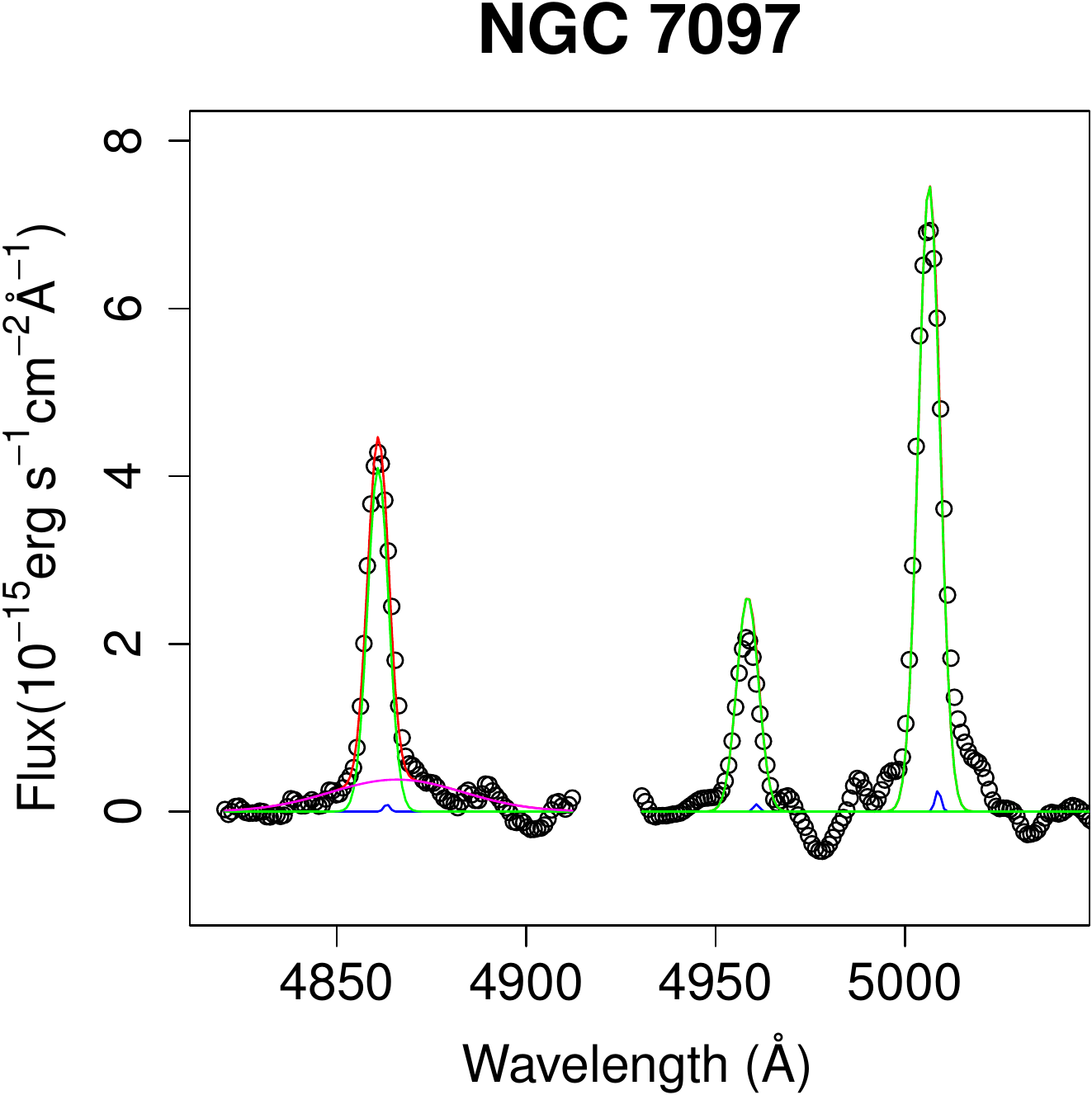}

\caption{H$\beta$ and [O III]$\lambda \lambda$4959, 5007 emission lines detected in eight galaxies of the sample. These fitted Gaussians have the same kinematics (FWHM$_1$, FWHM$_2$, $V_{r1}$ e $V_{r2}$) of the fitted Gaussians in H$\alpha$ and [N II]$\lambda \lambda$6548, 6583 emission lines. Again, Gaussians from set 1 are shown in blue, those from set 2 are shown in green. For a better presentation of the lines, we smoothed the observed spectra shown in this figure, although the fitting procedures were performed with the original gas spectra. \label{perfil_OIII_Hb}
}
\end{center}
\end{figure*}

For the other emission lines, we fitted two sets of Gaussian functions for their narrow component. In these cases, only the amplitudes of the Gaussians were set as free parameters. The radial velocities and the FWHM of the Gaussians of both sets 1 and 2 were fixed with the values found for the H$\alpha$ + [N II] line decomposition. In other words, the narrow components of all emission lines were fitted with the same FWHM and radial velocities. The theoretical ratios [O I]$\lambda$6300/[O I]$\lambda$6363 = 3.05 and [O III]$\lambda$5007/[O III]$\lambda$4959 = 2.92 \citep{2006agna.book.....O} were also fixed for each set. In addition to the H$\beta$ line, a broad component was also fitted, with the same radial velocity and FWHM of that fitted for the broad component of H$\alpha$. Also in this case, the amplitude was the only free parameter of the fitting procedures. Only in NGC 2663 we did fit first the [S II]$\lambda$6716 + [S II]$\lambda$6731 lines, because the simultaneous fitting of the NLR and the BLR in the H$\alpha$ + [N II]$\lambda \lambda$6548, 6583 lines range resulted in an H$\alpha$/H$\beta$ smaller value than the value calculated with photoionization models for NLRs of LINERs (H$\alpha$/H$\beta$ = 3.1 - \citealt{1983ApJ...264..105F,1983ApJ...269L..37H}).  The fitted results in H$\beta$ and [O III] in the same eight galaxies are shown in Fig. \ref{perfil_OIII_Hb}.  

Although the profiles should be different for each ionized species, we decided to apply the procedure described above in order to reduce the number of free parameters, avoiding unphysical results (e.g. H$\alpha$/H$\beta$ < 3.1 or even < 2.8 in some cases). Moreover, if the kinematics of the narrow component of H$\alpha$ were a free parameter, it could be affected by the broad line component. This could underestimate the flux from the BLR and, consequently, overestimate the narrow line component.  Since we are not interpreting the kinematics of the emission lines at all, the systematic errors in flux caused by this assumption will not be so important as an error caused by bad fitting results.

The cases of NGC 1399 and NGC 1404 will be discussed in section \ref{cD_cases}.

\subsection{The narrow line regions} \label{NLR_component}

The flux of the narrow components of the emission lines was calculated as the sum of the integrals of each Gaussian function of sets 1 and 2. The colour excess E(B-V) was calculated by means of the measured H$\alpha$/H$\beta$ ratio, using the extinction curve proposed by \citet{1989ApJ...345..245C} ($R_V$ = 3.1). In Table \ref{tab_f_NLR}, we show the reddened flux of the narrow component of the H$\alpha$ line, the measured H$\alpha$/H$\beta$ ratio, E(B-V) and the [N II]$\lambda$6583/H$\alpha$, ([S II]$\lambda$6716+[S II]$\lambda$6731)/H$\alpha$, [O I]$\lambda$6300/H$\alpha$, [O III]$\lambda$5007/H$\beta$ and [N I]$\lambda$5199/H$\beta$ line ratios. The H$\alpha$ luminosities of the narrow components, corrected for extinction effects, are shown in Table \ref{tab_cin_NLR}. The kinematics related to both sets of Gaussian functions are also shown in Table \ref{tab_cin_NLR}. The electronic density $n_e$ of the nuclear region was estimated with the [S II]$\lambda$6716/[S II]$\lambda$6731 ratio and assuming $T_e$ = 10000K. For this calculation of $n_e$, we used the NEBULAR package under the IRAF environment \citep{1995PASP..107..896S}, with the calculations proposed by \citet{1987JRASC..81..195D}.

We calculated the ionized gas mass of the NLR as

\begin{equation}
	M_{ion}=\frac{L(H\alpha)m_H}{\epsilon n_e}=\frac{2.2\times10^7L_{40}(H\alpha)}{n_e} M_\odot,
	\label{eqionizedmass}
\end{equation}
where $L_{40}(H\alpha)$ is the H$\alpha$ luminosity in units of 10$^{40}$ erg s$^{-1}$ and $\epsilon$ = 3.84$\times$10$^{-25}$ erg s$^{-1}$ cm$^3$ is the H$\alpha$ line emissivity (\citealt{2006agna.book.....O}, case B). The results are presented in Table \ref{tab_cin_NLR}. The values of $M_{ion}$ for the eight galaxies of the sample are very similar to each other (even the minimum value estimated for NGC 2663), varying from 3$\times$10$^4$ M$_\odot$ to 1.5$\times$10$^5$ M$_\odot$. 

\begin{table*}
 \scriptsize
 \begin{center}
 \begin{tabular}{@{}lcccccccc}
  \hline
  Galaxy name& $f$(H$\alpha$)$_n$ & (H$\alpha$/H$\beta$)$_n$ & E(B-V) & [N II]/H$\alpha$ & [S II]/H$\alpha$ & [O I]/H$\alpha$ & [O III]/H$\beta$ & [N I]/H$\beta$ \\
  \hline
  ESO 208 G-21 & 112$\pm$6 & 2.99$\pm$0.34 & -0.03$\pm$0.11 & 1.81$\pm$0.12 & 1.82$\pm$0.11 & 0.19$\pm$0.02 & 1.61$\pm$0.19 & 0.35$\pm$0.06 \\
  
  IC 1459 & 263$\pm$15 & 3.60$\pm$0.24 & 0.15$\pm$0.06& 2.55$\pm$0.18&1.49$\pm$0.09&0.39$\pm$0.03&1.94$\pm$0.09&0.61$\pm$0.03 \\
  
  IC 5181 & 86$\pm$3 & 3.26$\pm$0.17 & 0.05$\pm$0.05&1.50$\pm$0.07&1.08$\pm$0.04&0.16$\pm$0.02&2.03$\pm$0.10&0.43$\pm$0.03 \\
  
  NGC 1380 & 34$\pm$2 & 5.82$\pm$0.70 & 0.61$\pm$0.12&1.70$\pm$0.11&0.81$\pm$0.05&0.15$\pm$0.01&1.79$\pm$0.22 & 0.32$\pm$0.14 \\
  
  NGC 1399 & 1.7$\pm$0.3 & & & 1.14$\pm$0.28 & & & & \\
  
  NGC 1404 & 7.7$\pm$1 & & & 0.84$\pm$0.14 & & & & \\
  
  NGC 2663& 127$\pm$18 & 3.75$\pm$1.08 & 0.18$\pm$0.28 & 3.36$\pm$0.53 & 1.87$\pm$0.35 & 0.30$\pm$0.06 & 1.89$\pm$0.50 & 0.02$\pm$0.14 \\
  
  NGC 3136 & 79$\pm$2 & 5.47$\pm$0.96&0.55$\pm$0.17&1.82$\pm$0.07&1.09$\pm$0.04&0.16$\pm$0.03&2.24$\pm$0.41&0.21$\pm$0.12 \\
  
  NGC 4546 & 269$\pm$18 & 3.75$\pm$0.40& 0.18$\pm$0.10 & 2.05$\pm$0.17&0.70$\pm$0.05&0.42$\pm$0.04&3.19$\pm$0.30&0.28$\pm$0.08\\
  
  NGC 7097 & 90$\pm$2&3.15$\pm$0.09&0.02$\pm$0.03&1.16$\pm$0.03&1.36$\pm$0.03&0.32$\pm$0.01&1.88$\pm$0.07&0.04$\pm$0.04 \\
  \hline
 \end{tabular}
 \caption{Flux measurement for the narrow components of the nuclear emission lines. The H$\alpha$ flux is in units of 10$^{-15}$ erg s$^{-1}$ cm$^{-2}$.  \label{tab_f_NLR}
}
 \end{center}
\end{table*}

\begin{table*}
 \scriptsize
 \begin{center}
 \begin{tabular}{@{}lccccccccc}
  \hline
  Galaxy name & FWHM$_1$ & FWHM$_2$&$V_{r1}$&$V_{r2}$&log $L$(H$\alpha$)& $n_e$ & log $L_{bol}$ & $R_{Edd}$&$M_{ion}$ \\
   &(1) &(2) &(3) &(4) &(5) &(6) &(7) &(8)&(9) \\
  \hline
  ESO 208 G-21 & 132$\pm$9 & 502$\pm$15 &-6$\pm$2&14$\pm$4&40.13$\pm$0.19& 153$^{+87}_{-73}$ & 42.07 & -4.32&5.6$^{+10.1}_{-3.7}\times$10$^4$ \\
  
  IC 1459 & 305$\pm$12&748$\pm$23&13$\pm$4&-80$\pm$2&40.87$\pm$0.13& 658$^{+116}_{-100}$&43.09&-4.08&1.1$^{+0.6}_{-0.4}\times$10$^5$\\
  
  IC 5181 & 133$\pm$7&467$\pm$9 & 27$\pm$2 & 16$\pm$3&40.16$\pm$0.38& 395$^{+97}_{-83}$&42.45& -4.24&4.0$^{+5.6}_{-3.7}\times$10$^4$\\
  
  NGC 1380 & 134$\pm$23&505$\pm$9&-56$\pm$6&26$\pm$7&39.77$\pm$0.18 & 451$^{+181}_{-139}$ &42.30& -4.20&2.9$^{+3.0}_{-1.7}\times$10$^4$\\
  
  NGC 1399 &188$\pm$26 & & 54$\pm$9& & 37.81$\pm$0.14& &40.15&-7.24\\
  
  NGC 1404& 496$\pm$38 & & -15$\pm$13& &38.50$\pm$0.12 & &40.84&-5.91\\
  
  NGC 2663 & 375$\pm$43&1053$\pm$44&77$\pm$10&-23$\pm$59&40.90$\pm$0.25& $<$ 910 &42.79&-4.39& $>$1.1$\times$10$^4$ \\
  
  NGC 3136 & 145$\pm$13&511$\pm$6 & 152$\pm$4&-107$\pm$6&40.32$\pm$0.20 & 267$^{+93}_{-79}$&42.92&-3.86&1.7$^{+3.1}_{-1.0}\times$10$^5$\\
  
  NGC 4546 & 328$\pm$6&697$\pm$38& -23$\pm$1&44$\pm$8&40.38$\pm$0.25 & 874$^{+244}_{-189}$&42.94&-3.77&3.6$^{+3.7}_{-2.5}\times$10$^4$\\
  
  NGC 7097 & 51$\pm$24&391$\pm$4 &117$\pm$8&-141$\pm$8&40.52$\pm$0.05& 312$^{+48}_{-45}$&42.59& -4.25&7.7$^{+2.3}_{-1.8}\times$10$^4$ \\
  \hline
 \end{tabular}
 \caption{Columns 1-4: gas kinematics of sets 1 and 2 of the NLR, measured on the [N II]$\lambda$6583 emission line, in km s$^{-1}$. Column (5): total luminosity (NLR + BLR, when present) of H$\alpha$, in erg s$^{-1}$. Both components were corrected for reddening effects with the E(B-V) values shown in Table \ref{tab_f_NLR}, except for NGC 1399 and NGC 1404, since these objects did not reveal the H$\beta$ line. The uncertainties related to the luminosity of the H$\alpha$ line took into account the errors in the fluxes, in E(B-V) and in the distances of the galaxies (see Paper I). Column (6): electronic density of the NLR, in cm$^{-3}$. Column (7): bolometric luminosity of the nuclear region of the galaxy, in erg s$^{-1}$. The uncertanties in these values are about 0.48 dex. Column (8): Eddington ratio. In this case, the uncertanties are about 1 dex. Column (9) - Ionized gas mass of the NLR, in M$_\odot$. 
\label{tab_cin_NLR}}
 \end{center}

\end{table*}

As mentioned in section \ref{line_properties}, both sets of Gaussian functions are probably related to different regions of the NLR. One hypothesis is that the Gaussian with the smaller FWHM may actually be a correction to the NLR profile, which may contain asymmetries, as in Seyfert galaxies \citep{1997ApJS..112..391H}. Indeed, \citet{1997ApJS..112..391H} also fitted several Gaussian functions to the NLR. However, these authors did not propose any physical meaning for each Gaussian; they just used this methodology as a convenient fitting procedure. In seven objects, Gaussians from set 2 are more intense than those from set 1, at least in the H$\alpha$ line. The exception is NGC 4546, in which, within the uncertainties, both Gaussians result in the same flux. Gaussian functions from set 2 have FWHMs $>$ 390 km s$^{-1}$, values that are typical for an NLR \citep{2006agna.book.....O}. In NGC 1380 and NGC 3136, both sets may be related to multiple AGNs (NGC 1380 - Steiner et al. in preparation; NGC 3136 - Paper I). It should be noted that both galaxies have the largest reddening in their nuclear regions, possibly a consequence of mergers that may have occurred in these two objects. These events must have added a reasonable amount of gas and dust in the central regions of NGC 1380 and NGC 3136. 

Fig. 4 presented by \citet{2003ApJ...583..159H} showed that most LINERs have nuclear $n_e$ $<$ 400 cm$^{-3}$ (the average value for all the galaxies they studied is 281 cm$^{-3}$), which matches the results shown in Table \ref{tab_cin_NLR}. Only NGC 4546 and IC 1459 have nuclear $n_e$ $>$ 400 cm$^{-3}$, which implies, together with the higher $L_{bol}$ values (see section \ref{edd_ratio}), that these galaxies must have a larger reservoir of gas in their nuclear regions. In the case of NGC 2663, we only estimated a maximum value, which is quite large when compared to the other galaxies. However, the uncertainties for NGC 2663 are quite large ([S II]$\lambda$6716/[S II]$\lambda$6730 = 1.21$\pm$0.31; for this ratio, $n_e$ = 226 cm$^{-3}$). 

\subsection{The broad line regions} \label{BLR_component}

Broad H$\alpha$ components were detected in six galaxies of the sample. In all cases, FWHMs $>$ 2000 km s$^{-1}$. In galaxies where the broad component of H$\alpha$ was found, we supposed that this feature also exists in H$\beta$. The fitting procedures in H$\beta$ were performed with the same radial velocity and the FWHM measured for the H$\alpha$ lines. The ratio between the broad components of H$\alpha$ and H$\beta$ is expected to be higher than the respective ratio between the narrow components, since, in the BLR, these lines are emitted in hot and partially ionized zones, causing an increase in the intensity of H$\alpha$ by collisional effects \citep{2006agna.book.....O}. It is not clear how to determine the reddening effect caused by dust in BLRs, since we cannot properly model the importance of these collisional effects. Thus, the H$\alpha$ luminosities of the broad components, shown in Table \ref{tab_f_BLR}, were corrected for the reddening effects with the E(B-V) of the narrow components, also shown in Table \ref{tab_f_BLR}. Probably, this is underestimated for the BLR, although a fraction of the reddening of the BLR may be caused by the same dust component that quenches the light from the NLR.

\begin{figure*}
\begin{center}
\includegraphics[scale=0.4]{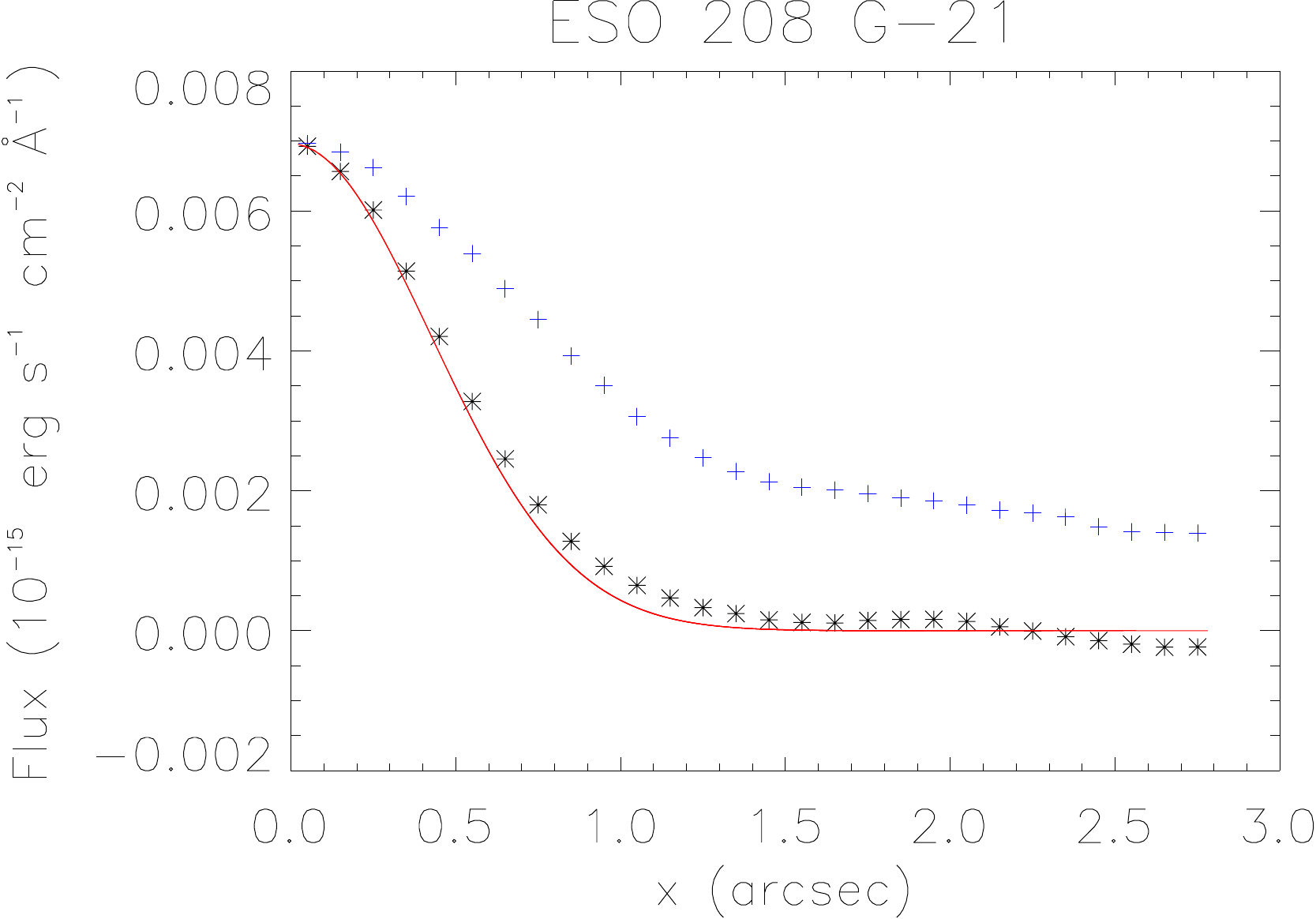}
\includegraphics[scale=0.4]{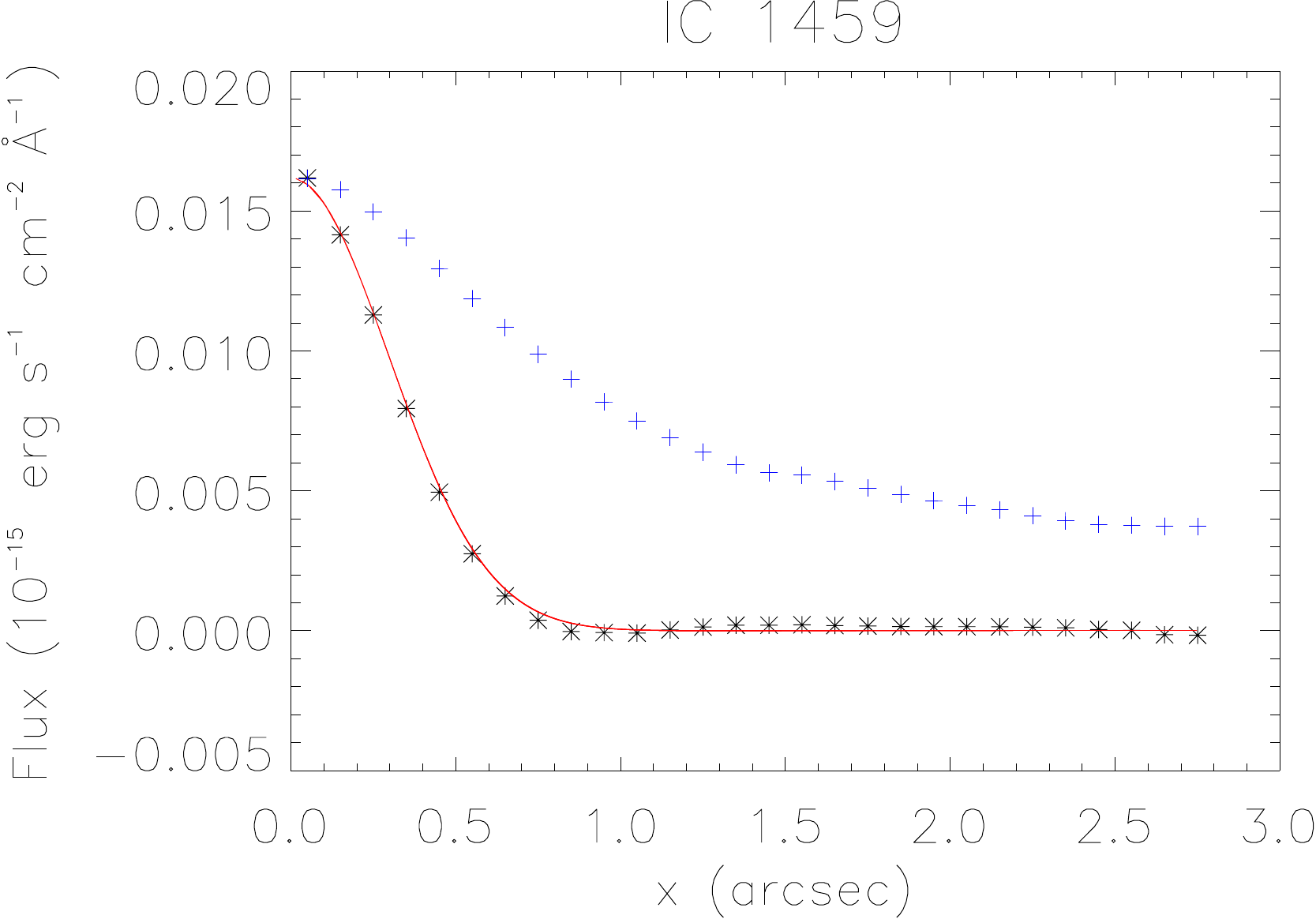}
\includegraphics[scale=0.4]{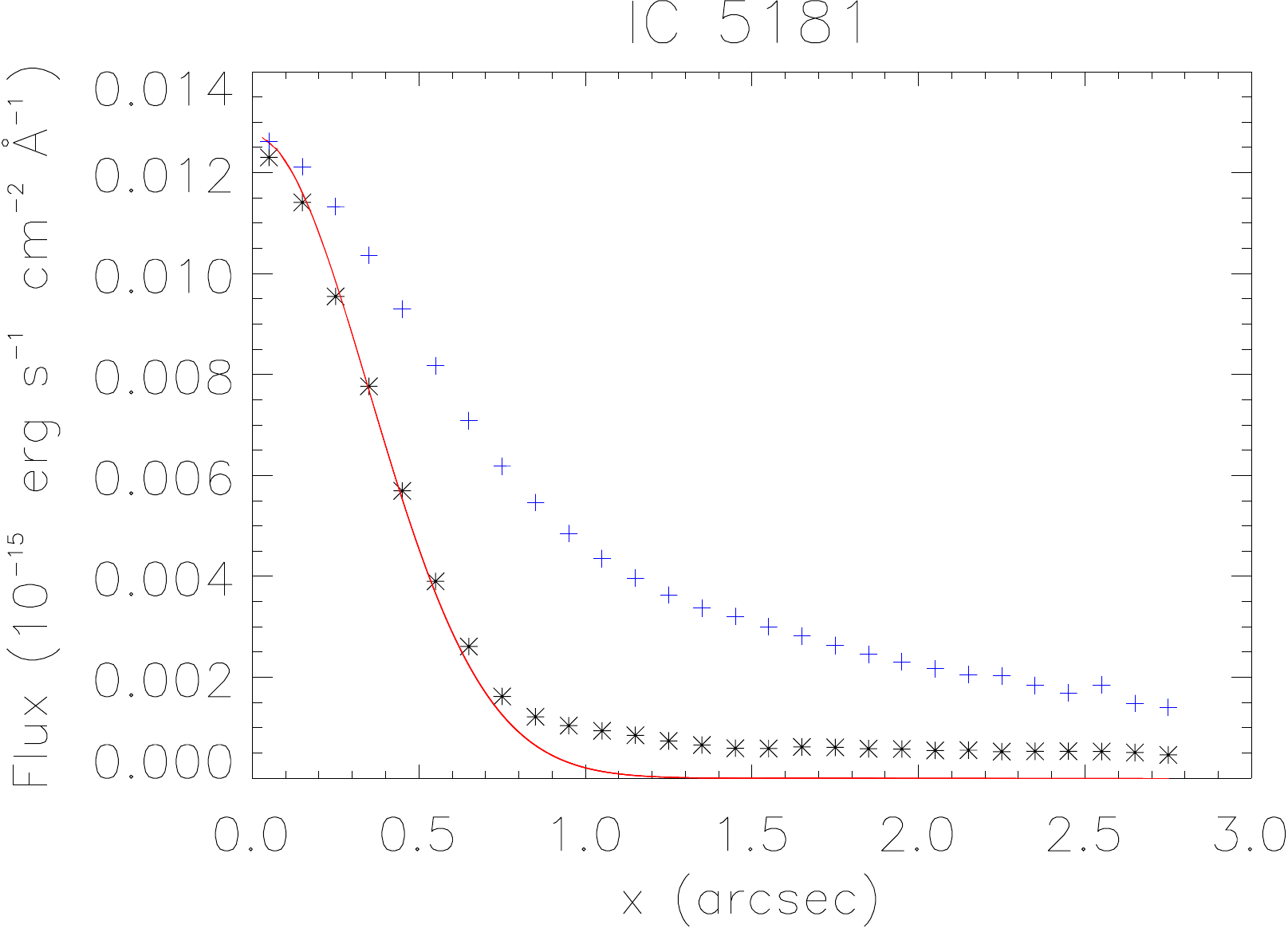}
\includegraphics[scale=0.4]{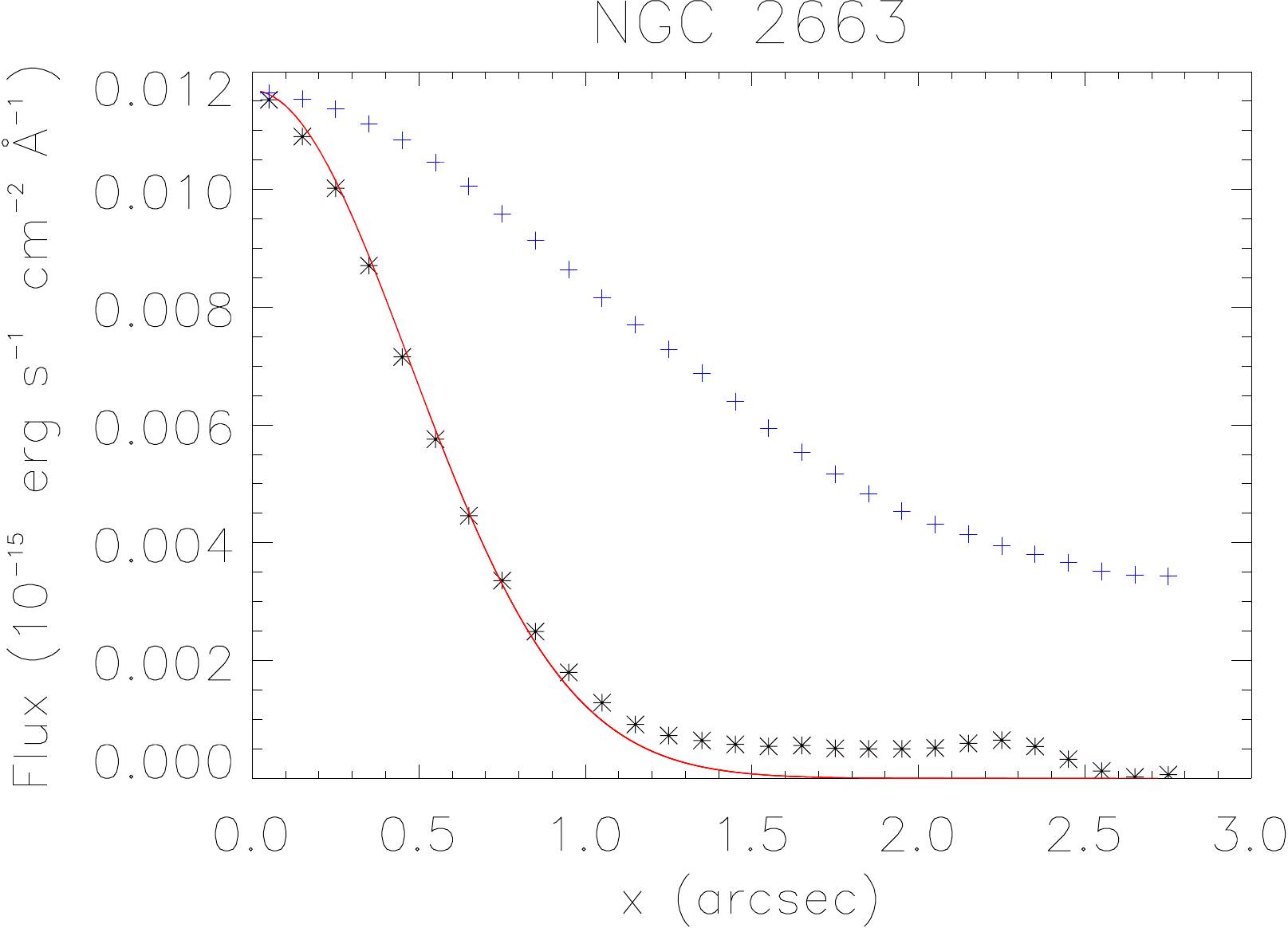}
\includegraphics[scale=0.4]{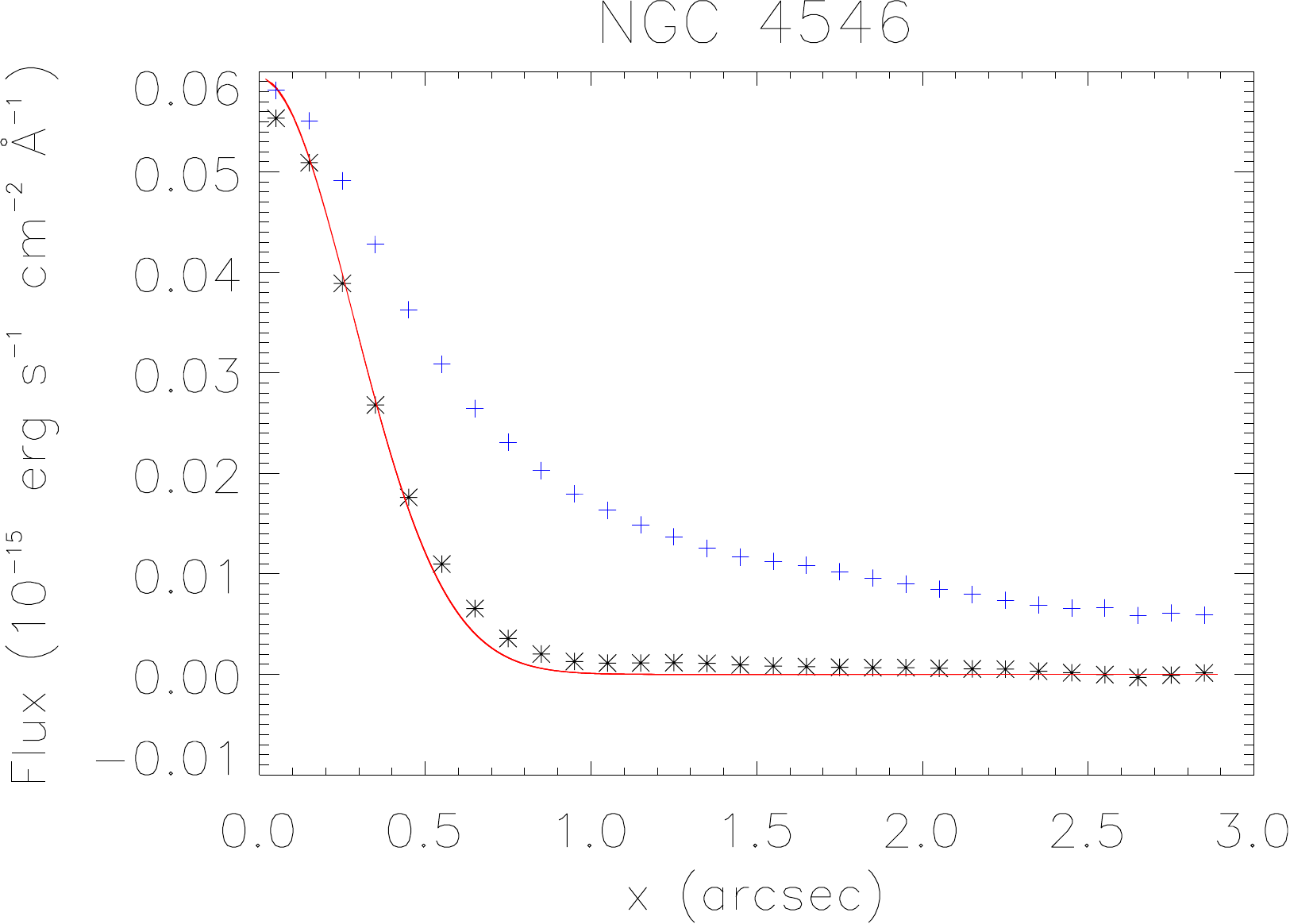}
\includegraphics[scale=0.4]{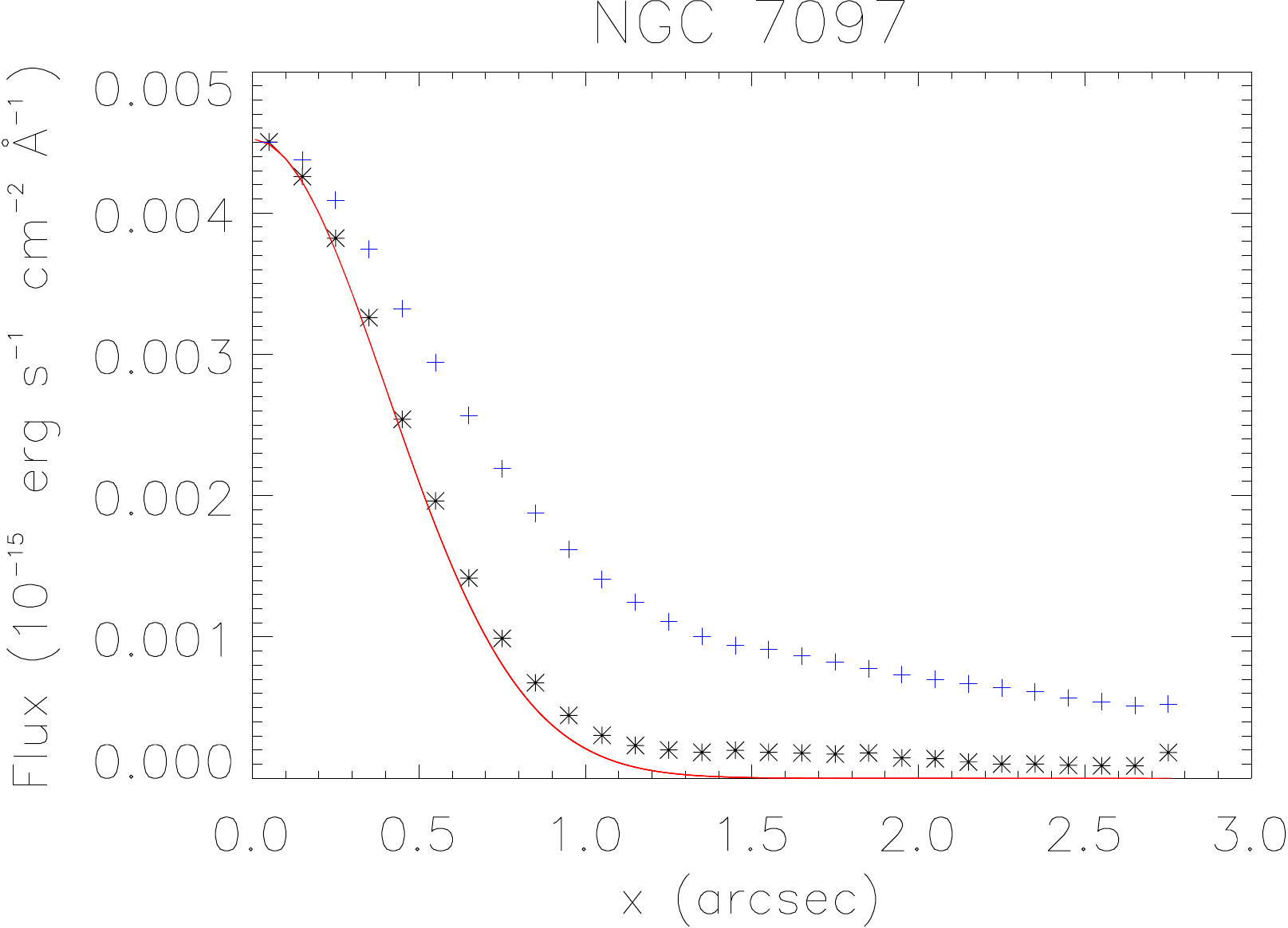}

\caption{Radial profiles of images extracted from the red wings of the broad component of H$\alpha$ of six galaxies of the sample. The red line corresponds to the PSFs of the gas cubes, whose FWHMs are shown in Table \ref{tab_f_BLR}, and the blue crosses are related to the radial profile of the stellar component, extracted from an image of the stellar continuum. The maximum stellar flux was renormalised to the maximum emission of the BLR, so a convenient comparison may be done. The comparison between the stellar and the broad component profiles in five galaxies suggests that the BLR is real and not an effect of a bad starlight subtraction.   \label{perfil_BHa}
}
\end{center}
\end{figure*}

\citet{1997ApJS..112..391H} argued that contamination by late-type stars, in particular features of type K and M giant stars, may generate a bump near H$\alpha$, which may mimic a broad emission line component. Therefore, one should be very careful in starlight subtraction procedures when looking for BLR features, specially in LLAGNs. However, an advantage of the gas cubes is that we are able to make an image of the red wing of the broad component of H$\alpha$, which is less contaminated by the NLR emission. Fig. \ref{perfil_BHa} shows the radial profiles of the these images, taken from six galaxies of the sample where a broad component is clearly detected. To establish a comparison, we also plotted the radial profiles of the stellar components of the six galaxies, taken from the stellar continuum regions of the original data cubes. Note that, in every case, the broad component profiles are much narrower than the stellar profiles. Indeed, the FWHMs of the broad component spatial profiles, shown in Table \ref{tab_f_BLR}, are very similar to the FWHMs of the seeing of the data cubes, whose values were presented in Paper I. Thus, these FWHMs may also represent a direct measurement of the PSF of the gas cubes. The only exception is NGC 2663, where the FWHM of the broad component profile is higher than the FWHM of the seeing. One hypothesis is that the image extracted from the red wing of the broad component of this object was contaminated by the NLR emission, which is quite broad when compared to that of the other galaxies of the sample.  

The results described above imply that the broad components are emitted by an unresolved object, hence corresponding to the BLRs and, consequently, confirming the presence of AGNs in, at least, six galaxies of the sample.

The galaxies NGC 1380, NGC 1399, NGC 1404 and NGC 3136 did not reveal any sign of BLR.

\begin{table*}
 \scriptsize
 \begin{center}
 \begin{tabular}{@{}lcccccc}
  \hline
  Galaxy name & $f$(H$\alpha$)$_b$ & (H$\alpha$/H$\beta$)$_b$  & FWHM(H$\alpha$)$_b$&$V_r$(H$\alpha$)$_b$& log $L$(H$\alpha$)$_b$&FWHM(PSF)\\
       &                     &                          & (km s$^{-1}$)        &                    &                                & (arcsec) \\
  \hline
  ESO 208 G-21 & 281$\pm$0.04 & 10.3$\pm$2.1 & 2455$\pm$66 & 459$\pm$23&  39.99$\pm$0.21 &1.00\\
  
  IC 1459 & 324$\pm$63&17.8$\pm$3.9&2786$\pm$172&518$\pm$44& 40.61$\pm$0.15& 0.70\\
  
  IC 5181 & 87$\pm$7&9.3$\pm$1.4&2127$\pm$55&514$\pm$29&39.87$\pm$0.39&0.82\\
  
  NGC 2663 & 456$\pm$62&51$\pm$48&2952$\pm$135&605$\pm$49&40.79$\pm$0.32&1.11 \\
  
  NGC 4546 & 174$\pm$57&4.3$\pm$1.6&2519$\pm$251&279$\pm$48&39.97$\pm$0.29&0.66\\
  
  NGC 7097 & 182$\pm$6&10.9$\pm$0.9&2538$\pm$39&420$\pm$16& 40.34$\pm$0.05&0.95\\
  \hline
 \end{tabular}
 \caption{Flux and kinematic measurements taken from the broad component. The last column shows the FWHMs, in arcsec, of the broad component profiles shown in Fig. \ref{perfil_BHa}. These values may be related to the PSF of the gas cubes, although the case of NGC 2663 may be peculiar (see text for details). The H$\alpha$ fluxes are in units of 10$^{-15}$ erg s$^{-1}$ cm$^{-2}$, the velocities in km s$^{-1}$ and the luminosities of the broad component of H$\alpha$ in erg s$^{-1}$. The H$\alpha$ luminosities were corrected for the reddening effects with the E(B-V) values presented in Table \ref{tab_f_NLR}. One should have in mind that this correction may be underestimated, since we considered the colour excesses measured for the NLR. The uncertainties related to the luminosity of the H$\alpha$ line took into account the errors in the fluxes, in E(B-V) and in the distances of the galaxies (see Paper I).  \label{tab_f_BLR}
}
\end{center}
\end{table*}

\subsection{Bolometric luminosity and Eddington ratio} \label{edd_ratio}

Bolometric luminosity ($L_{bol}$) and the Eddington ratio ($R_{edd}$) are useful parameters to characterize the LLAGNs' emission. Both parameters suggest a sequence among the types of LLAGNs \citep{2008ARA&A..46..475H}. Seyferts have the highest values for these parameters, followed by LINERs and TOs \citep{2008ARA&A..46..475H}. Ho furthermore proposed that type 1 AGNs are more luminous than type 2 AGNs. Bolometric luminosity is calculated by integrating all the SED. However, a small number of galaxies contain multi-wavelength observations, so relations between $L_{bol}$ and the X-ray luminosity \citep{2008ARA&A..46..475H,2010ApJS..187..135E} or emission line luminosities, as [O III]$\lambda$5007 \citep{2005ApJ...634..161H} or  H$\alpha$ \citep{2008ARA&A..46..475H}, must be applied. An issue with the H$\alpha$ line is that its non-nuclear emission is quite important in LINERs, leading one to overestimate $L_{bol}$. On the other hand, according to \citet{2003MNRAS.346.1055K}, the [O III] line is an important tracer of nuclear luminosity, since it is intense and produced in the NLR. In galaxies with AGNs, [O III] emissions emerging from H II regions are relatively small \citep{2003MNRAS.346.1055K}.  

\citet{2010ApJS..187..135E} calculated $L_{bol}$ for seven galaxies by integrating their respective SEDs. The [O III] luminosities of these objects were taken from \citet{1997ApJS..112..315H}. With both parameters in hand, we calculated the median of the $L_{bol}$/$L$([O III]) ratio as $\sim$ 584, which should be reasonable within 0.48 dex. Through this relation, we then estimated $L_{bol}$ for the galaxies of the sample, except for NGC 1399 and NGC 1404, which contain only [N II]+H$\alpha$ emission (see section \ref{cD_cases}). In these cases, we used the relation proposed by \citet{2008ARA&A..46..475H}, given by $L_{bol}$/$L$(H$\alpha$) $\sim$ 220 with an intrinsic scatter of $\sim$ 0.4 dex. Despite this scatter being smaller than the one estimated for $L_{bol}$/$L$([O III]), one should have in mind that the use of the H$\alpha$ line for estimating $L_{bol}$ is more affected by systematical errors. Estimated values of $L_{bol}$ are shown in Table \ref{tab_cin_NLR}.

The Eddington ratio is given by

\begin{equation}
	R_{edd} = \frac{L_{bol}}{L_{Edd}} = \frac{L_{bol}}{1.3\times10^{38}\times M_{SMBH}}
	\label{eddingtonratio}
\end{equation}
where $L_{Edd}$ is the Eddington luminosity. The SMBH masses were estimated with the $M_{SMBH}-\sigma$ relation proposed by \citet{2009ApJ...698..198G}, using the $\sigma$ measurements of \citet{ricci2013}. $R_{Edd}$ values for the galaxies of the sample are presented in Table \ref{tab_cin_NLR}. Since the intrinsic scatter of the $M_{SMBH}-\sigma$ relation is 0.48 dex \citep{2009ApJ...698..198G} and our confidence in $L_{bol}$ is about 0.44 dex, then the error in $R_{Edd}$ should be $\sim$ 1 dex. 

\begin{figure*}
\begin{center}
\includegraphics[scale=0.86]{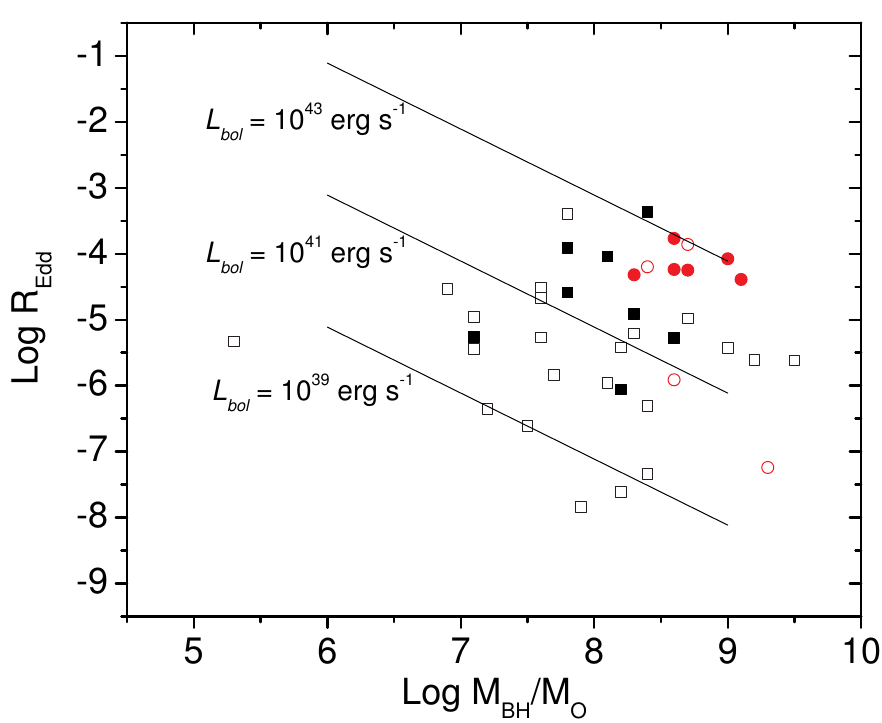}
\caption{Eddington ratio x SMBH mass. The hollow (filled) black squares correspond to type 2 (1) AGNs from the \citet{2010ApJS..187..135E} sample. The hollow (filled) red circles are related to type 2 (1) AGNs from our sample. The black lines correspond to $L_{bol}$ = 10$^{39}$, 10$^{41}$ and 10$^{43}$ erg s$^{-1}$. Most type 1 AGNs from the \citet{2010ApJS..187..135E} sample are in the same region of this plot (higher $L_{bol}$ and $R_{Edd}$). NGC 1380 and NGC 3136, although type 2 AGNs, must contain multiple photoionization sources that contribute to rise $L_{bol}$ values in these objects. In NGC 1399 and NGC 1404, their $L_{bol}$ are lower when compared to the other galaxies of our sample. \label{ReddxMbh}
}
\end{center}
\end{figure*}

The galaxies whose BLRs were detected have $R_{Edd} \sim $ 10$^{-4}$ and $L_{bol} >$ 10$^{42}$ erg s$^{-1}$. Comparing our results with those of \citet{2010ApJS..187..135E}, type 1 AGNs are located in the same region of the $R_{Edd}$ x $M_{SMBH}$ graph (Fig. \ref{ReddxMbh}). Despite NGC 3136 and NGC 1380 being type 2 AGNs (Paper I and section \ref{BPT_diagrams} of this work), they are located in the same region of type 1 AGNs in Fig. \ref{ReddxMbh}. This may happen because these objects probably have multiple ionization sources (Paper I; Steiner et al. in preparation), which may increase the estimated $L_{bol}$ for both galaxies. One can see that type 1 AGNs drawn from \citet{2010ApJS..187..135E} are more luminous and have higher $R_{Edd}$ values. This corroborates the fact that type 1 AGNs are more luminous than type 2 AGNs, as proposed by \citet{2008ARA&A..46..475H}. Another interesting result is that all the galaxies of the sample have $R_{Edd} < $ 10$^{-3}$, implying that their AGNs should have SEDs typical of LINERs \citep{2008ARA&A..46..475H}, with an absent big blue bump and a redder continuum in the optical region (specially in the range between 10000 \AA\ and 4000 \AA).

\subsection{Diagnostic diagrams} \label{BPT_diagrams}

\begin{figure*}
\includegraphics[width=84.7mm,height=72.6mm]{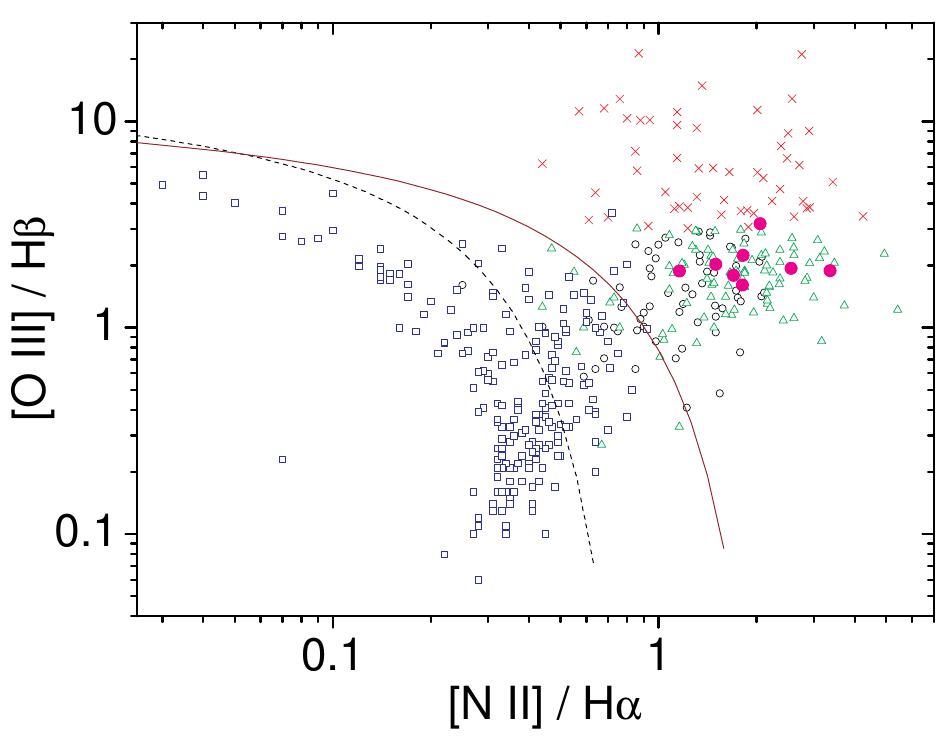}
\includegraphics[width=84.7mm,height=72.6mm]{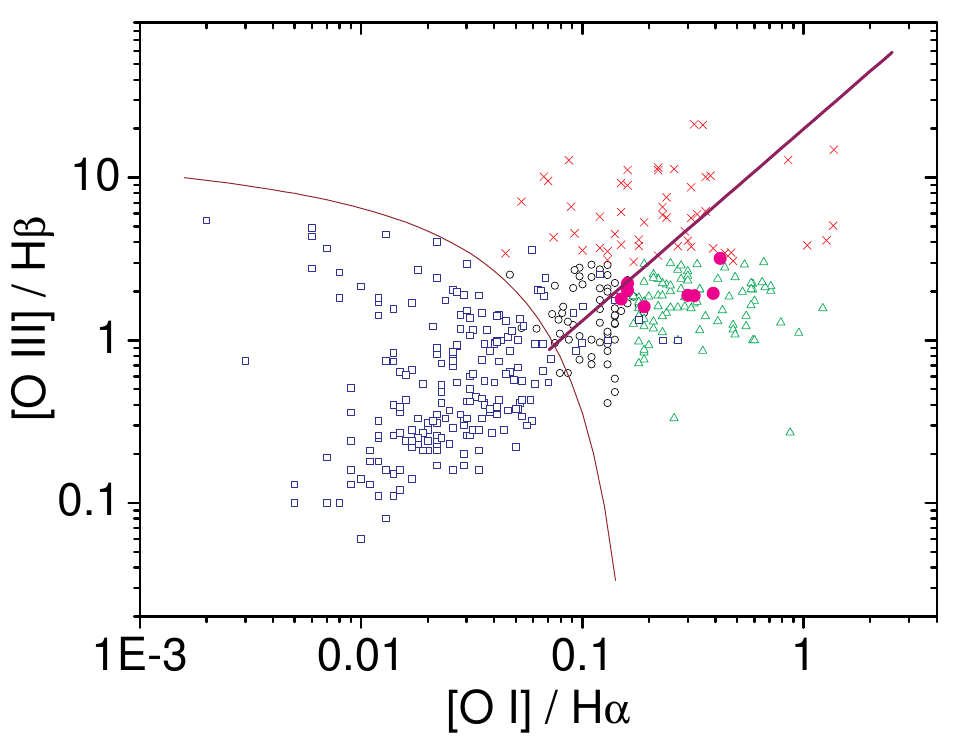}
\includegraphics[width=84.7mm,height=72.6mm]{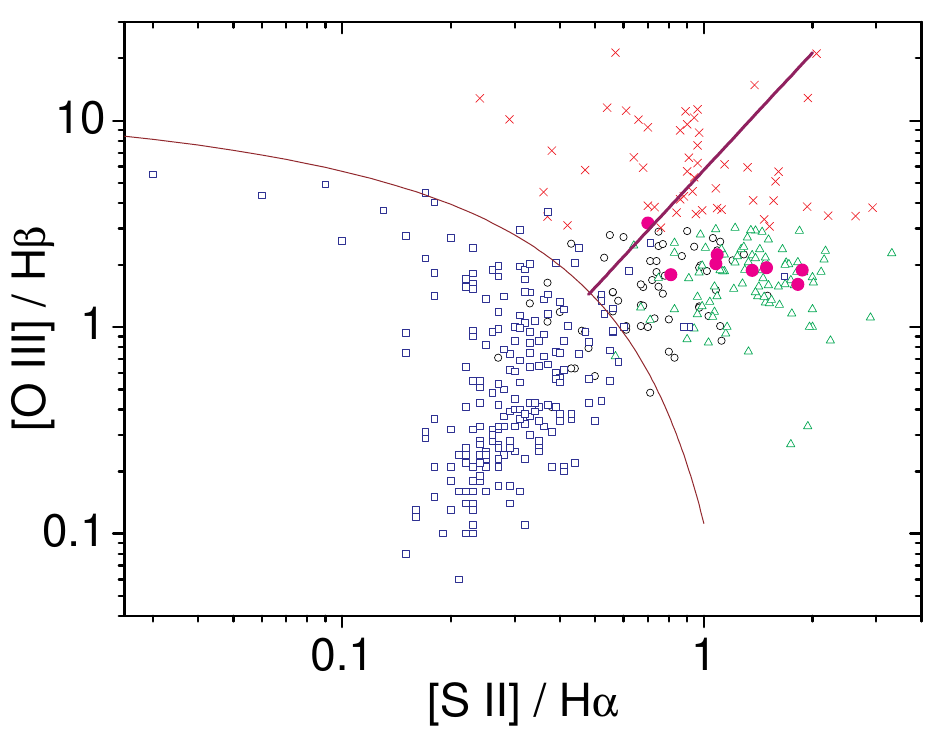}
\caption{Diagnostic diagrams. Eight galaxies of our sample are represented by the filled magenta circles. Palomar survey galaxies were also inserted in the graphics. According to the classification proposed by \citet{1997ApJS..112..315H}, the red crosses are for Seyferts, green triangles are for LINERs, open black circles are for OTs and blue squares are for H II regions. The thin brown line is for the maximum starburst line proposed by \citet{2001ApJ...556..121K}, the dashed black line is for the empirical division between H II regions and AGNs \citep{2003MNRAS.346.1055K} and the thick purple line is for the LINER-Seyfert division suggested by \citet{2006MNRAS.372..961K}.\label{BPTdiagrams} 
 }
\end{figure*}

In the diagnostic diagrams shown in Fig. \ref{BPTdiagrams}, we inserted eight galaxies of our sample together with the galaxies of the Palomar survey \citep{1997ApJS..112..315H}. These eight galaxies may be classified as LINERs, although IC 5181, NGC 1380 and NGC 3136 may be defined, within the uncertainties, as TOs. In NGC 1380, two H II regions and three AGN candidates, which may account for the nuclear line emissions, are discussed in detail in Steiner et al. (in preparation). In Paper I, PCA Tomography revealed that at least two point-like objects are detected in the central region of NGC 3136. In Paper III, we will show that, besides these two objects, NGC 3136 has several ionization sources in the central region. Also within the errors, NGC 4546 may be classified as Seyfert. Moreover, this galaxy has the densest nuclear region in the sample, as well as the highest $R_{edd}$ and $L_{bol}$ values. Thus, it is expected that NGC 4546 is closer to the Seyfert classification than the other galaxies of our sample. Finally, four galaxies (IC 1459, ESO 208 G-21, NGC 2663 and NGC 7097) are genuine LINERs.

\subsection{Two special cases: NGC 1399 and NGC 1404} \label{cD_cases}

\begin{figure*}
\begin{center}
\includegraphics[width=70mm,height=60mm]{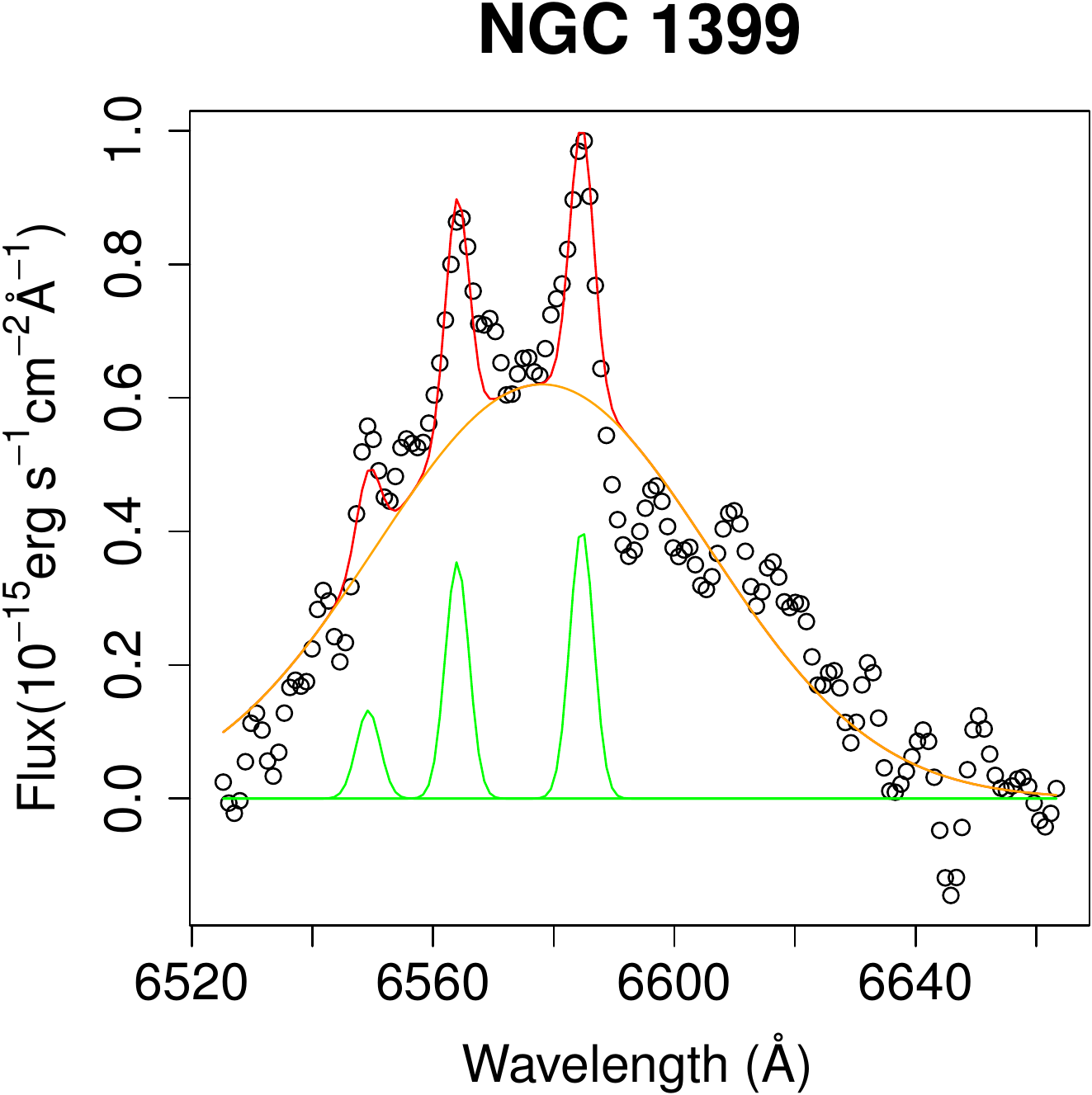}
\includegraphics[width=70mm,height=60mm]{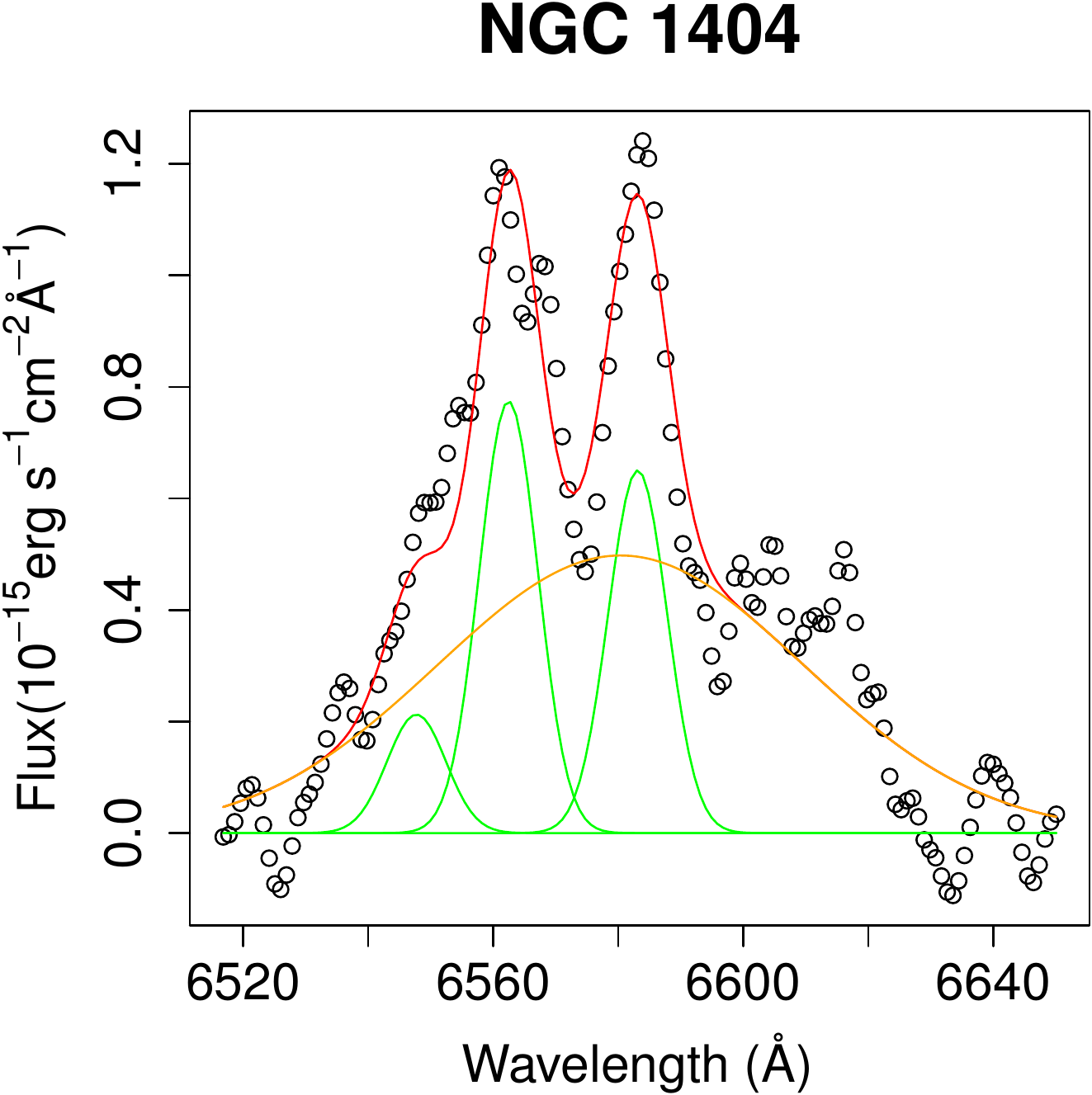}

\caption{H$\alpha$ + [N II] emission lines detected in NGC 1399 and NGC 1404. Note that both objects contain a broad component superposed on the emission lines, shown in orange, which must be caused by bad starlight subtraction rather than by a BLR. Green Gaussian functions are related to the fitting procedure of the lines while the red lines are related to the sum between the green and the orange profiles. \label{perfil_cD_NII_Ha}
}
\end{center}
\end{figure*}

In Paper I, PCA Tomography applied to the data cubes of NGC 1399 and NGC 1404 did not reveal any sign of ionized gas in their central regions. We proposed that emission lines could exist, but their variances would be of the same order of the variance of the noise contained in the data cubes. The question now is: Can we detect emission lines after the subtracting the stellar spectra derived from spectral synthesis? Indeed, in the spectra extracted from their nuclear regions, H$\alpha$ and [N II] lines can be seen in both galaxies. However, the line decomposition is not simple in either case. In Fig. \ref{perfil_cD_NII_Ha}, we may note that, in both galaxies, the lines are superposed on a broad component. In these cases, this component must be associated with a bad starlight subtraction rather than with a BLR (see Appendix \ref{starlight_results}). With that in mind, we decomposed the H$\alpha$ and [N II] lines with one Gaussian function per emission line, plus another Gaussian function to account for this broad feature. As done before, we fitted all three Gaussians with the same radial velocity and FWHM and with the [N II]$\lambda6583$/[N II]$\lambda6548$ ratio fixed at 3.06. The results are shown in Fig. \ref{perfil_cD_NII_Ha}. Although the narrow component of H$\alpha$ may be also a result of an imperfect subtraction of the stellar component, the [N II] doublet undoubtedly suggests a nebular emission. The level of detection of the lines in terms of peak amplitude of fit residual level for NGC 1399 is 4.5 for H$\alpha$ and 5.1 for [N II]$\lambda$6583. For NGC 1404, the level of detection is 6.8 for H$\alpha$ and 5.7 for [N II]$\lambda$6583. Moreover, only the integrals of the emission line Gaussian functions were taken into account for in the flux measurements, as shown in Table \ref{tab_f_NLR}. The kinematic features of the lines and the H$\alpha$ luminosity are presented in Table \ref{tab_cin_NLR}. It is worth mentioning that H$\alpha$ luminosities were not corrected for extinction effects in these galaxies, since the H$\beta$ line was not detected in both spectra.



The H$\alpha$ emission lines indicate, for NGC 1399 and NGC 1404 respectively, $L_{bol}$ $\sim$ 10$^{40}$ erg s$^{-1}$ and $L_{bol}$ $\sim$ 10$^{41}$ erg s$^{-1}$. Both results match the $L_{bol}$ estimated from the X-ray luminosities of these galaxies. Assuming $L_{bol}$ = 16L$_X$ for LLAGNs \citep{2008ARA&A..46..475H} and knowing that $L_X$ $\sim$ 10$^{40}$ erg s$^{-1}$ \citep{2011ApJ...731...60G} for NGC 1404 and $L_X$ $<$ 9.7$\times$10$^{38}$ erg s$^{-1}$ \citep{2005ApJ...635..305O} for NGC 1399, then $L_{bol}$ $\sim$ 1.6 $\times$ 10$^{41}$ and $L_{bol}$ $<$ 1.5$\times$10$^{40}$ erg s$^{-1}$ for NGC 1404 and NGC 1399, respectively. These results imply that, in both galaxies, the X-ray and H$\alpha$ emissions may be related to AGNs. 

We also analysed the H$\alpha$ and [N II] emission lines of these galaxies with the WHAN diagram \citep{2011MNRAS.413.1687C}, which compares the [N II]/H$\alpha$ ratio with the equivalent width of the H$\alpha$ line. To do so, we measured the equivalent width of H$\alpha$, resulting in  $EW(H\alpha)$ = 0.06 $\pm$ 0.01 and 0.24 $\pm$ 0.03 \AA, for NGC 1399 and NGC 1404, respectively. According to the classification criteria of the WHAN diagram, both are passive galaxies, i.e., there are no detectable emission lines. In fact, NGC 1404 could almost be classified as a retired galaxy, whose gas emission is accounted for by pAGBs stellar populations. NGC 1399, on the other hand, is a classic radio galaxy \citep{2008MNRAS.383..923S}. These authors estimated the jet power for the AGN of this object as $P_{jet} \sim$ 10$^{42}$ erg s$^{-1}$. Comparing it with $L_{bol}$ $\sim$ 10$^{40}$ erg s$^{-1}$, it follows that most of the energy released by the AGN of NGC 1399 must be mechanical. In NGC 1404, in addition to a point-like source detected in X-ray, \citet{2005ApJ...635..305O} also reported an extended X-ray emission, which may emerge from circumnuclear regions, as observed in some Seyfert 2 galaxies. The spectra extracted from the gas cubes must be affected by systematic errors, which emerged from a bad starlight subtraction. On the other hand, under LLAGN assumption, the H$\alpha$ luminosities match the X-ray emission of these galaxies. Probably, the high systematical uncertainties associated with the emission lines measurements do not allow an accurate analysis of these galaxies in the context of the WHAN diagram (one should have in mind that the errors presented for $EW(H\alpha)$ are purely statistical).

\section{Discussion and Conclusions} \label{sec:conc}

Overall, the results presented in this paper are consistent with those from Paper I. We found that all 10 galaxies of the sample have some nuclear activity. Moreover, we detected emission lines in NGC 1399 and NGC 1404, although both H$\alpha$ luminosities are quite low. This may cause a variance of the emission lines of the same order of magnitude as the variance of the noise in the data cubes of both galaxies. This explains the non-detection of their gas emission with PCA Tomography in Paper I. NGC 1380 and NGC 3136 contain the highest E(B-V) values among the galaxies of the sample. In Paper I, we proposed that the correlation between the emission lines and the red region of the continuum in the eigenspectra that revealed the AGNs could be caused by dust extinction. In ESO 208 G-21, IC 5181, NGC 1380, NGC 3136, NGC 4546 and NGC 7097, the Na D absorption lines were also correlated with the red region of the continuum, thus indicating an extinction (dust) associated with the interstellar gas. Probably in NGC 1380 and NGC 3136, a fraction of correlation between the red region of the continuum and the emission lines in the AGN eigenspectrum is also caused by the nebular extinction.

Within the uncertainties, all the galaxies of the sample may be classified as LINERs. Nevertheless, other classifications should not be ignored. For instance, the AGN of NGC 4546 is on the limit between LINERs and Seyferts. In fact, its Eddington ratio is the highest among the galaxies of the sample. Besides, its bolometric luminosity is only lower, within the errors, than the $L_{bol}$ of the AGN of IC 1459. Thus, a Seyfert classification is plausible for NGC 4546, according to the diagnostic diagram and to what is predicted by the $L_{bol}$ and $R_{Edd}$ distributions obtained from the Palomar survey (see fig. 9 of \citealt{2008ARA&A..46..475H}). 

In six galaxies of the sample, we detected broad H$\alpha$ and H$\beta$ emissions, which corroborates the presence of an AGN in these objects. In all cases, FWHMs $>$ 2000 km s$^{-1}$. \citet{1997ApJS..112..391H} found similar FWHM values for this feature in galaxies from the Palomar survey (see their Table 1). Since these features are weak in LINERs, one should be cautious when characterizing these components. Gaussian fits performed in the [N II]+H$\alpha$ region are similar to what \citet{1997ApJS..112..391H} have done. A difficulty highlighted by \citet{1997ApJS..112..391H} is that, since the narrow components do not show exactly a Gaussian profile, broader wings from the NLR may also contaminate the BLR. This may be happening for NGC 2663, whose NLR Gaussian profiles are the broadest among the galaxies of the sample. Besides, the PSF measured with image from the red wing of the broad H$\alpha$ component is larger than the seeing estimated for the observation of this galaxy (see Paper I). Although the seeing of the observation and the PSF of the data cube are not the same thing, one should expect similar FWHMs for both parameters. Another systematical effect which may affect the analysis of the BLR is a bad starlight subtraction. Notwithstanding, the spatial profiles extracted from the images of the red wings of the BLR reveal point-like objects in these six objects. This agrees with an AGN interpretation. Since the broad components have $V_r$ $>$ 280 km s$^{-1}$, the red wings that emerge from the BLR have a huge fraction of the intensity of this feature.    

In NGC 1399 and NGC 1404, the H$\alpha$ + [N II] emission lines were detected in their nuclear region. A comparison between H$\alpha$ luminosities and X-ray data from both objects suggests that NGC 1399 and NGC 1404 must contain AGNs with very low luminosities, although their $EW(H\alpha)$ measurements indicate that these emissions are too weak to be produced by AGNs. In fact, the detection of a UV point-like source and also strong radio jets in NGC 1399 \citep{2005ApJ...635..305O,2008MNRAS.383..923S} leave no doubt about the existence of an AGN in this galaxy. In NGC 1404, according to \citet{2011ApJ...731...60G}, the combination between X-ray and IR (2MASS) data results in a bolometric luminosity that is high enough for this galaxy to contain an AGN. 

Below, we summarize the main conclusions of this work.

\begin{itemize}

\item	With well-established methodology of spectra analysis, we show that the main conclusions of Paper I were duly recovered. This shows that PCA Tomography may be used as an useful tool to extract informations from data cubes.
\item	We show that all the galaxies of the sample have emission lines in their nuclear regions. Among the 10 galaxies, in two (NGC 1399 and NGC 1404) PCA Tomography was not able to detect the emission lines. With spectral synthesis techniques, however, we show that both galaxies have very low intensity emission lines. This indicates that, for very low luminosities AGNs, stellar spectral synthesis may be more effective than PCA Tomography in order to detect emission lines related, for instance, to AGNs. 
\item	In six galaxies of the sample, a broad H$\alpha$ component is detected, which evidences the existence of a BLR in these objects. This feature proves that these six galaxies have an AGN as central objects and a likely photoionization source. Multi-Gaussian decomposition of the H$\alpha$+[N II] set shows that these six galaxies possess a NLR with two kinematic features: a narrower one with 50 $<$ FWHM $<$ 500 s$^{-1}$ and an intermediate one with 400 $<$ FWHM $<$ 1000 km s$^{-1}$.
\item	Among the four galaxies without a detected BLR, two may contain multiple objects in their centre: NGC 1380 and NGC 3136. These galaxies also revealed a diffuse emission that may be related to H II regions.
\item	In the other two galaxies, NGC 1399 and NGC 1404, the H$\alpha$+[N II] emission lines have very low intensity. However, X-ray and radio data indicate that they also must contain AGNs. 
\item	Judging from the emission line luminosities, all 10 galaxies have $L_{bol}$ between 10$^{40}$ erg s$^{-1}$ and 10$^{43}$ erg s$^{-1}$, which corresponds to $R_{edd}$ between 10$^{-6}$ and 10$^{-3}$. These Eddington ratios are significantly smaller than the ones found for Seyfert galaxies, but resemble those of LINER-like emissions \citep{2008ARA&A..46..475H}. 

\end{itemize}

\section*{Acknowledgments}

This paper is based on observations obtained at the Gemini Observatory, which is operated by the Association of Universities for Research in Astronomy, Inc., under a cooperative agreement with the NSF on behalf of the Gemini partnership: the National Science Foundation (United States), the National Research Council (Canada), CONICYT (Chile), the Australian Research Council (Australia), Minist\'{e}rio da Ci\^{e}ncia, Tecnologia e Inova\c{c}\~{a}o (Brazil) and Ministerio de Ciencia, Tecnolog\'{i}a e Innovaci\'{o}n Productiva (Argentina). IRAF is distributed by the National Optical Astronomy Observatory, which is operated by the Association of Universities for Research in Astronomy (AURA) under cooperative agreement with the National Science Foundation.

T.V.R and R.B.M. also acknowledge FAPESP for the financial support under grants 2008/06988-0 (T.V.R.), 2012/21350-7 (T.V.R.) and 2012/02262-8 (R.B.M.). We also thank the anonymous referee for valuable suggestions that improved the quality of this paper.

\bibliographystyle{mn2e}
\bibliography{bibliografia}

\appendix

\section{Results of the stellar spectral synthesis performed for the sample galaxies} \label{starlight_results}

In this appendix, we present the results from the stellar spectral synthesis performed on the sample galaxies. The idea is to show the behaviour of the fitting procedures in spectra with both high and low S/N ratios. In Figs. \ref{starlight_graf_1}, \ref{starlight_graf_2} and \ref{starlight_graf_3}, we present the spectra, the fitted results and the residuals for a representative spaxel of the galaxies' centre and for a spaxel located at the upper-right region of the FOVs, $\sim$ 2 arcsec away from the galaxies' centre. 

\renewcommand{\thefigure}{\arabic{figure}\alph{subfigure}}
\setcounter{subfigure}{1}

\begin{figure*}
\begin{center}
\includegraphics[width=70mm,height=55mm]{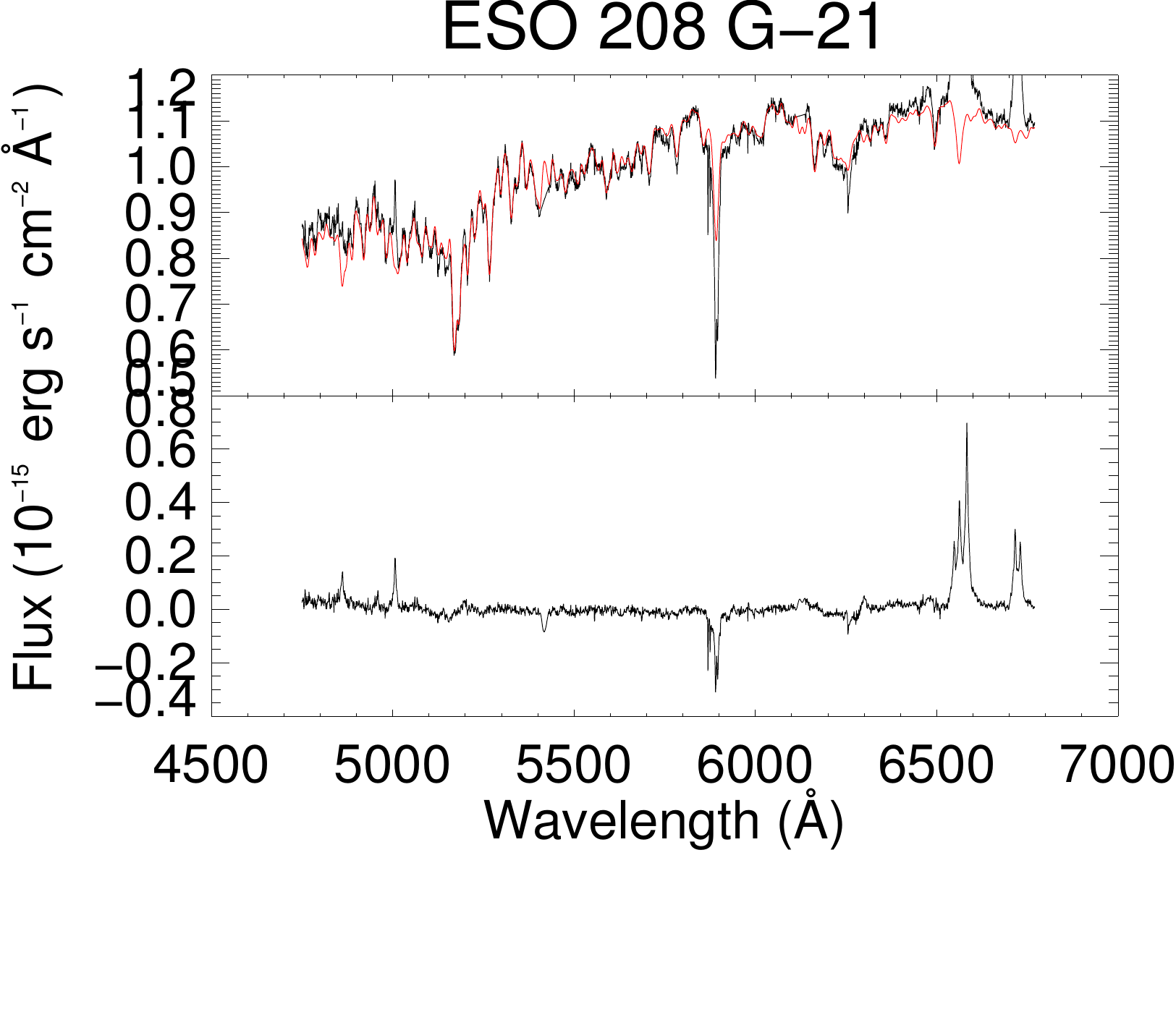}
\includegraphics[width=70mm,height=55mm]{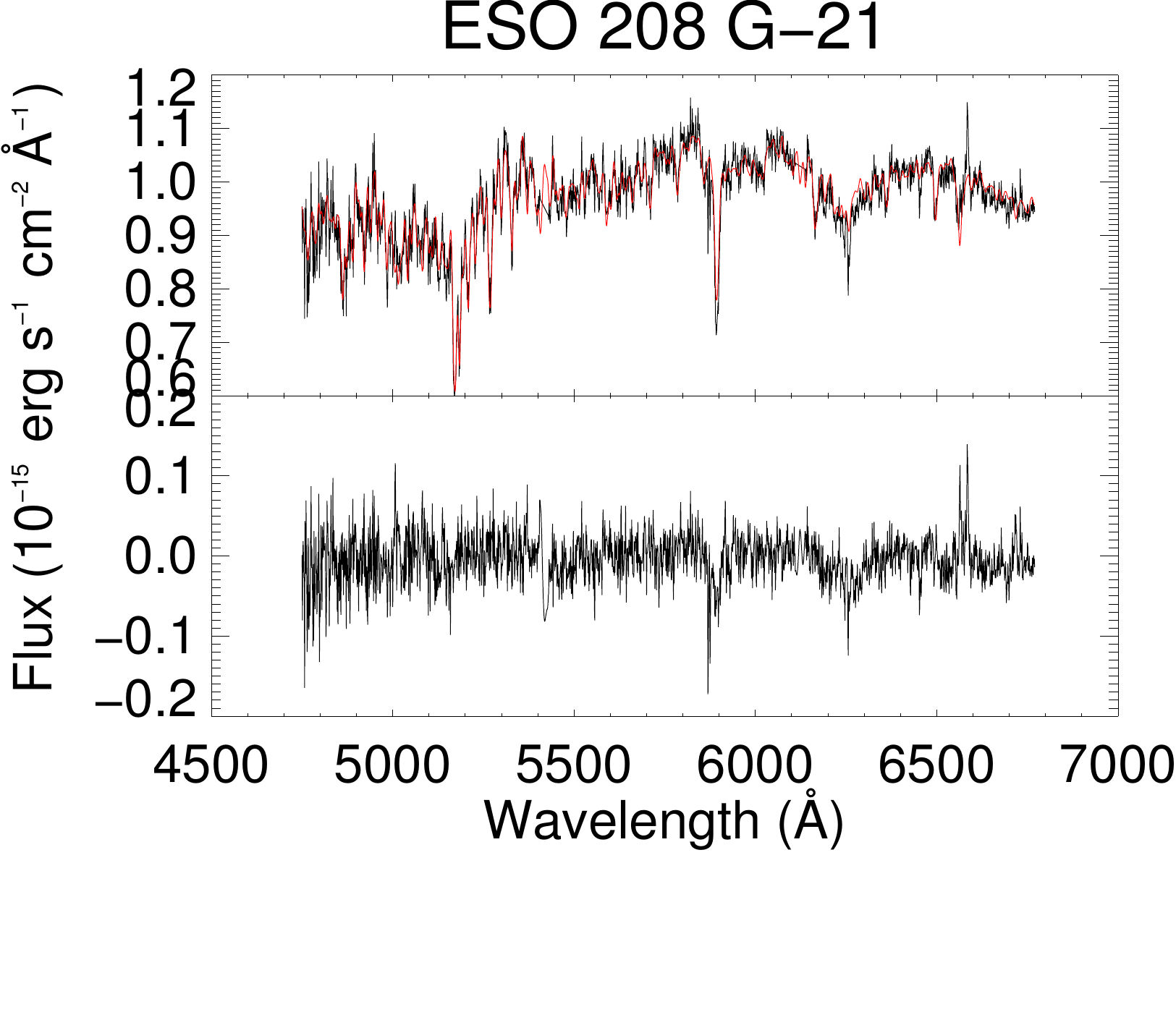}
\includegraphics[width=70mm,height=55mm]{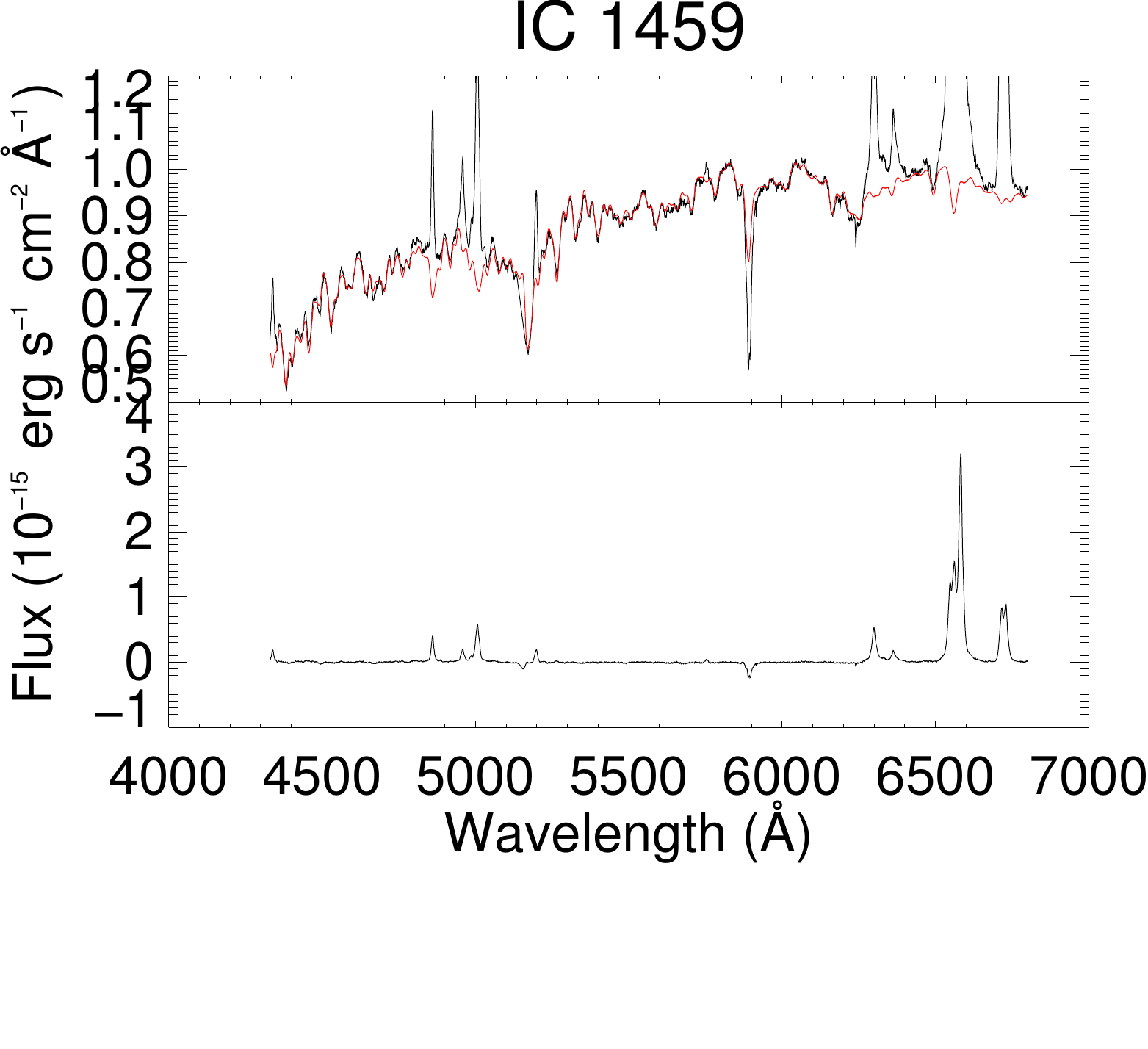}
\includegraphics[width=70mm,height=55mm]{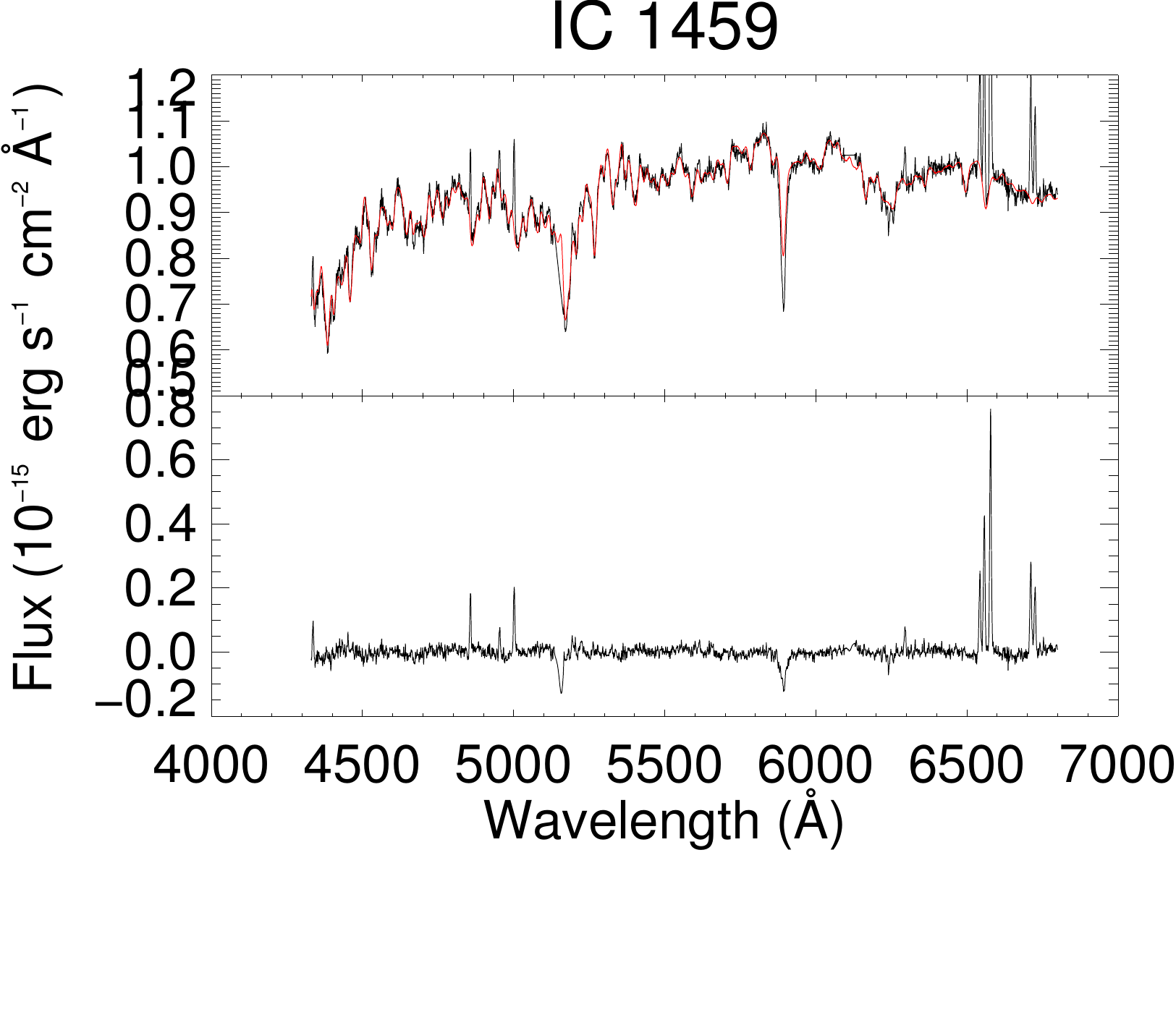}
\includegraphics[width=70mm,height=55mm]{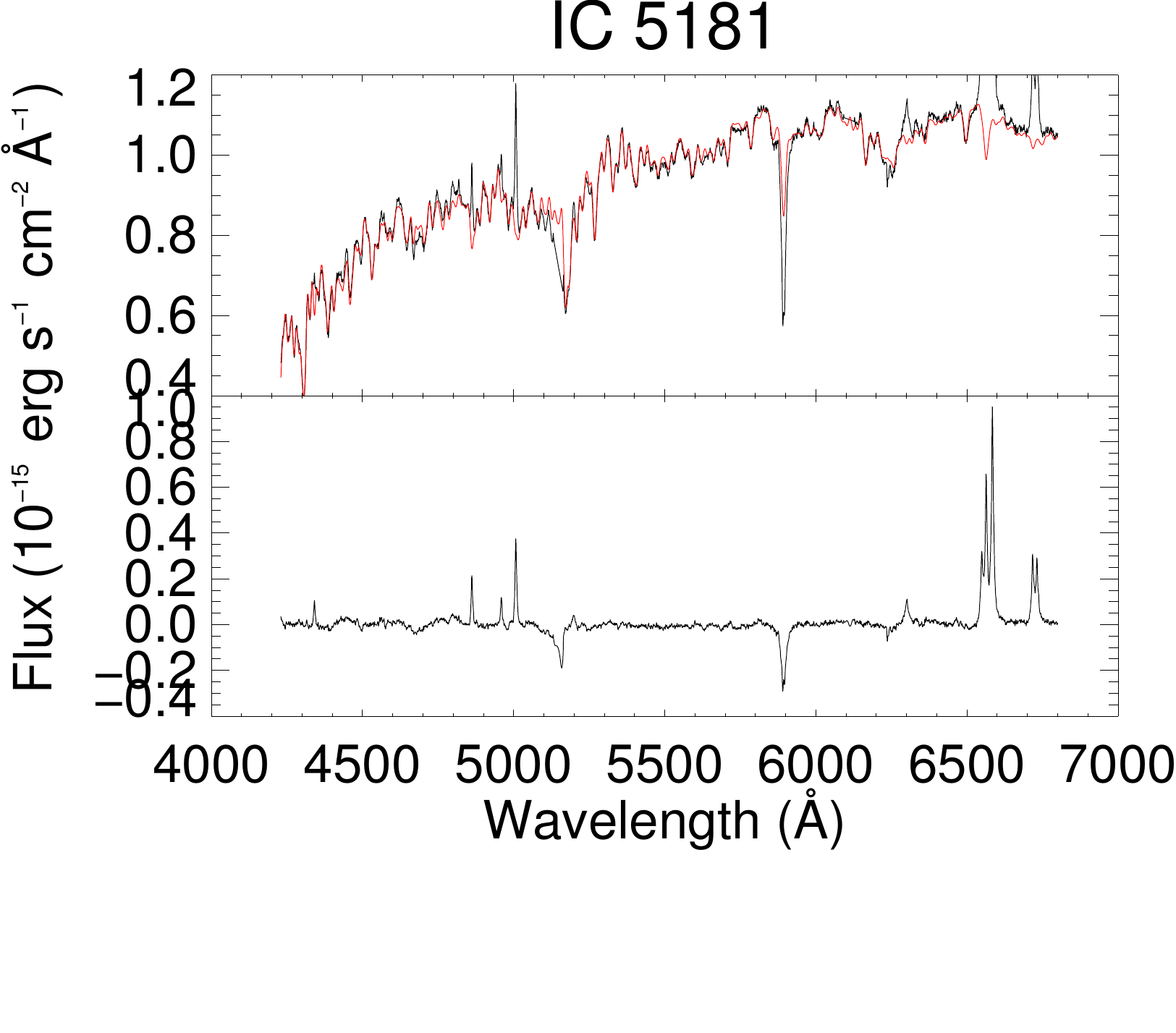}
\includegraphics[width=70mm,height=55mm]{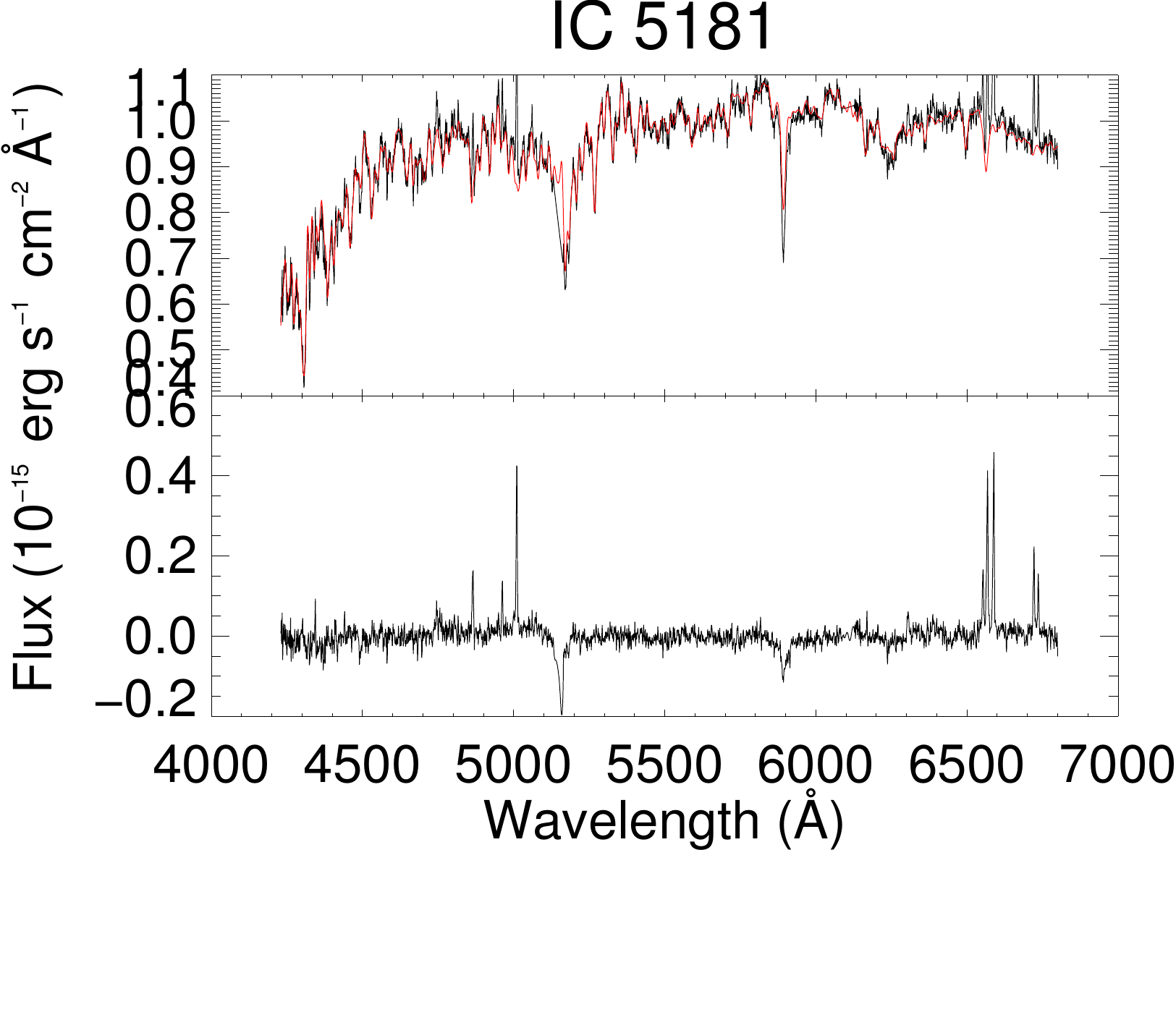}
\includegraphics[width=70mm,height=55mm]{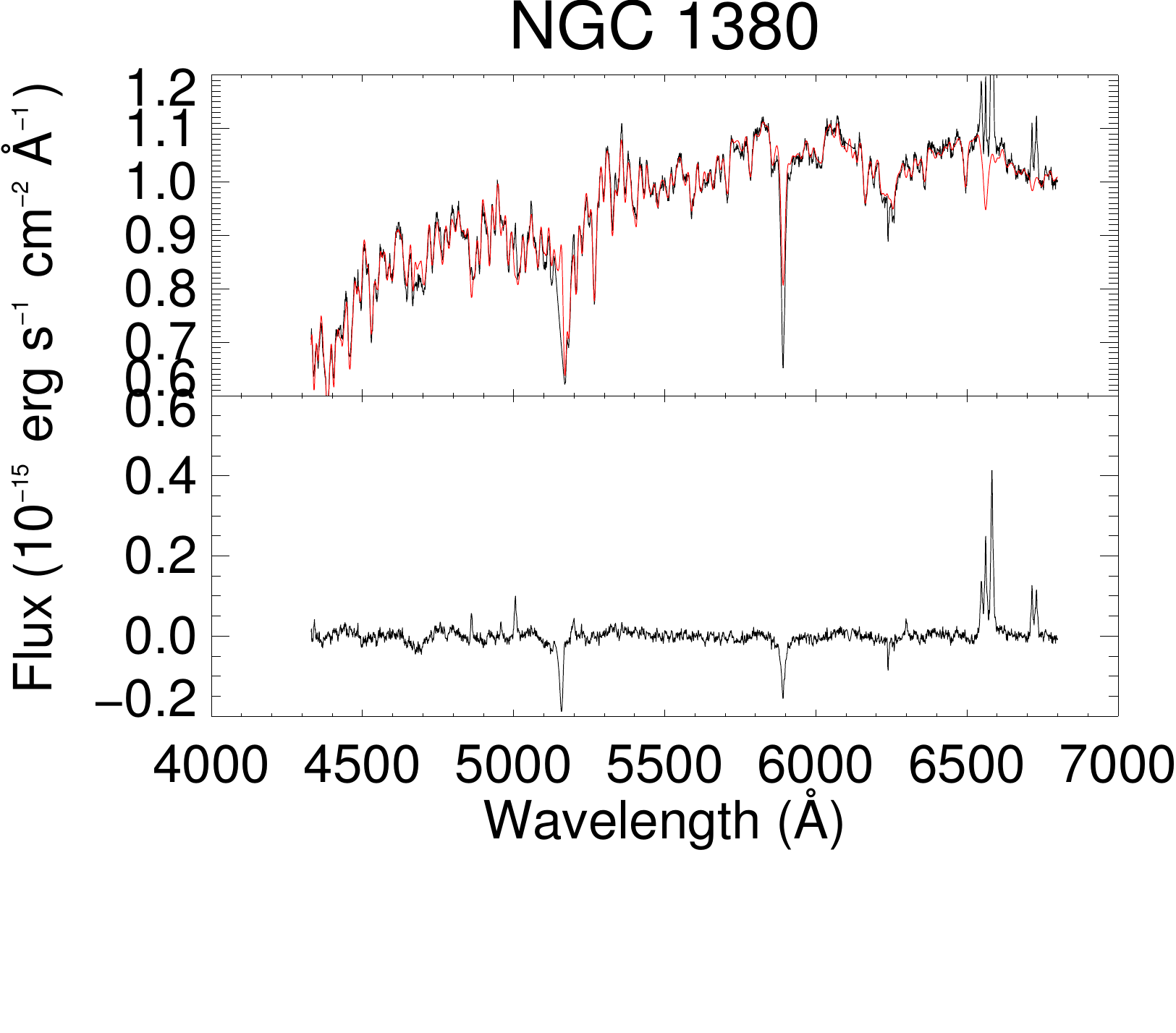}
\includegraphics[width=70mm,height=55mm]{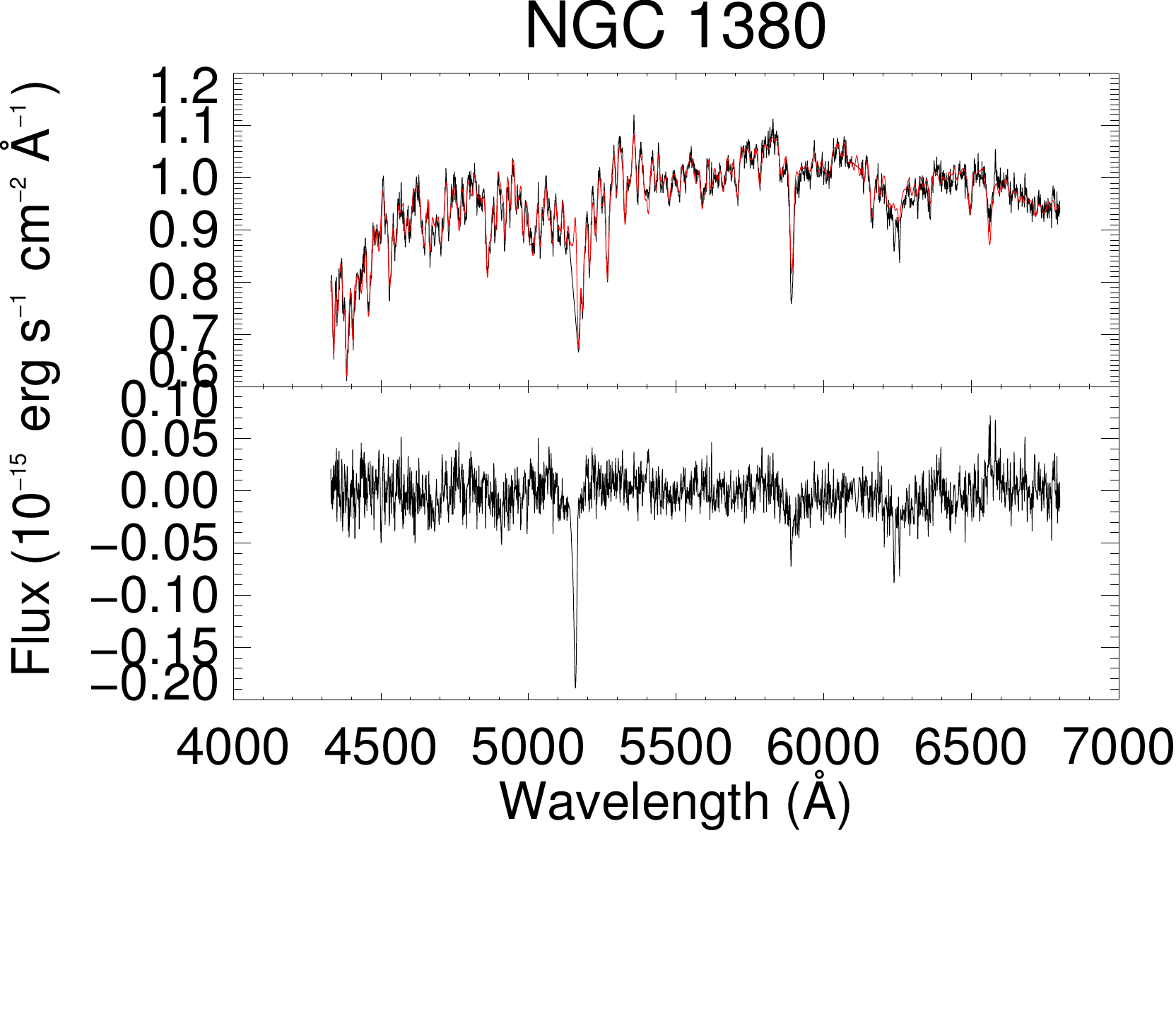}

\caption{Left: results from a representative spaxel of the central region (high S/N ratio) of galaxies ESO 208 G-21, IC 1459, IC 5181 and NGC 1380. Right: results from a spaxel located at the upper-right position of the FOV, $\sim$ 2 arcsec away from the galaxies' centre (low S/N ratio). For each galaxy, we show the spectrum (black) and the fitted result (red) at the top and the residuals at the bottom. \label{starlight_graf_1}
}
\end{center}
\end{figure*}

\addtocounter{figure}{-1}
\addtocounter{subfigure}{1}

\begin{figure*}
\begin{center}
\includegraphics[width=70mm,height=55mm]{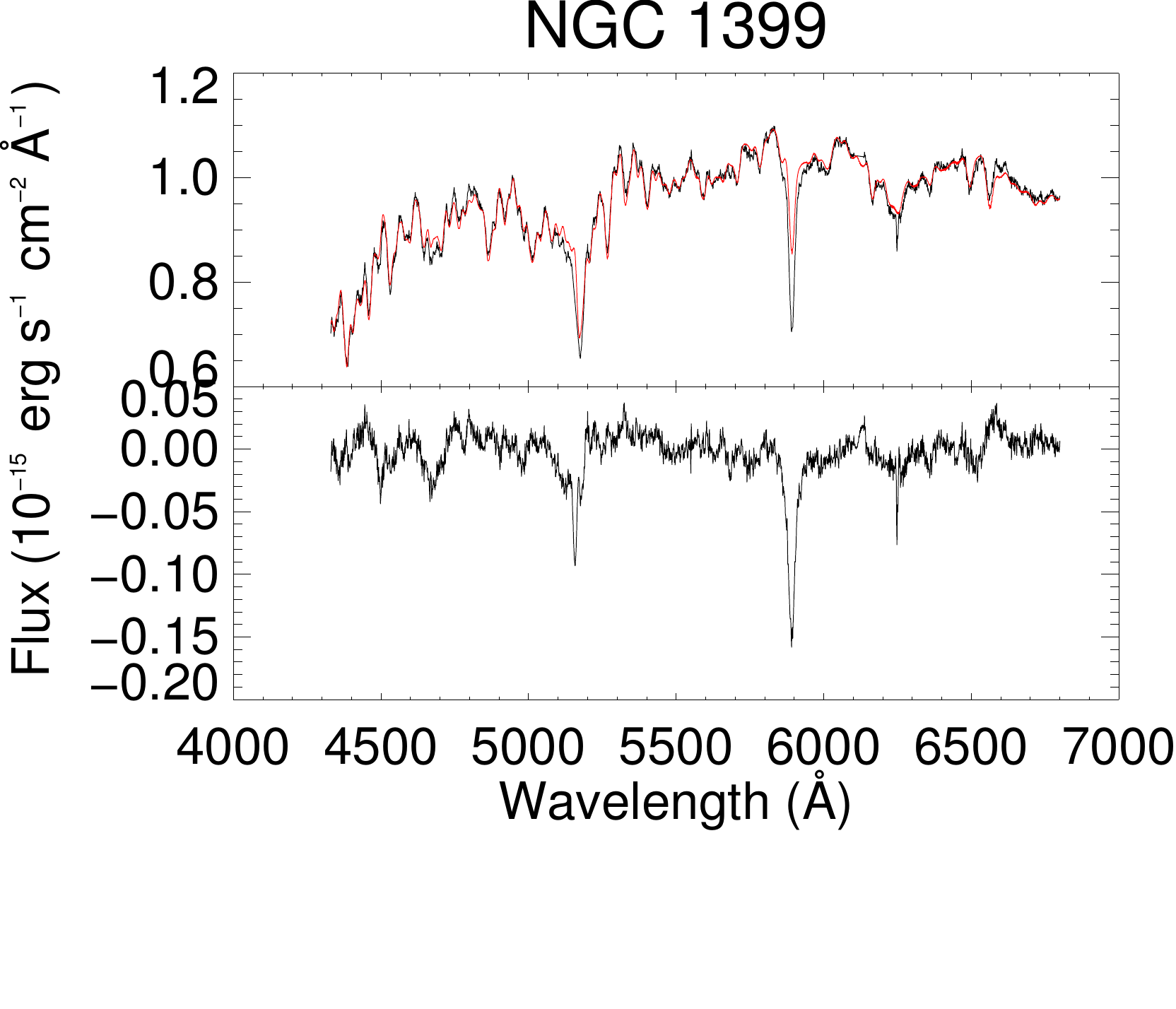}
\includegraphics[width=70mm,height=55mm]{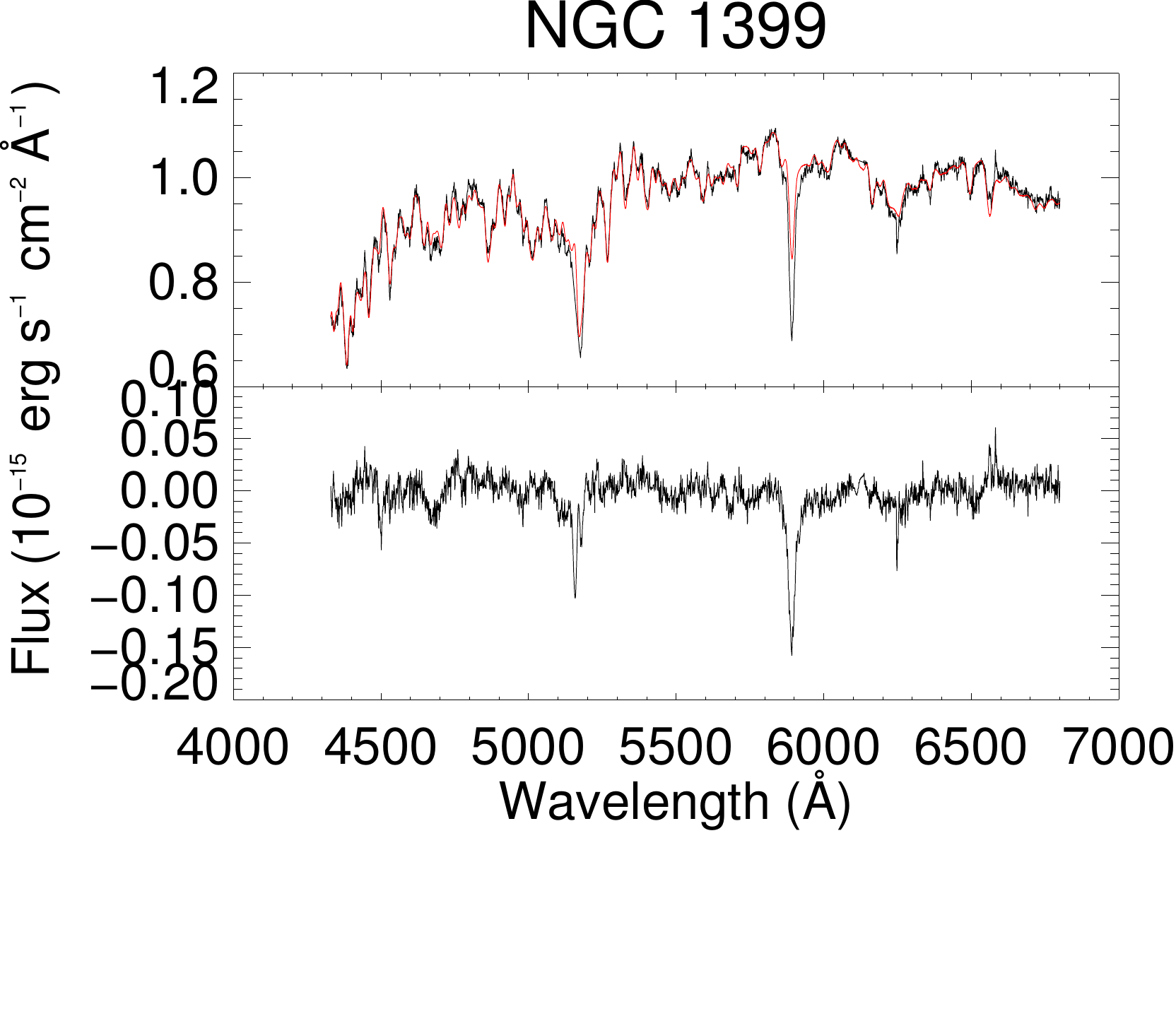}

\includegraphics[width=70mm,height=55mm]{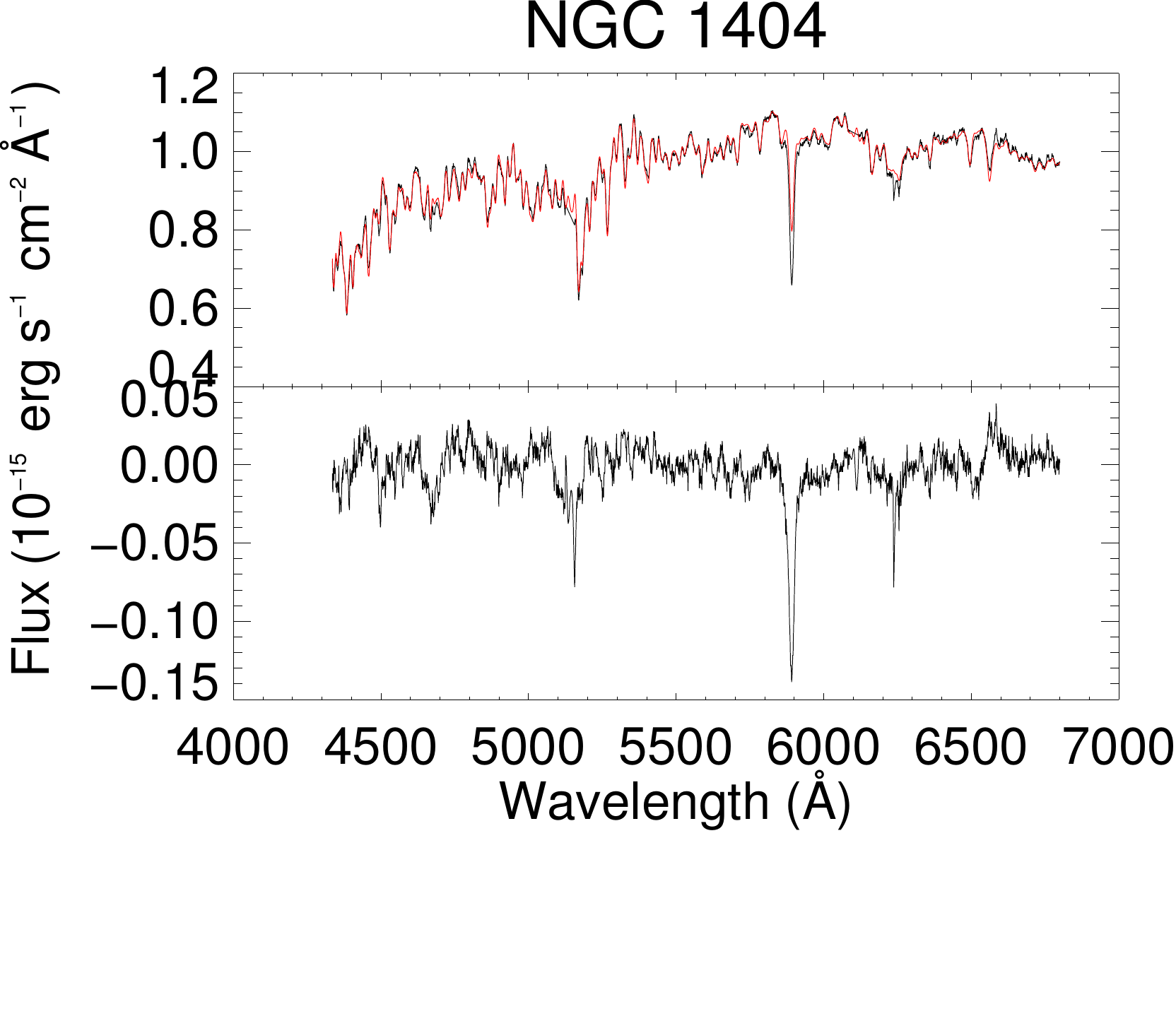}
\includegraphics[width=70mm,height=55mm]{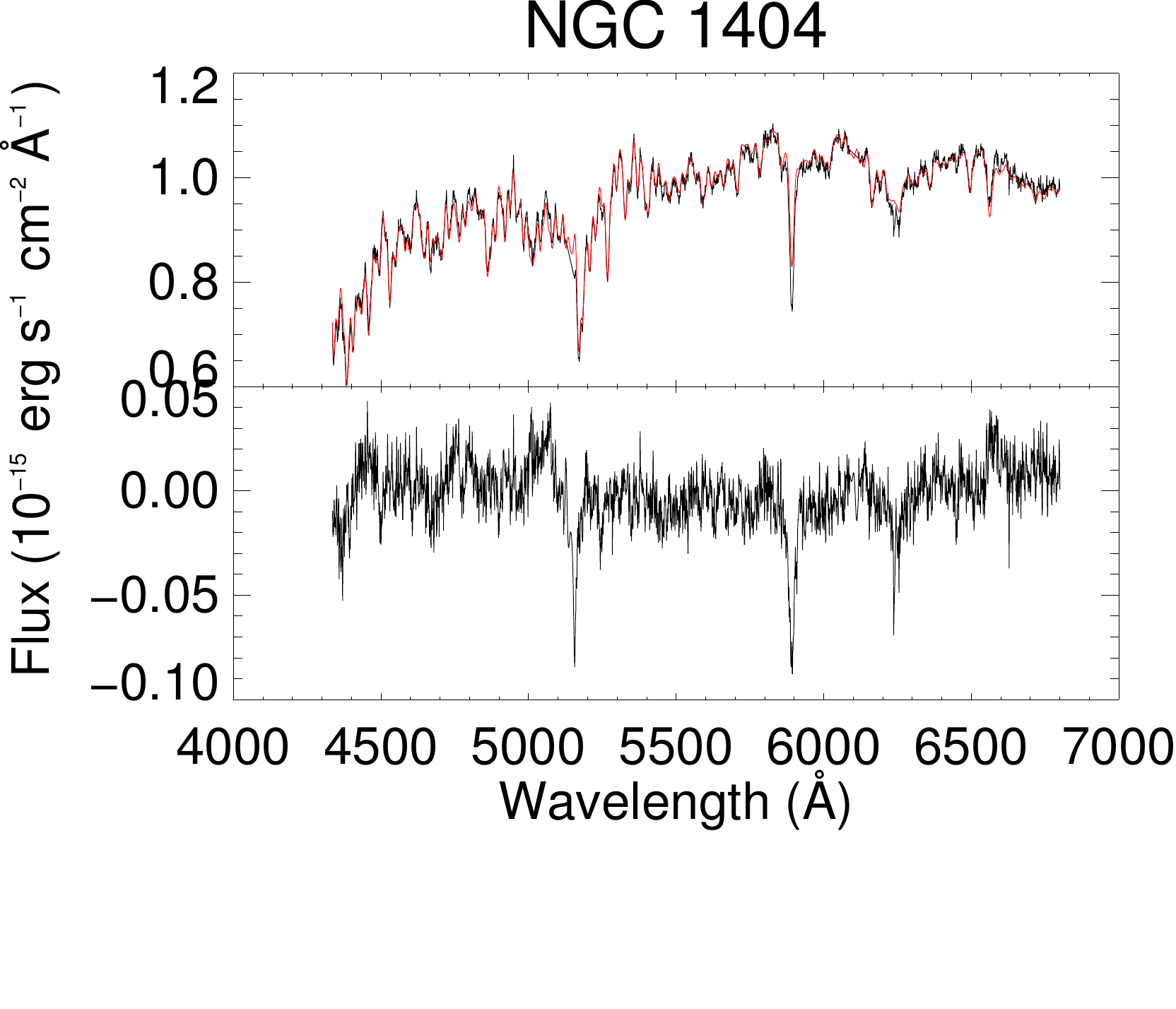}
\includegraphics[width=70mm,height=55mm]{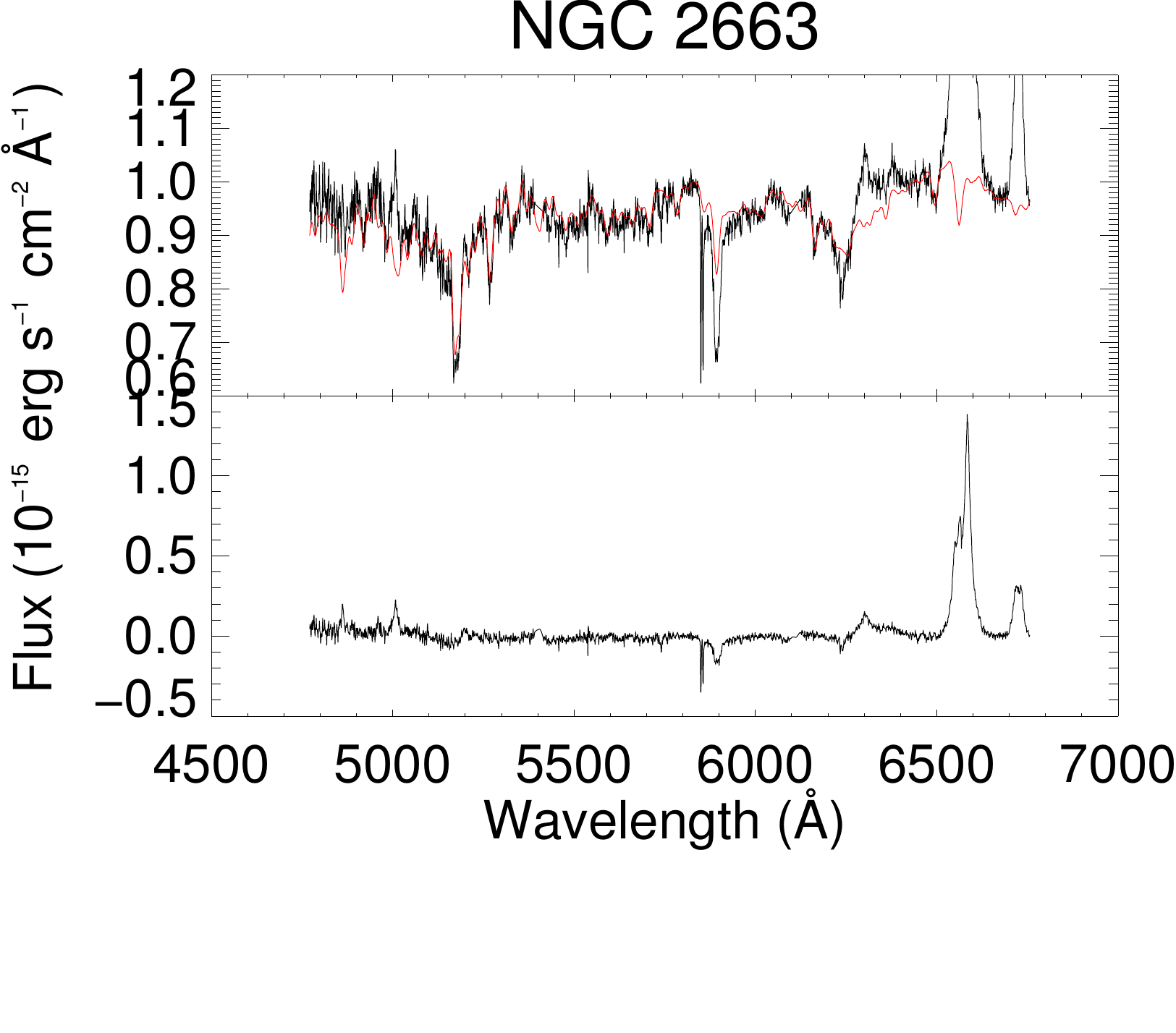}
\includegraphics[width=70mm,height=55mm]{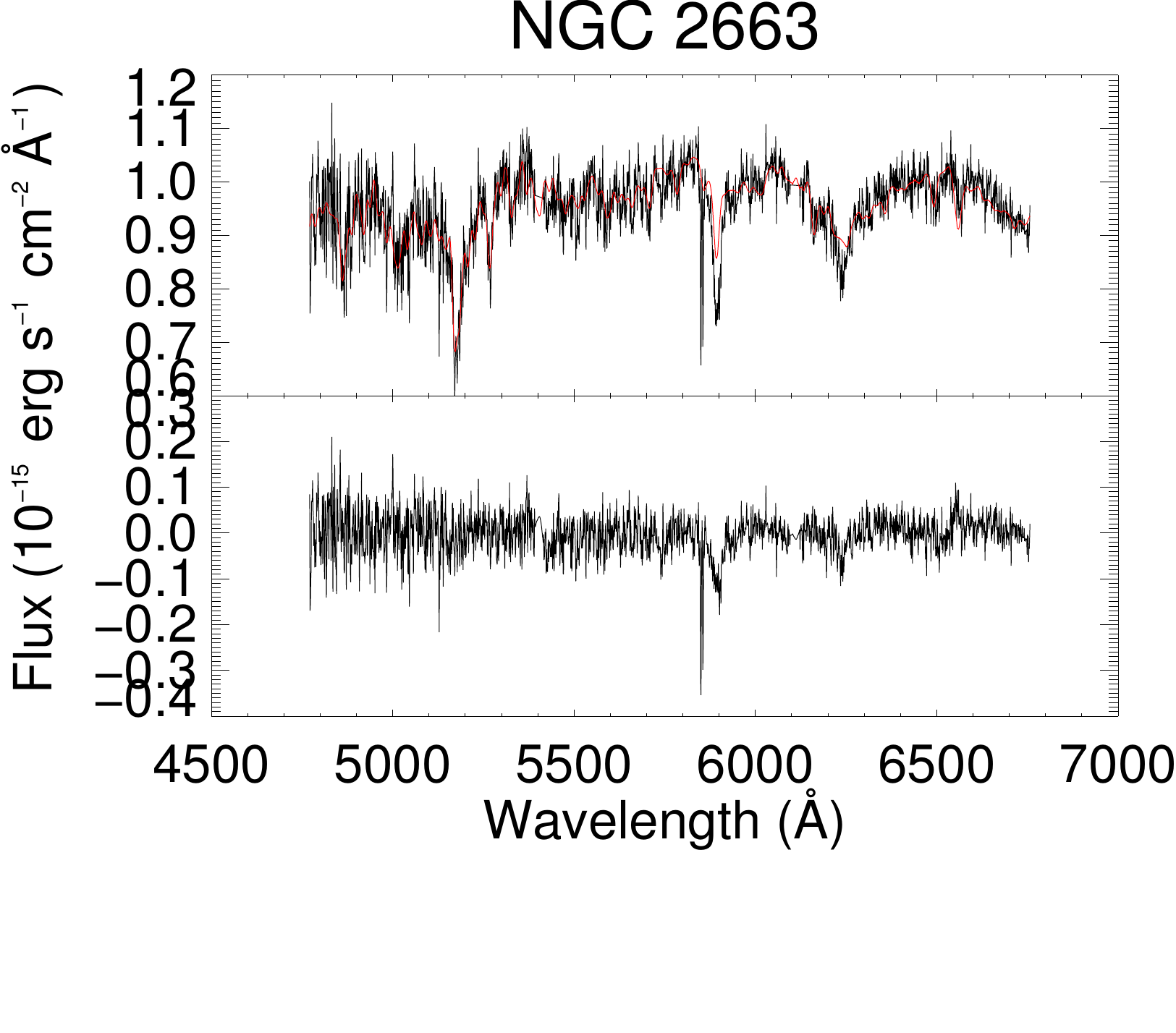}

\caption{The same as in Fig. \ref{starlight_graf_1} but for galaxies NGC 1399, NGC 1404 and NGC 2663.  \label{starlight_graf_2}
}
\end{center}
\end{figure*}

\addtocounter{figure}{-1}
\addtocounter{subfigure}{1}

\begin{figure*}
\begin{center}
\includegraphics[width=70mm,height=55mm]{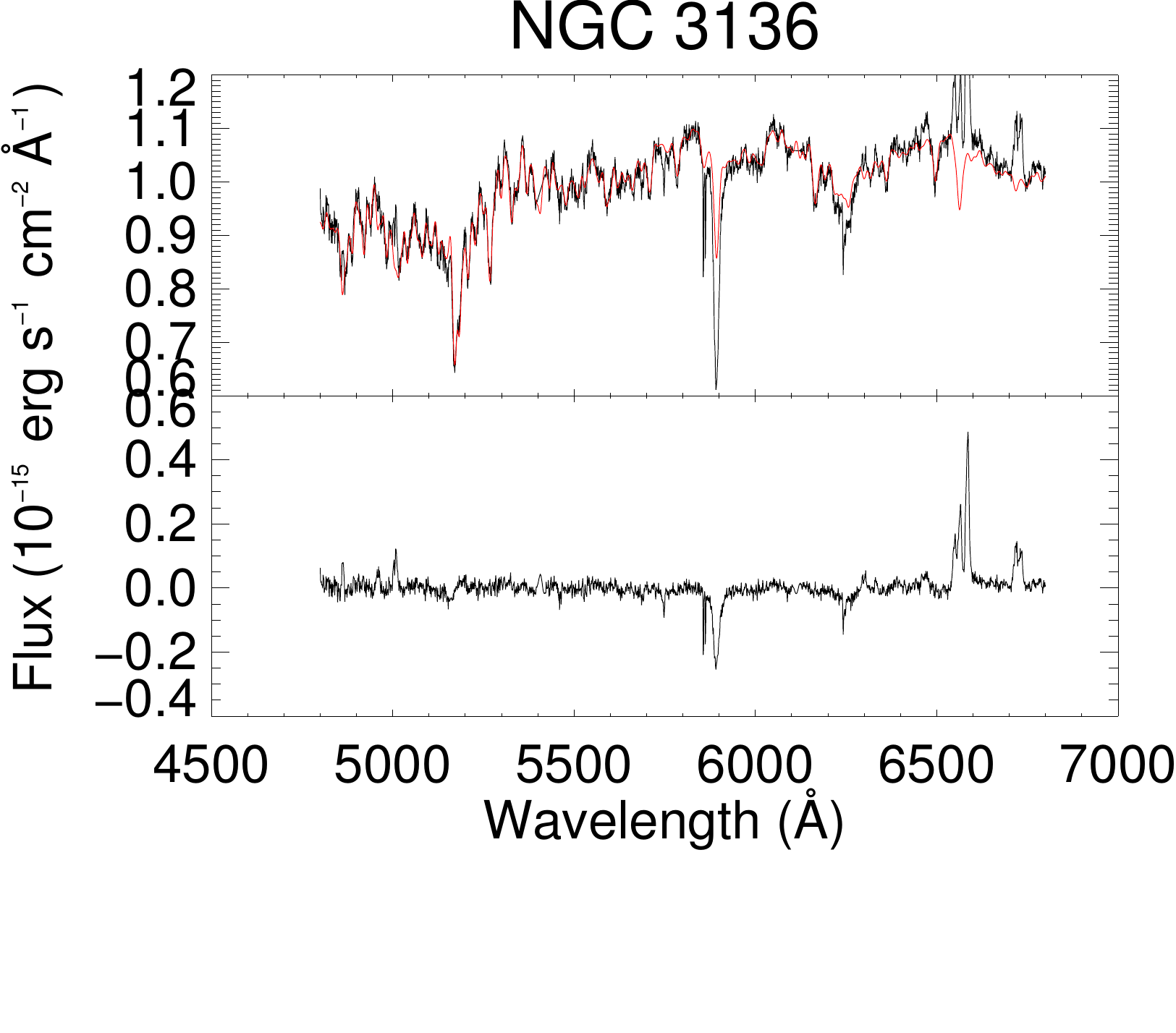}
\includegraphics[width=70mm,height=55mm]{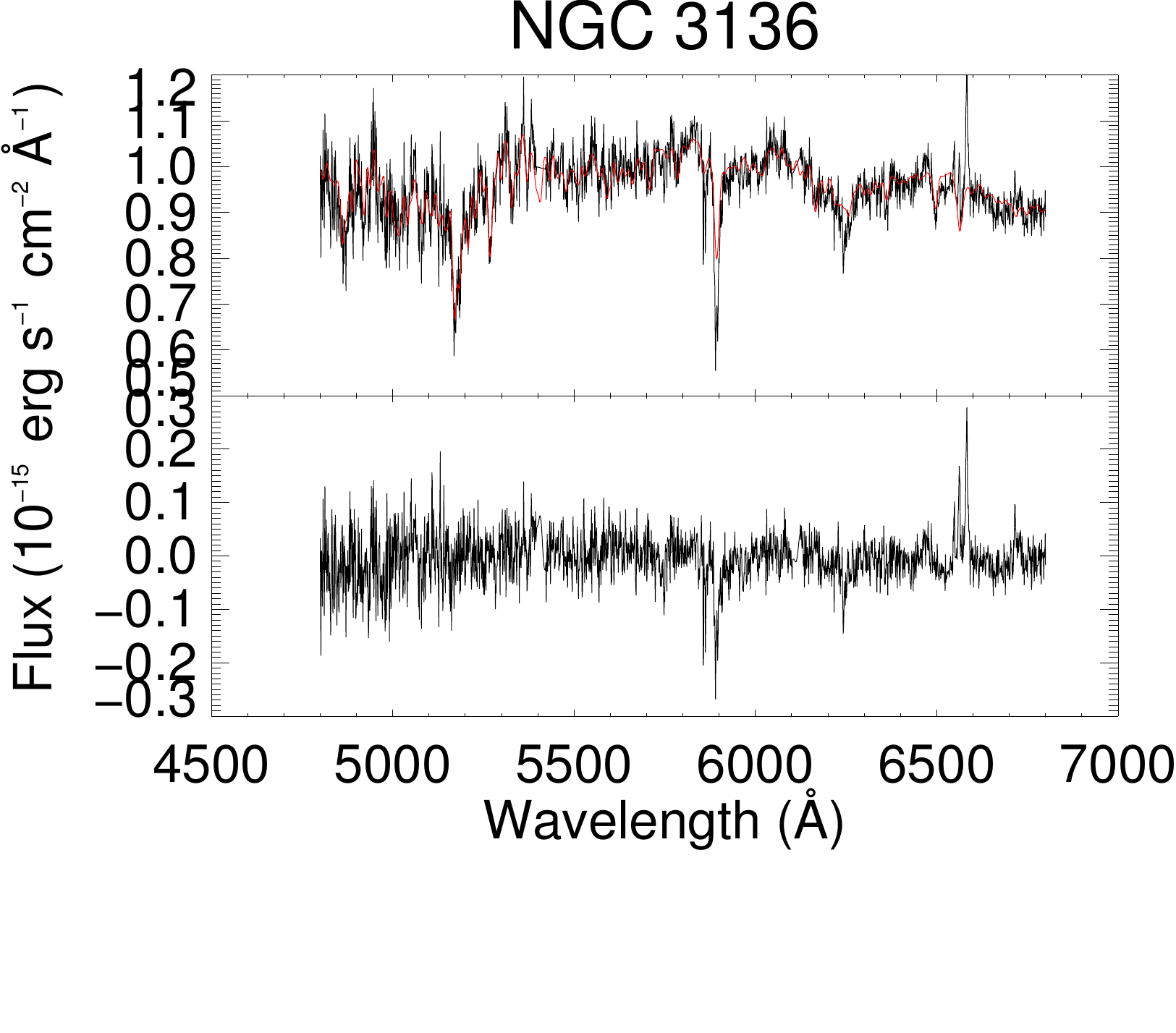}
\includegraphics[width=70mm,height=55mm]{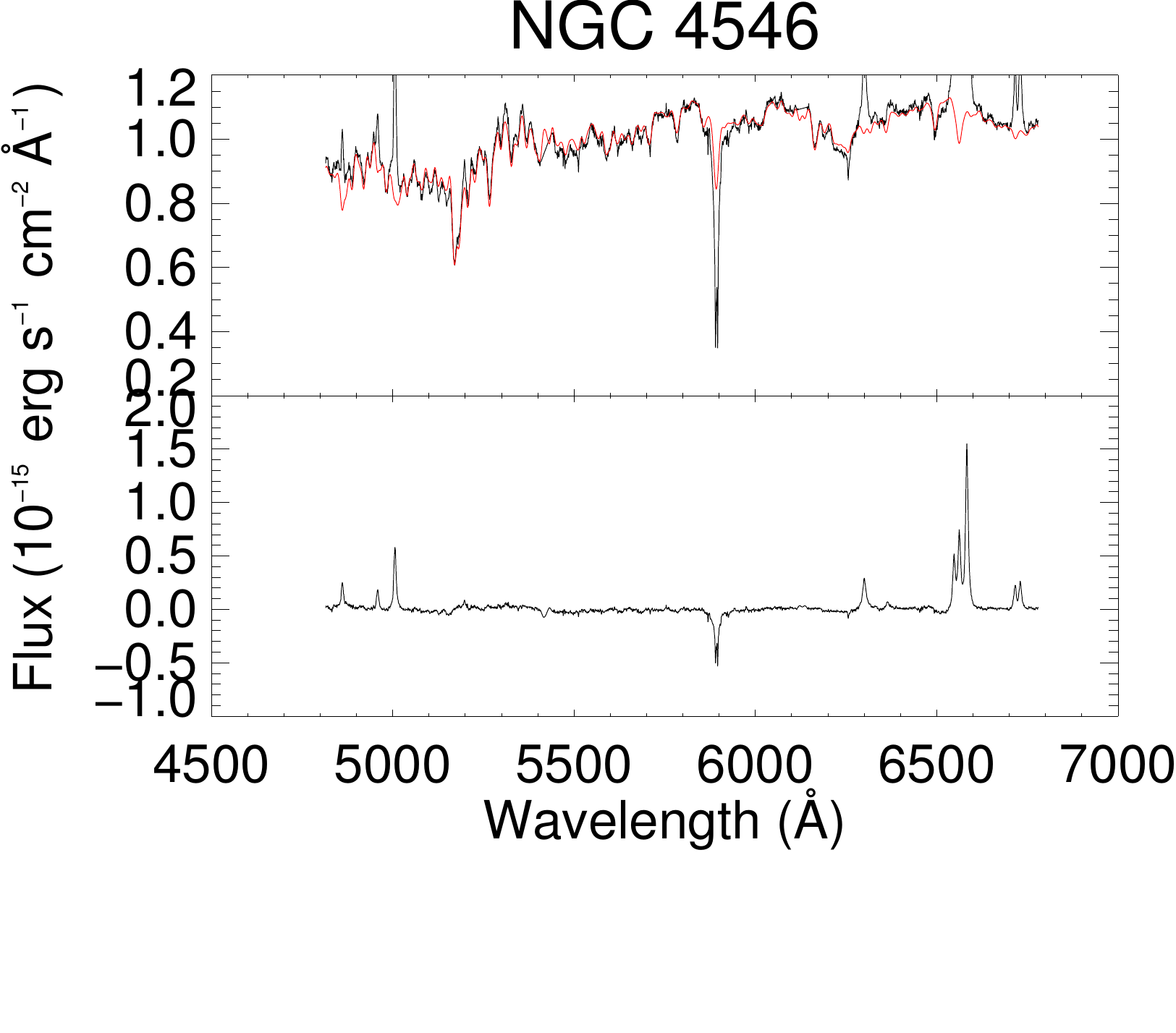}
\includegraphics[width=70mm,height=55mm]{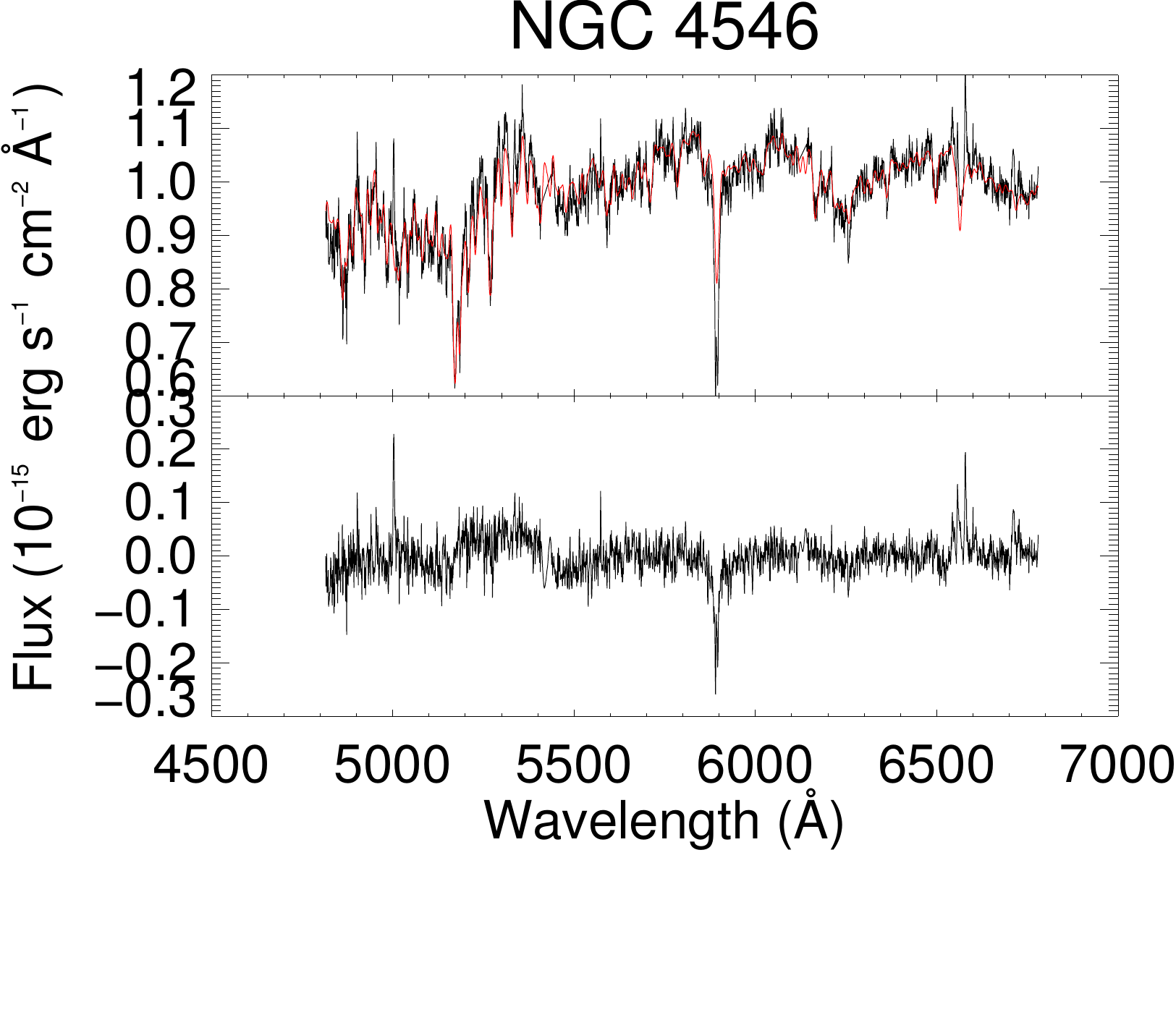}
\includegraphics[width=70mm,height=55mm]{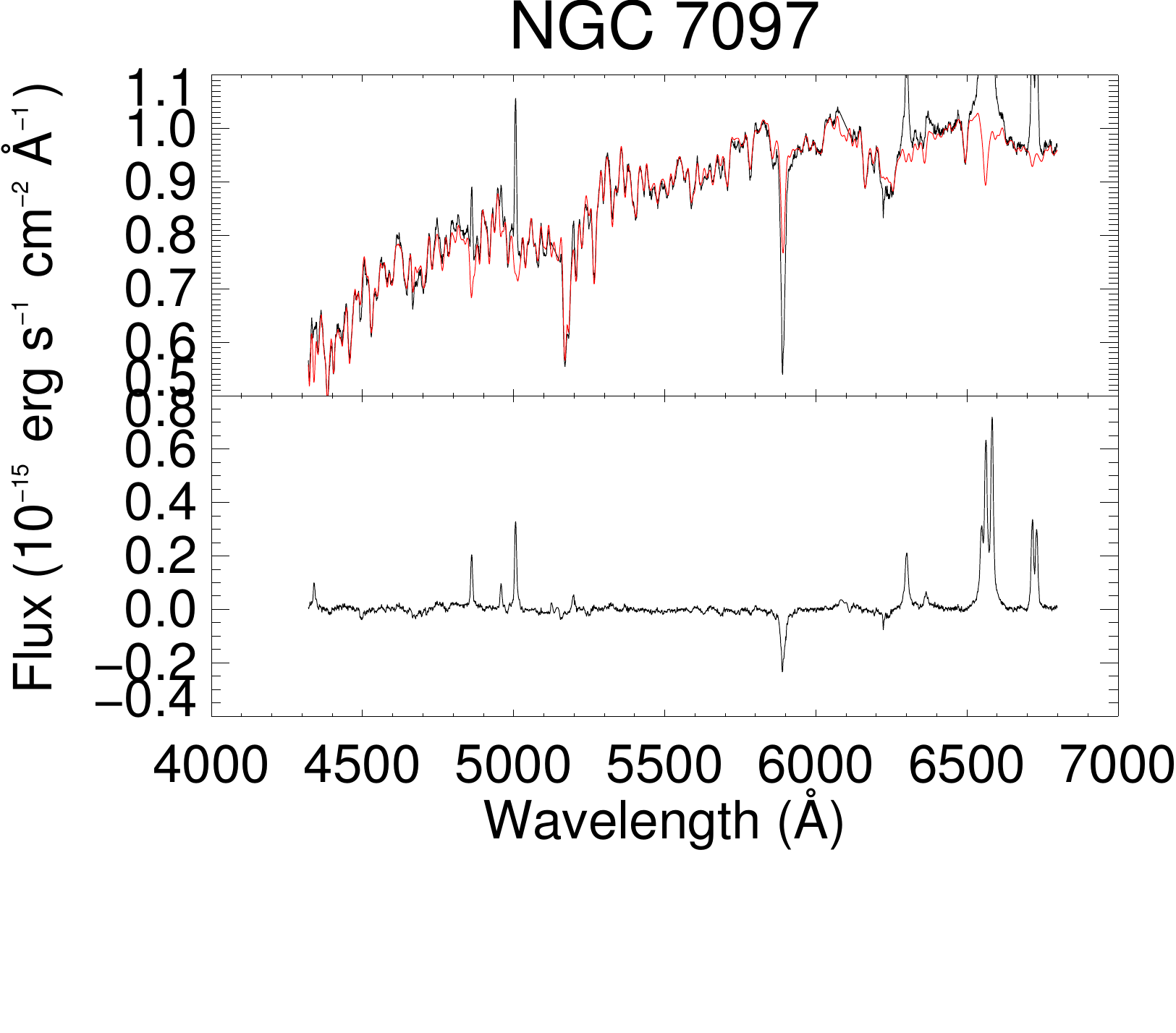}
\includegraphics[width=70mm,height=55mm]{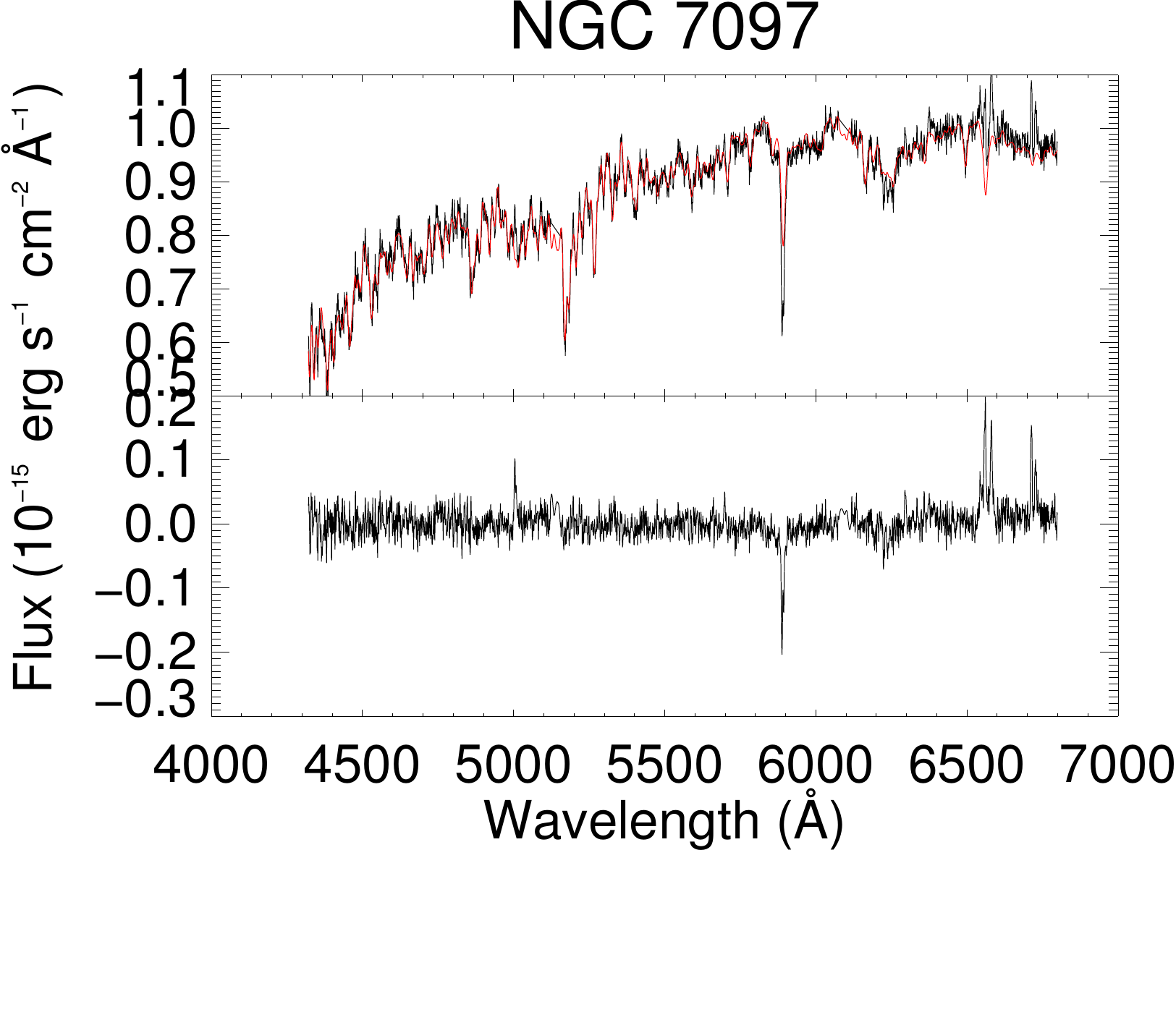}

\caption{The same as in Fig. \ref{starlight_graf_1} but for galaxies NGC 3136, NGC 4546 and NGC 7097  \label{starlight_graf_3}
}
\end{center}
\end{figure*}

\renewcommand{\thefigure}{\arabic{figure}}

Some issues may be noticed in both high and low S/N spectra. Since GMOS-IFU observations are made with three CCDs, gaps between them contain no information at all. In basic reduction procedures, a simple interpolation is performed to avoid a `hole' in the spectra of the data cubes. Therefore, no information is contained in these spectral regions. For the objects from the Gemini programme GS-2008A-Q-51 (see Table 2 in Paper I), these gaps may be seen in $\sim$ 5400 and in $\sim$ 6150 \AA. Galaxies observed under the Gemini programme GS-2008B-Q-21 (see Table 2 in Paper I) have their gaps in $\sim$ 5150 and in $\sim$ 6100 \AA. Note that the gaps produce features in the residuals that resemble absorption lines. Moreover, the spectral synthesis procedures does not fit well the Mg absorption lines, which are very sensitive to $\alpha$-enhancement effects, in the region around 5150 \AA. However, this mismatch may be caused by the fact that one of the gaps is right on the Mg absorption (see that the worst cases, namely IC 1459, IC 5181, NGC 1380, NGC 1399 and NGC 1404, were observed under the Gemini programme GS-2008B-Q-21). A bad subtraction in this spectral region affects the [N I]$\lambda\lambda$5198, 5200 emission lines. Combined to the weakness of the [N I] emission, this effect is responsible for the higher uncertainties estimated for the lines. Another systematic feature present in all residual spectra is associated with the [Na I]$\lambda\lambda$5890, 5896 absorption lines. Although a fraction of these lines may be related to the stellar populations, they actually trace the interstellar medium as well, something that is not taken into account by the spectral synthesis. It is important to mention that all these regions were properly masked for the spectral synthesis procedures. 

Another point is that the galaxies may contain stellar populations with ages $<$ 3 Gyr. In order to quantify if young stellar populations play a role in the sample galaxies, we performed the stellar synthesis in all spectra of the data cubes with a SSP library based on the \citet{2003MNRAS.344.1000B} models. This library has stellar populations with ages from 10$^6$ up to 10$^{10}$ yr and metallicities from 0.004 up to 0.05. One of the reasons for why we did not use this library for the analysis of the nuclear spectra is that this library does not take into account the $\alpha$ enhancement process. Also, when fitting the spectra that are affected by the featureless continuum, a young population is inserted by the {\sc starlight} code in order to find a better solution for the overall spectrum. However, when a young population is inserted to compensate the effects of the featureless continuum, the H$\beta$ absorption line is overestimated. Thus, a subtraction of the stellar population from the nuclear spectra will produce an H$\beta$ emission that is higher than the real value. In all cases, we obtained H$\alpha$/H$\beta$ < 3.1 when analysing the nuclear spectra obtained after the starlight subtraction with the \citet{2003MNRAS.344.1000B} models. Notwithstanding, this procedure allowed us to infer the presence of young stellar populations in the sample galaxies. In ESO 208 G-21, IC 1459, IC 5181, NGC 1380, NGC 3136, NGC 4546 and NGC 7097, the light fraction of populations with ages $\sim$ 6.3$\times$10$^6$ yr, considering all spectra from the data cubes, is about 5\%. This result may be somehow related to the featureless continuum but we suspect that it is more likely associated with the stellar spectral basis that does not reproduce the blue emission in ETGs \citep{2011MNRAS.413.1687C}. Only for NGC 2663, the light fraction is $\sim$ 30\%, which indicates that this object may have a young stellar population in its central region. A more detailed analysis of the stellar archaeology of this sample will be published in a forthcoming paper.

Individually, we call the attention to three objects. In NGC 1399 and NGC 1404, their residual spectra, shown in Fig. \ref{starlight_graf_2}, show a bump around H$\alpha$. As discussed in section \ref{cD_cases}, this feature may mimic a broad component. The case of NGC 2663 is more critical. In the spectral region between 6200 and 6750\AA, the resulting stellar population did not fit adequately the observed spectra. Therefore, we decided to run the {\sc starlight} code twice for this galaxy's data cube (see section \ref{dados_PaperII} for more details). In Fig. \ref{starlight_graf_2}, we combined the results from both runs in each spectrum (high and low S/N ratio). Note that the fitted spectrum seems to agree with the stellar components, specially in the H$\alpha$ and [S II] lines. 

\end{document}